\documentclass[a4paper,11pt]{article}
\pdfoutput=1

\usepackage{jheppub}

\usepackage[T1]{fontenc}
\usepackage{xcolor}
\usepackage[small]{subfigure}
\usepackage[normalem]{ulem}
\usepackage{amsmath}
\usepackage{hyperref}
\usepackage[all]{hypcap}
\usepackage{mathtools}
\usepackage[small]{subfigure}
\usepackage{orcidlink}

\usepackage{tabularx}

\usepackage{mathrsfs}
\usepackage{booktabs}

\usepackage{mwe}
\newcommand{\imineq}[2]{\vcenter{\hbox{\includegraphics[height=#2ex]{#1}}}}

\newcommand*{\myalign}[2]{\multicolumn{1}{#1}{#2}}

\newcommand*\pFqskip{8mu}
\catcode`,\active
\newcommand*\pFq{\begingroup
\catcode`\,\active
\def ,{\mskip\pFqskip\relax}%
\dopFq}
\catcode`\,12
\def\dopFq#1#2#3#4#5{%
{}_{#1}F_{#2}\biggl[\genfrac..{0pt}{}{#3}{#4};#5\biggr]%
\endgroup}

\newcommand{\nb}[0]{{\bar{n}}}
\newcommand{\nl}{n_{\ell}}

\newcommand{\MSb}{\overline{\mathrm{MS}}}

\newcommand{\LQCD}{\Lambda_\mathrm{QCD}}

\newcommand{\dd}{\mathrm{d}}

\newcommand{\bra}[1]{\left\langle #1 \right|}
\newcommand{\ket}[1]{\left| #1 \right\rangle}
\newcommand{\eps}{\varepsilon}
\newcommand{\cB}{\mathcal{B}}
\newcommand{\ord}{\mathcal{O}}
\newcommand{\bare}{\mathrm{bare}}
\newcommand{\nn}{\nonumber}
\newcommand{\as}{\alpha_s}
\newcommand{\tr}{\mathrm{tr}}
\newcommand{\rd}{\mathrm{d}}
\newcommand{\rc}{\mathrm{c}}
\newcommand{\cL}{\mathcal{L}}
\newcommand{\Gamt}{\Gamma_{\!t}}
\newcommand{\mathd}{\mathrm{d}}

\newcommand{\MI}[2]{\text{MI}^\text{#1L}_{#2}}

\DeclareRobustCommand{\eq}[1]{Eq.~\eqref{eq:#1}}
\DeclareRobustCommand{\eqs}[2]{Eqs.~\eqref{eq:#1} and \eqref{eq:#2}}
\DeclareRobustCommand{\eqsm}[2]{Eqs.~\eqref{eq:#1}\,--\,\eqref{eq:#2}}
\DeclareRobustCommand{\eqsss}[3]{Eqs.~\eqref{eq:#1}, \eqref{eq:#2} and \eqref{eq:#3}}
\DeclareRobustCommand{\fig}[1]{Fig.~\ref{fig:#1}}

\DeclareRobustCommand{\sec}[1]{Sec.~\ref{sec:#1}}

\DeclareRobustCommand{\subsec}[1]{Sec.~\ref{subsec:#1}}
\DeclareRobustCommand{\subsecs}[2]{Secs.~\ref{subsec:#1} and \ref{subsec:#2}}
\DeclareRobustCommand{\app}[1]{Appendix~\ref{app:#1}}
\DeclareRobustCommand{\apps}[2]{Appendices~\ref{app:#1} and \ref{app:#2}}
\DeclareRobustCommand{\rcite}[1]{Ref.\,\cite{#1}}
\DeclareRobustCommand{\rcites}[1]{Refs.\,\cite{#1}}
\DeclareRobustCommand{\tab}[1]{Table~\ref{tab:#1}}

\newcommand{\MIone}{d}
\newcommand{\MItwo}{e}
\newcommand{\MIthree}{f}
\newcommand{\MIfour}{g}
\newcommand{\MIfive}{c}
\newcommand{\MIsix}{m}
\newcommand{\MIseven}{b}
\newcommand{\MIeight}{l}
\newcommand{\MInine}{h}
\newcommand{\MIten}{i}
\newcommand{\MItone}{j}
\newcommand{\MIttwo}{r}
\newcommand{\MItthree}{o}
\newcommand{\MItfour}{k}
\newcommand{\MItsix}{q}
\newcommand{\MItseven}{a}
\newcommand{\MIteight}{p}
\newcommand{\MItnine}{n}
\newcommand{\MItwenty}{s}
\newcommand{\MIttone}{t}

\title{Three-loop jet function for boosted heavy quarks}

\preprint{\begin{flushright} FR-PHENO-2024-008\end{flushright}\vspace*{-2cm}}

\author[a]{Alberto M.~Clavero\orcidlink{0009-0004-7406-4776},}
\author[b]{Robin Br\"user,}
\author[a]{Vicent Mateu\orcidlink{0000-0003-0902-5012}}
\author[b]{and Maximilian Stahlhofen\orcidlink{0000-0002-2613-9014}}

\affiliation[a]{Departamento de F\'isica Fundamental e IUFFyM,\\Universidad de Salamanca, E-37008 Salamanca, Spain}
\affiliation[b]{Albert-Ludwigs-Universit\"at Freiburg, Physikalisches Institut, D-79104 Freiburg, Germany}

\emailAdd{albertomarcla@usal.es}
\emailAdd{Robin.Brueser@physik.uni-freiburg.de}
\emailAdd{vmateu@usal.es}
\emailAdd{maximilian.stahlhofen@physik.uni-freiburg.de}

\abstract{We compute the inclusive jet function for boosted heavy quarks
to $\mathcal{O}(\alpha_s^3)$. The jet function is defined and calculated in the framework of boosted Heavy-Quark Effective Theory (bHQET). It describes the effect of radiation collimated in narrow jets arising from energetic heavy quarks on observables probing the jet invariant mass $M$
in the region where $M^2 - m^2 \ll m^2$, with $m$ the heavy quark mass. This kinematic situation is relevant e.g.\ in boosted top (pair) production at high-energy colliders.
We have verified that our result satisfies non-Abelian exponentiation and checked that our calculation reproduces the known cusp and non-cusp anomalous dimensions of the jet function to $\mathcal{O}(\as^3)$. We also confirmed that the $n_\ell^2\as^3$ contribution, where $n_\ell$ is the number of massless quark flavors, agrees with the prediction from renormalon calculus.
Our computation provides the last missing piece to obtain the N$^3$LL$^\prime$ resummed (self-normalized) thrust distribution used for the calibration of the top quark mass parameter in parton-shower Monte Carlo generators. Our result also contributes to the invariant mass distribution of reconstructed top quarks at N$^3$LL$^\prime$, which can be employed for a precise top mass determination at future lepton colliders. As a by-product, we obtain the relation between the pole and short-distance jet-mass schemes at $\mathcal{O}(\alpha_s^3)$. Finally, we estimate the non-logarithmic contribution to the four-loop jet function based on renormalon dominance.}

\begin{document}
\maketitle
\flushbottom

\section{Introduction}\label{sec:intro}
The top quark is the heaviest fundamental particle discovered so far, and plays a central role in testing the validity of the Standard Model (SM). Due to its large mass, and the correspondingly large (Yukawa) coupling to the Higgs field in the SM, the top quark critically affects the electroweak vacuum. In fact, the precise value of its mass determines whether the electroweak vacuum is stable or not, see \rcites{Degrassi:2012ry,Buttazzo:2013uya,Branchina:2013jra,Bednyakov:2015sca,Andreassen:2017rzq}. Moreover, the uncertainty of the top mass substantially affects the uncertainties of (indirect) determinations of other SM precision parameters like the $W$ or Higgs masses, see e.g.\ \rcite{Haller:2018nnx}. A precise top mass is therefore crucial for many consistency checks of the SM.

Since the lifetime of a top quark is so short that it decays before it can hadronize, many experimental analyses try to determine its mass using kinematic information from its decay products, as customarily done for resonances. Such so-called direct measurements of the top quark mass have been employed at the Tevatron and the LHC
reaching a precision of $300\,$MeV~\cite{ATLAS:2024dxp}, and prospects indicate that this uncertainty can be further reduced, see e.g.\ \rcite{Schwienhorst:2022yqu}. The mass determined by direct measurements does however neither correspond to the on-shell (or pole) mass nor to any other well-defined scheme within quantum field theory. It is rather the mass parameter inherent to
the parton-shower Monte Carlo used to generate the distributions that are compared to the experimental data.
This mass parameter is dubbed $m_t^{\rm MC}$ and, according to \rcite{Hoang:2020iah}, appears to be to some extent related to the MSR mass $m_t^{\rm MSR}(R\!=\!1\,{\mathrm{GeV}})$~\cite{Hoang:2008yj,Hoang:2017suc}. A quantitative relation between $m_t^{\rm MC}$ and $m_t^{\rm MSR}$ is the purpose of the calibration studies carried out in \rcites{Butenschoen:2016lpz,Dehnadi:2023msm}. There are also indirect top mass measurements at the LHC, see e.g.\ \rcite{ATLAS:2019guf} (based on the strategy devised in \rcite{Alioli:2013mxa}), but they are less precise and typically
utilize the pole scheme for the top quark mass, which is afflicted by an $\ord(\LQCD)$
renormalon ambiguity~\cite{Smith:1996xz}.

Event shapes at $e^+e^-$ colliders have been extensively used to explore the gauge structure of the strong interactions, tune Monte Carlo generators, learn about hadronization, and determine the strong coupling $\as$ with high precision. The theoretical predictions for massless quarks including the resummation of large logarithms have reached an impressive level of precision, often primed next-to-next-to-next-to-leading logarithmic (N$^3$LL$^\prime$) and in some cases even N$^4$LL accuracy.\footnote{The primed counting of logarithmic accuracy implies that at N$^n$LL$^\prime$ ($n\geq 1$) the fixed-order ingredients to the corresponding factorized cross section are included at N$^n$LO, and not only at N$^{n-1}$LO as for N$^n$LL.} For $e^+e^-$ event shapes with massive quarks in the Born-level final state (primary), and/or in the real and virtual corrections (secondary), the computations have reached the two-loop level, enabling N$^3$LL precision, see \rcites{Fleming:2007qr,Fleming:2007xt,Gritschacher:2013tza,Pietrulewicz:2014qza,Hoang:2019fze,Bris:2020uyb,Bachu:2020nqn,Bris:2024bcq}.
Fixed-order QCD results at $\mathcal{O}(\alpha_s)$ and $\mathcal{O}(\alpha_s^2)$ have also been obtained in \rcites{Nason:1997nw,Bernreuther:1997jn,Brandenburg:1997pu,Rodrigo:1999qg,Lepenik:2019jjk}.

In the case of the production of boosted primary top-quark pairs, it was shown in \rcites{Fleming:2007qr, Fleming:2007xt} that using a class of event shapes related to the hemisphere invariant masses the top mass (in a suitable short-distance scheme) can be determined with an uncertainty smaller than $\Lambda_{\mathrm{QCD}}$. Here, hemispheres are defined with respect to the plane normal to the thrust axis, and the momenta of all particles within a hemisphere are taken into account to compute its invariant mass. The maximal sensitivity to the top mass is obtained in the peak region of the hemisphere mass distribution.
In this kinematic region the highly energetic top quarks can be described using two boosted copies of Heavy Quark Effective Theory~\cite{Manohar:2000dt} (denoted as bHQET) matched to (massive) Soft Collinear Effective Theory (SCET)~\cite{Bauer:2000ew, Bauer:2000yr, Bauer:2001yt, Bauer:2002nz, Bauer:2001ct, Beneke:2002ph}. This effective field theory (EFT) framework allows to resum large logarithms,%
\footnote{For the production of a pair of (very) boosted tops with large invariant mass (of the top pair) combined soft-gluon and small-mass resummation can be performed as demonstrated in the context of $pp$ collisions in \rcite{Pecjak:2016nee}. That kinematic regime is however different from the one considered in the present work, where we consider the peak region of the invariant mass distribution of a single top jet.} %
consistently include the decay width of the top, and account for soft hadronization power corrections from first principles. In particular, the corresponding event-shape cross sections factorize at leading order (``leading power'') in the EFT expansion, i.e.\ the systematic expansion in powers of $m/Q$, $\Gamt/m$, and $\hat{s}/m$, where $Q$ is the center-of-mass energy of the collision, $m$ is the top mass, $\Gamt$ is the top width, and $\hat{s} \sim \Gamt \gg \LQCD$ represents a measure for the offshellness of the top quarks, cf.\ \eq{shat}. In this regime, the momentum of the heavy quark is parametrized as $p=mv+k$, where $v^2=1$ and $k$ denotes the residual momentum, which in the heavy quark's rest frame scales like $k^\mu \sim\Gamma_{\!t} \ll m$. Quantum modes with momenta scaling like $p_c^\mu \sim k^\mu$ represent dynamic degrees of freedom in the EFT and are referred to as ultra-collinear. Boosting to the center-of-mass (c.o.m.) frame generates a strong hierarchy among their light-cone momentum components, see \subsec{JF_def}. Quantum modes with such virtuality define the two jet functions. There is also genuinely soft wide-angle radiation, with momenta which already in the c.o.m.\ frame scale like $p_s^\mu \sim m\Gamt/Q\ll \Gamt$. Modes with such virtuality define the soft function.

The prototype of a (dijet) factorization theorem (valid to all orders in $\as$)
involving these jet and soft functions
is the one for the double hemisphere invariant mass differential distribution~\cite{Fleming:2007qr}:
\begin{align}
\frac{\mathd^2 \sigma^{(\mathrm{dijet})}}{\mathd M_1^2 \mathd M_{2}^2} ={}& \sigma_0\, H^{(n_\ell+1)}_Q\!\bigl(Q, \mu\bigr)\, H_m\biggl(m, \frac{Q}{m}, \mu\biggr)\!
\int \! \mathd \ell^{+} \mathd \ell^{-} \,
S^{(n_\ell)}\bigl(\ell^{+}, \ell^{-}, \mu\bigr)
\label{eq:double_hemisphere_invariant} \\
&\times
B^{(n_\ell)}_n\!\biggl(\frac{M_1^2-Q \ell^{+}}{m}-m, \Gamt, \mu\!\biggr)\,
B^{(n_\ell)}_{\bar n}\!\biggl(\frac{M_2^2-Q \ell^{-}}{m}-m, \Gamt, \mu\!\biggr).\nn
\end{align}
This formula (valid for $\hat s \sim \Gamt\ll m$) contains elements of the corresponding SCET factorization theorem (valid for $\hat{s}\sim m$): the massless hard and soft functions, $H_Q$ and $S$, which are known at three and two loops, respectively~\cite{Matsuura:1987wt,Matsuura:1988sm,Gehrmann:2005pd,Moch:2005id,Baikov:2009bg,Lee:2010cg,Gehrmann:2010ue,Kelley:2011ng,Monni:2011gb,Hornig:2011iu}. This hard function arises from integrating out the hard modes scaling like $k^\mu\sim Q \gg m$. The 0-jettiness projection of the soft function, which suffices to obtain the thrust cross section, has recently been computed to three-loop order in \rcite{Baranowski:2024vxg}. The two functions $B_n$ and $B_{\bar n}$ are the bHQET jet functions describing each one of the back-to-back boosted heavy (top) quarks as well as the collinear radiation accompanying them.
The subscript of the jet functions indicates the respective heavy (anti)quark jet direction. By charge conjugation, the two jet functions in \eq{double_hemisphere_invariant} are equal, and we denote $B\equiv B_n=B_{\bar n}$. It is the main aim of this work to compute this jet function at three loops, i.e.\ at $\ord(\as^3)$. The superscripts (in brackets) on the various factorization functions indicate the number of quark flavors treated as massless in their computation. In particular, the running of $\alpha_s$, which is eventually evaluated at the characteristic scale of each function in the resummed version of \eq{double_hemisphere_invariant}, is governed by this number. In the case of top pair production we have $n_\ell=5$. Only the hard mass mode matching factor $H_m$ lacks a superscript, since it lives at the top mass threshold and can therefore be expressed in terms of $\alpha_s$ evaluated with either $n_\ell$ or ($n_\ell+1$) active flavors without generating large logarithms. For better readability we will suppress this superscript in the following.

The matching factor $H_m$ encodes the effect of (virtual) soft [\,$p_m^s\sim (m,m, m)$\,] and collinear [\,$p_m^n\sim (m^2/Q,Q, m)$, $p_m^{\bar n}\sim (Q, m^2/Q, m)$\,] mass-modes,%
\footnote{Here we refer to the light-cone momentum components $p^\mu=(p^+,p^-,p_\perp)$ as introduced in \subsec{JF_def}.} %
which have been integrated out as their virtuality scales like $m^2 \gg \Gamt^2$.
This function can be extracted from the leading term of the small-mass (high-energy) expansion of the
massive quark form factor known at three loops~\cite{Fael:2022miw}, see \rcites{Hoang:2015vua,Mitov:2006xs,Becher:2007cu}. The fixed-order expression for $H_m$ however contains large rapidity logarithms of $Q/m$. In the EFT approach the origin of these logarithms is explained by the fact that the relevant soft and collinear mass-modes live at roughly the same virtuality ($\sim m$), but are separated in rapidity. The corresponding factorization reads~\cite{Hoang:2015vua}
\begin{equation}
\label{eq:Hmfactors}
H_m\biggl(m,\frac{Q}{m},\mu\biggr) = H_{m,n}\biggl(m,\mu,\frac{Q}{\nu}\biggr) H_{m,\nb}\biggl(m,\mu,\frac{Q}{\nu}\biggr) H_{m,s}\biggl(m,\mu,\frac{m}{\nu}\biggr),
\end{equation}
where $\nu$ denotes the rapidity renormalization scale, and $H_{m,n}$, $H_{m,\nb}$ and $H_{m,s}$ represent the contributions from the (two types of) collinear and soft mass-modes, respectively.\footnote{For symmetric rapidity regulators, like the one of \rcite{Chiu:2012ir}, we have $H_{m,n}= H_{m,\nb}$.}
This factorized form enables the resummation of the large rapidity logarithms in $H_m$ via rapidity renormalization group equations~\cite{Chiu:2012ir}, which is currently known at the NNLL$^\prime$ level~\cite{Hoang:2015vua}.%
\footnote{Note that $H_{m,n} \times B_n$ represents the leading-power contribution of the SCET (``mass-mode'') quark jet function with full mass dependence in the small-$\hat{s}$ limit, see \rcite{Hoang:2019fze}.}

In bHQET, the heavy-quark (top) mass is no longer a dynamical scale and the theory naturally has $n_\ell$ $(=5)$ active quark flavors. It is however still crucial to define the bHQET jet function in a renormalon-free scheme. While the ($u=1/2$) renormalon contained in the soft function can be removed by gap subtractions, see \rcite{Hoang:2007vb}, there is yet another $\mathcal{O}(\Lambda_{\mathrm{QCD}})$ ambiguity hiding in $B$. Therefore, one should switch from the pole to a short-distance mass scheme which respects the power counting of the EFT. Although the jet-mass scheme~\cite{Jain:2008gb} constructed from the perturbative expression of $B$ was initially proposed, it has lately become more standard to use the MSR mass~\cite{Hoang:2008yj,Hoang:2017suc}.
The main reason is that it naturally generalizes the well-known $\MSb$ scheme, which allows to infer its relation to the pole mass at four~loops~\cite{Marquard:2015qpa,Marquard:2016dcn}.
Furthermore, its renormalization group (RG) evolution and relation to the $\MSb$ mass can be obtained, (along with the strong-coupling running and threshold matching) from the public code REvolver~\cite{Hoang:2021fhn}, making it a very convenient choice.
Nevertheless, given the resemblance of the jet-mass to the R-gap scheme initially proposed in \rcite{Hoang:2008fs} (and its variations discussed in \rcite{Bachu:2020nqn}), which is used to subtract the leading renormalon of the soft function appearing in event-shape cross sections, studying this short-distance mass may yield further insight on the advantages of using one scheme versus the other.
Concretely, different low-scale short-distance masses approach the asymptotic behavior differently and depending on the situation one of them might be more appropriate than the others. In any case, using different mass schemes may help to better asses the perturbative uncertainties.

In this article we increase the precision of the inclusive bHQET jet function $B$, previously known at two loops~\cite{Jain:2008gb}, by one order.
In contrast to jet functions based on jet algorithms,
the inclusive jet function is defined without any phase-space constraints except for probing the invariant mass of the jet.
Hence, concerning collinear radiation, the jet function $B$ is completely inclusive.
Our results can be used to obtain the di-hemisphere mass, thrust, heavy jet mass and \mbox{C-parameter} (in its so-called C-jettiness version, see Refs.~\cite{Gardi:2003iv,Lepenik:2019jjk}) cross sections in the peak region at N$^3$LL$^\prime$ order once $H_m$ and the di-hemisphere and \mbox{C-parameter} soft functions are known at this level,
or even N$^4$LL if the corresponding cusp and non-cusp anomalous dimensions are computed at five and four loops, respectively.
For Monte Carlo top mass calibration based on 2-jettiness it is customary to use distributions that are self-normalized to the peak region~\cite{Butenschoen:2016lpz,Dehnadi:2023msm}.
In this case, both hard functions cancel in the ratio at leading power in the EFT expansion, and hence our result represents the last missing piece to achieve N$^3$LL$^\prime$ precision.
This will improve the calibration between $m_t^{\rm MC}$ and $m_t^{\rm MSR}$ and eventually the precision of (indirect) top mass determinations
from boosted-top production at a future lepton collider, see \rcite{Bachu:2020nqn} for a preliminary analysis at N$^3$LL.

This article is organized as follows:
In \sec{theoretical_setup} we define the jet function in bHQET and discuss some of its properties as well as its RG evolution. We also show there how to implement the finite decay width of an unstable heavy quark and an appropriate short-distance mass scheme, and establish useful relations among the coefficients appearing in the perturbative expansion of various versions of the jet function. Section~\ref{sec:computation} describes our three-loop calculation. The results are shown in \sec{new_results}, which includes the renormalized three-loop (position-space) jet function, along with the bare one- and two-loop contributions for arbitrary spacetime dimension, and the relation between pole and jet-mass schemes to $\ord(\as^3)$. An estimate of the non-logarithmic contribution to the four-loop jet function based on renormalon dominance is given in \sec{estimate}. Our conclusions and a brief outlook are contained in \sec{conclusion}. The momentum-space expression of the jet-function, the relevant anomalous dimensions, the results for our three-loop master integrals, as well as some technical discussions are relegated to the appendices.

\section{Theoretical setup}
\label{sec:theoretical_setup}

\subsection{Definitions and jet function properties}
\label{subsec:JF_def}
We start by presenting
the basic definitions for a stable massive quark whose mass is defined in the pole scheme. A finite width can be easily introduced later through convolution with a Breit-Wigner distribution or by shifting the invariant mass into the complex plane, as shall be explained later in this section.
Switching to a short-distance mass scheme also corresponds to a shift in the argument of the jet function followed by a perturbative expansion in powers of the strong coupling and is therefore similarly straightforward, see \rcite{Fleming:2007xt}.
We work in bHQET at leading power, i.e.\ at first order in the EFT expansion, to describe the collinear radiation inside a heavy-quark jet. The (three-dimensional) jet direction is denoted by $\vec{n}$ and we define a corresponding light-like four-vector $n^\mu = (1,\vec{n})$. We also introduce the anti-collinear vector $\bar{n}^\mu$ which serves to decompose an arbitrary four-vector $p^\mu$ in the so-called light-cone components: $p^\mu = p^- n^\mu/2 + p^+ \bar{n}^\mu/2 + p_\perp^\mu$ with \mbox{$n^2=\bar{n}^2=0$}, \mbox{$n\cdot \bar{n}=2$} and $p_\perp\! \cdot n = p_\perp \!\cdot \bar{n} =0$.
The bHQET Lagrangian is derived from the $n$-collinear sector of the (massive) SCET Lagrangian by integrating out particle modes with \mbox{(hard-)collinear} momenta ($p^2\sim m^2$) leaving only ultracollinear dynamical degrees of freedom ($p^2\sim \Gamt^2$),\footnote{Even though our results apply in principle to any heavy quark flavor, to avoid confusion with the Gamma function, we use $\Gamt$ to denote the width of the heavy quark.} which are soft in the rest frame of the heavy quark. Since the purely collinear sectors of SCET correspond to boosted versions of QCD~\cite{Bauer:2000yr} this construction is completely analogous to the procedure of matching QCD to HQET. In a physical scattering process, the ultracollinear modes can, however, also couple to other colored particles outside the $n$-collinear sector (which from the point of view of the heavy quark appear to be boosted in the $\bar{n}$ direction). In the EFT approach these interactions
are described (at leading power) by the
$n$-ultracollinear Wilson line defined as
\begin{equation}
W_n(z)
=
\overline{\mathrm{P}} \exp \biggl[- ig_s \!\!\int_0^{\infty}\! {\rm d} t\, \bar{n} \cdot
A_n(\bar{n} t+ z)
\biggr].
\label{eq:def_Wilson}
\end{equation}
Here $g_s$ is the bare gauge coupling, $A_{n,\mu} \equiv A_{n,\mu}^a T^a$ denotes the $n$-ultracollinear gluon field, and $\overline{\mathrm{P}}$ the anti-path ordering of the associated color generators ($T^a$). The argument of the exponential is dimensionless since the integration variable $t$ has dimensions of an inverse mass, while gluon field and bare coupling in dimensional regularization with $d=4-2\varepsilon$ have mass dimension $[M]^{1-\varepsilon}$ and $[M]^\varepsilon$, respectively. Following \rcites{Fleming:2007xt,Jain:2008gb} we define the bare forward-scattering matrix element%
\footnote{We will not renormalize any fields. Subdivergences in $\cB^\bare$ are absorbed entirely by the bare coupling $g_s$, see \sec{JF_renormalization}.}
\begin{equation}
\mathcal{B}^\bare (\hat{s}) \equiv\! \frac{-
i}{4 \pi N_c m} \!\int\! \rd^d z\, e^{i r \cdot z} \!\bra{ \,0\, }
T \bigl\{ \bar{h}_{v} (0) W_n (0) W_n^{\dag} (z) h_{v} (z) \bigr\}\! \ket{\,0\,}\!
= - \frac{1}{\pi m \hat{s}} \,+ \ord(\as)\,,
\label{eq:def_calB}
\end{equation}
where $\hat s \equiv 2\, v \cdot r$,
$N_c$ is the number of colors ($N_c=3$ in QCD), $h_v$ denotes the bHQET field operator, $m$ the pole mass, and $v^\mu$ the (time-like) four-velocity ($v^2=1$) of the heavy quark. For the $e^+e^- \to t\,\bar{t}$ process, kinematics would imply $n\cdot v = m/Q$, $\bar n\cdot v=Q/m$ and $v^\mu_\perp=0$ (up to small reparametrizations). As discussed below, the jet function is however independent of these boost invariants and we will in practice set $n\cdot v = \bar{n}\cdot v=1$ for simplicity of our calculation. The function $\cB^\bare$ is gauge invariant by construction and has dimensions of an inverse mass squared since the Wilson lines are dimensionless and the heavy-quark fields have mass dimension $[M]^{3/2-\varepsilon}$. Its argument, the variable $\hat s$, is related to the invariant mass of the jet $M$ as
\begin{equation}
\hat s= \frac{M^2 - m^2}{m}\sim \Gamt\ll m\,.
\label{eq:shat}
\end{equation}
In \eq{shat} we also indicate the scale hierarchies assumed in our EFT framework. In practice, this implies that we consider the peak region in the jet invariant-mass distribution of the heavy-quark (top) jet, see \rcites{Fleming:2007qr,Fleming:2007xt}.

In (b)HQET, the hard(-collinear) modes of the heavy quark (with characteristic momentum square $p^2\sim m^2$) have been integrated out. Hence, heavy quarks cannot be produced/annihilated. The heavy flavor is thus not treated as active anymore, but rather as a background color source. Therefore, only the $n_\ell$ flavors of quarks lighter than the heavy one are responsible for the running of the strong coupling $\as$ (in a minimal subtraction scheme like $\MSb$). In our computation we treat all such lighter quarks as massless (see \rcite{Bris:2024bcq} for the contribution of massive lighter quarks to the two-loop bHQET jet function). To avoid double counting the contributions from infrared (IR) momentum regions already taken into account by the soft function, it is necessary to subtract the so-called zero-bin~\cite{Manohar:2006nz} from $\cB^\bare$ as described in detail in \rcite{Fleming:2007xt}. In dimensional regularization, which we employ for our calculations, all zero-bin contributions to $\cB^\bare$ are proportional to
scaleless integrals and therefore vanish. Their only effect is to turn all IR divergences in the matrix element $\cB^\bare$ into ultraviolet (UV) ones, which are eventually absorbed along with the original UV divergences by the jet function renormalization factor, see \sec{JF_renormalization}.

The bHQET jet function $B$ is defined through the imaginary part of the matrix element in \eq{def_calB} as follows\footnote{Here and in the rest of \subsec{JF_def} we drop the label ``bare'' because all relations are valid for bare and renormalized quantities.}
\begin{equation}
B (\hat s) \equiv \mathrm{Im} \bigl[ \mathcal{B} (\hat{s} + i \eta) \bigr] = \frac{1}{m} \delta(\hat{s}) + \ord(\as)\,,
\label{eq:def_B}
\end{equation}
where $\eta$ is a positive infinitesimal.
One may also consider $\hat{s}$ a complex variable such that $\cB(\hat{s})$ is analytic in the whole complex plane except for a branch cut coinciding with the positive real axis, which shows up for the first time at $\ord(\as)$. (At tree-level there is a simple pole sitting at the origin.)
This analytic property can be used to invert the above relation, yielding the following dispersive integral
\begin{equation}
\mathcal{B} (\hat s)
=\! \int \frac{\rd \hat s'}{\pi} \, \frac{B (\hat s')}{\hat s' -\hat s}
\,.
\label{eq:from_calB_to_B}
\end{equation}
Assuming from now on (again) real $\hat{s}$, $\cB(\hat{s})$ is an ordinary complex function for $\hat{s}<0$,
whereas $B(\hat{s})$ is distribution-valued. Concretely, it is a linear combination of Dirac delta and plus distributions, as given in \eq{exp_calB}, and has support only for $\hat{s}\ge 0$.%
\footnote{Implementing a finite width $\Gamt$ amounts to replacing $\eta \to \Gamt$ in \eq{def_B}, turning $B(\hat{s},\Gamt)$ into an ordinary function of $\hat{s}$ with support on the entire real axis.}

To turn convolutions like the ones in \eq{double_hemisphere_invariant} into simple products, which is convenient e.g.\ for an efficient solution of the jet function's renormalization group equation (RGE), it is useful to work with the Fourier transform of $B$, also referred to as the ``position-space'' jet function, defined as
\begin{equation}
\tilde{B}(x)\equiv \!\int \rd \hat s\,e^{-i (x - i \eta) \hat s} B(\hat s)\,,\qquad
B(\hat s)= \frac{1}{2 \pi} \!\int \rd x\,e^{ix\hat{s}} \tilde{B}(x-i\eta)\,,
\label{eq:FT_invFT_B}
\end{equation}
where we have also displayed the inverse transformation. The $\eta$ prescription on the left integral is necessary for real-valued $x$. For complex $x$ the integral converges if ${\rm Im}(x)<0$, otherwise one needs to analytically continue to the upper half of the complex plane. The position-space jet function $\tilde{B}(x)$ (with dimensions of an inverse mass) is analytic in the entire complex $x$ plane except for a branch cut coinciding with the positive imaginary axis (which also appears at one loop for the first time).
If a finite width is taken into account, the position-space jet function is defined for real $x$ only.

As argued and verified explicitly at two loops in \rcite{Jain:2008gb}, the jet function is subject to non-Abelian exponentiation~\cite{Gatheral:1983cz,Frenkel:1984pz,Gardi:2013ita}. This statement is based on the fact that $\tilde{B}(x)$ can also be expressed as a vacuum expectation value of Wilson line exponentials~\cite{Jain:2008gb},
\begin{align}
\tilde{B}(x) = \frac{1}{N_c m}
\bra{ \,0\, } \tr \Bigl[\,
\overline{T} \bigl\{ W_n^\dagger (z) W_v(z) \bigr\} \,
T \bigl\{ W_v^\dagger (0) W_n(0) \bigr\}\!
\Bigr] \ket{\,0\,}
,
\label{eq:BWilsonCorr}
\end{align}
where $z^0=2x$ and $\vec{z}=0$, the time-like Wilson line $W_v$ is defined analogous to the light-like $W_n$ in \eq{def_Wilson} but with $\bar{n}$ replaced by the heavy quark velocity $v$, and the trace is to be taken in color space. As consequence, the result for the position-space jet function takes the exponentiated form
\begin{equation}
\tilde{B}(x) = \frac{1}{m} \exp \Bigl[\tilde{b} (x)\Bigr],
\label{eq:exp_theorem}
\end{equation}
where the exponent $\tilde{b}(x)$ exclusively contains ``fully connected'' color factors~\cite{Gardi:2013ita}. This class of color factors corresponds to that of Feynman diagrams that remain connected once all Wilson lines are removed from the graph. In particular, this implies for an Abelian gauge theory (like QED) without light fermions that the exponent $\tilde{b}(x)$ is one-loop exact. For the non-Abelian case we are concerned with, exponentiation implies that there are no terms proportional to $\as^nC_F^{n}$ with $n\geq 2$ in $\tilde{b}(x)$.\footnote{Here and in the following $C_F=(N_c^2-1)/(2N_c)$ and $C_A=N_c$ denote the quadratic Casimir in the fundamental and adjoint representations of the color group SU($N_c$), respectively, and $T_F=1/2$ fixes the normalization of the fundamental color generators $T^a$.} Likewise, at $\ord(\as^3)$, the color structure $C_F^2C_A$ must be absent\footnote{Accordingly, these color factors are absent in the corresponding cusp and non-cusp anomalous dimensions, as can be seen in the known expressions for these quantities, see \app{expansion}.} and only diagrams involving a fermion loop with four gluon attachments, like the one in
\fig{FDEx1}, can contribute to the $C_F^2T_Fn_\ell$ term of $\tilde{b}(x)$.
These findings, which hold for the bare and renormalized position-space jet function, can be used as checks of our explicit results, see \sec{new_results}.

\subsection{Jet function renormalization and resummation}
\label{sec:JF_renormalization}
To regulate UV and IR divergences in loop diagrams we work in dimensional regularization with $d=4-2\varepsilon$. For the renormalization of the jet function and the strong coupling we employ the $\MSb$ scheme, while the heavy-quark mass $m$ is defined (for the time being) in the pole scheme.
The divergences in the bare jet function are removed by convolving it with a
(distribution-valued)
renormalization factor $Z_B(\hat{s},\mu$), which has support for $\hat{s} \ge 0$ and depends on the $\MSb$ renormalization scale $\mu$:
\begin{equation}
B^{\bare}(\hat s)
\equiv\!
\int
\! \rd \hat s' \,Z_B(\hat s-\hat s',\mu) \, B(\hat s',\mu)\,.
\label{eq:def_ZB}
\end{equation}
Hence, the renormalized jet function is $\mu$-dependent. The inverse renormalization factor, $Z_B^{-1}(\hat{s}, \mu)$, also has support for $\hat{s}\ge0$ and is
defined through
\begin{equation}
\int
\! \rd\hat s' \, Z_B^{-1}(\hat s-\hat s',\mu) \, Z_B(\hat s',\mu)=\delta(\hat s)\,,
\label{eq:relation_Z_invZ}
\end{equation}
such that
\begin{equation}
B(\hat s,\mu)
=\!
\int
\! \rd\hat s' \, Z_B^{-1}(\hat s-\hat s',\mu) B^{\bare}(\hat s')\,.
\label{eq:inv_ZB}
\end{equation}
Inserting \eq{inv_ZB} in \eq{from_calB_to_B}, we find that the renormalization of $B$ and $\mathcal{B}$ takes the same form:
\begin{equation}
\mathcal{B} (\hat s,\mu) = \int \rd \hat s' \, Z_B^{- 1} (\hat s',\mu) \, \mathcal{B}^{\bare} (\hat s - \hat s')\,.
\label{eq:Z_for_calB}
\end{equation}

For the (dimensionless) renormalization factor of the strong coupling $\as$ and its anomalous dimension, i.e.\ the QCD beta function, we define
\begin{align}\label{eq:alpBare}
\as^\bare=\frac{g_s^2}{4\pi}={}&
\biggl(\frac{\mu^2e^{\gamma_E}}{4 \pi}\biggr)^{\!\!\varepsilon}
Z_\alpha[\as(\mu)] \, \as(\mu)\,, &\mu \dfrac{\rd \as}{\rd \mu}&\equiv -2 \varepsilon \as +\beta_{\mathrm{QCD}}(\as)\,,
\\
\beta_{\mathrm{QCD}} (\as) ={}& -\! 2 \as \hat{\beta} (\as)\,, &\hat{\beta}(\as) &= \sum^\infty_{l = 0} \beta_l \Bigl( \frac{\as}{4 \pi} \Bigr)^{\!l +1}\,,\nonumber
\end{align}
where $\gamma_E$ is the Euler-Mascheroni constant.
The renormalized coupling $\as \equiv \as(\mu)$ is thus dimensionless.
For later convenience, we have factored out $-2\as$ in the four-dimensional QCD beta function $\beta_\mathrm{QCD}$ such that
$\hat{\beta}(\as) = \ord(\as)$. %
The terms in the perturbation series of $\hat{\beta}(\as)$ and $Z_\alpha(\as)$ that are relevant for our calculation are given in \apps{expansion}{consistency}, respectively.

The convolutions in \eqsm{def_ZB}{inv_ZB} become simple products in position space:
\begin{equation}
\tilde{B}^{\bare}(x)
=
\tilde{Z}_B(x,\mu)\,
\tilde{B}(x,\mu)\,,
\label{eq:def_ZB_FT}
\end{equation}
where
$\tilde{Z}_B(x,\mu) \equiv \int {\rm d}\hat s\,e^{-i\hat sx} Z_B (\hat s,\mu)$ is dimensionless.
The Fourier transform of the inverse renormalization factor $Z_B^{-1}(\hat s, \mu)$ is simply $1/\tilde Z_B(x,\mu)$.
One can linearize \eq{def_ZB_FT} by taking logarithms on both sides:
\begin{equation}
\log \bigl[m\tilde{B}^{\bare}(x)\bigr]
=
\log\bigl[ \tilde{Z}_B(x,\mu)\bigr] +
\log \bigl[m\tilde{B}(x,\mu)\bigr].
\label{eq:def_ZB_FT_log}
\end{equation}
This is particularly useful for determining $\tilde{Z}_B$, because the $1/\eps^n$ poles in $\log\bigl[ \tilde{Z}_B\bigr]$ must directly cancel those in $\log \bigr[m\tilde B^{\bare}\bigr]$ order by order in $\as$. In \app{consistency}, we derive consistency conditions, an RGE, and a closed form for the position-space renormalization factor $\tilde Z_B$ that, to the best of our knowledge, has not been provided before.

For the expansion of the renormalized jet function in powers of $\as$, and the associated bare and renormalized matrix elements in powers of $\alpha^\bare_s$ and $\as$, respectively, we define
\begin{align}
m\mathcal{B} (\hat s, \mu) ={}& \sum^\infty_{l=0} \biggl[ \dfrac{\as(\mu)}{4 \pi}\biggr]^l \sum^{2l}_{k=0}\mathcal{B}_{lk} L^k(\hat s,\mu) \,,
&m\mathcal{B}^\bare\! (\hat s) ={}& \dfrac{1}{\pi(-\hat s)} \sum^\infty_{l=0} \biggl[\frac{e^{-\varepsilon\gamma_E}\as^\bare}{(-\hat s)^{2\varepsilon}(4 \pi)^{1-\varepsilon}\!}\biggr]^l\! \mathcal{B}^\bare_{l}\!,\nonumber
\\
\label{eq:exp_calB}
mB (\hat s, \mu) ={}& \sum^\infty_{l=0} \biggl[ \dfrac{\as(\mu)}{4 \pi}\biggr]^l \! \sum^{2l-1}_{k=-1} \!B_{lk}\mathcal{L}^k(\hat s,\mu) \,,
&\! m\tilde{B} (x, \mu) ={}& \sum^\infty_{l=0} \biggl[ \dfrac{\as(\mu)}{4 \pi}\biggr]^l \! \sum^{2l}_{k=0} \tilde B_{lk}\tilde L^k(x,\mu) \,,\!
\end{align}
adapting the notation of \rcite{Jain:2008gb}:
\begin{align}
L^k(\hat s,\mu)={}& -\!\dfrac{1}{\pi \hat s} \log^k \Bigl(- \dfrac{\mu}{\hat s}\Bigr)\,,
&\tilde{L}(x,\mu)& = \log(ie^{\gamma_E}x\mu)\,, \label{eq:plus_log}\\
\mathcal{L}^{k\geq0}(\hat s,\mu)={}&\dfrac{1}{\mu} \biggl[ \dfrac{\theta(\hat s) \log^k(\hat s / \mu)}{\hat s / \mu} \biggr]_+ \,,
&\mathcal{L}^{-1}(\hat s,\mu)&= \delta (\hat s - \eta)\,,\nonumber
\end{align}
where $\eta$ is again a positive infinitesimal.
Useful properties of the logarithms and plus distributions in \eq{plus_log}, along with the definition of the latter, are presented in \app{plus_distributions}.

Taking the logarithm of the position-space jet function reduces not only the number of different color factors due the non-Abelian exponentiation theorem, but also the maximal power of the logarithms at each perturbative order:%
\footnote{The coefficients $\tilde b_{j2}$ and $\tilde b_{j1}$ contain the $j$-loop correction to the cusp and non-cusp anomalous dimensions of the jet function respectively.
The $\tilde b_{j,k\geq3}$ are non-zero due to the single-logarithmic running of the coupling, i.e.\ vanish for $\beta_n \to 0$, see \subsec{relations}.}
\begin{equation}
\log\bigl[m\tilde{B} (x, \mu)\bigr] = \tilde b(x, \mu)=\sum^\infty_{l=1} \biggl[ \dfrac{\as(\mu)}{4 \pi}\biggr]^l \sum^{l+1}_{k=0}\tilde b_{lk}\, \tilde L^k(x,\mu) \,.
\label{eq:B_exp_series}
\end{equation}
The coefficients $\mathcal{B}_{lk}$, $B_{lk}$, $\tilde B_{lk}$, and $\tilde b_{lk}$ are real and dimensionless, and knowing one of these sets allows to deduce the others. Furthermore, one can express the logarithmic coefficients (i.e.\ $\mathcal{B}_{l,k\geq1}$, $B_{l,k\geq0}$, $\tilde B_{l,k\geq1}$, and $\tilde b_{l,k\geq1}$) in terms of the non-logarithmic ($\mathcal{B}_{l0}$, $B_{l,-1}$, $\tilde B_{l0}$, and $\tilde b_{l0}$) and anomalous dimension coefficients. All these relations are worked out in \subsec{relations}.

The only missing pieces at $\mathcal{O}(\as^3)$ are therefore the $C_F C_A^2$, the $C_FC_A T_F n_\ell$ and $C_F^2T_Fn_\ell$ terms of $\tilde b_{30}$, since: i)~all $\tilde b_{3,i\geq1}$ can be obtained from \eqs{nonRec}{Rec},
ii)~the terms proportional to $C_F^3$ and $C_F^2C_A$ are zero according to the non-Abelian exponentiation theorem, iii)~the $C_FT_F^2n_\ell^2$ piece was determined in \rcite{Gracia:2021nut} through a large-$\beta_0$ computation. The analytic expressions of those three terms are the main result of this article. Based on renormalon dominance \rcite{Gracia:2021nut} also provides an estimate for the $n_\ell$ coefficient and the flavor-number-independent term of $\tilde b_{30}$.
As we shall see, the exact results obtained here lie within the uncertainties of the estimates given there.

From the $\mu$ independence of $B^\bare$ and $\mathcal{B}^\bare$ one obtains the RGE of their renormalized counterparts:
\begin{align}
\mu \dfrac{\rd B(\hat s,\mu)}{\rd \mu} = \!\!\int
\! \rd \hat s' \,\gamma_B(\hat s',\mu)\, B(\hat s - \hat s',\mu)\,
, \quad
\mu \dfrac{\rd \mathcal{B}(\hat s,\mu)}{\rd \mu} =\!\! \int
\! \rd \hat s' \,\gamma_B(\hat s',\mu)\, \mathcal{B}(\hat s-\hat s',\mu)\,.
\label{eq:RGE_for_B}
\end{align}
The momentum-space anomalous dimension is given by
\begin{align}
\label{eq:def_anomalous_dim}
\gamma_B(\hat s,\mu) =& \int
\! \rd \hat s' \, Z_B (\hat s-\hat s',\mu)\, \dfrac{\rd \:}{\rd\mu} Z_B^{-1}(\hat s',\mu)
=
-\!\int
\! \rd \hat s' \,Z_B^{-1} (\hat s-\hat s',\mu) \,\dfrac{\rd \:}{\rd\mu} Z_B(\hat s',\mu)\\
=&-\! 2\Gamma^\rc[\as(\mu)] \,\cL^{(0)}(\hat s,\mu)
+ \gamma_B^\text{nc}[\as(\mu)] \,\cL^{(-1)}(\hat s,\mu)
\,.\nn
\end{align}
As seen in the second line, $\gamma_B$ is a distribution with support for positive values of its argument.
Here $\Gamma^\rc$ and $\gamma_B^{\rm nc}$ are dubbed the (light-like) cusp and non-cusp anomalous dimensions, respectively. Both can be written as an expansion in powers of $\as$
\begin{equation}
\Gamma^{\rm c}(\as)
=
\sum^\infty_{l=0} \Bigl( \dfrac{\as}{4 \pi}\Bigr)^{\!l+1} \Gamma^{\rm c}_l\,,
\qquad \qquad
\gamma_B^\text{nc}(\as)
=
\sum^\infty_{l=0} \Bigl( \dfrac{\as}{4 \pi}\Bigr)^{\!l+1} \gamma^B _l\,,
\label{eq:expansion_gamma}
\end{equation}
where the relevant coefficients are again collected in \app{expansion}.
It is often more convenient to write the jet-function RGE in Fourier space:
\begin{equation}
\dfrac{\rd \log \bigl[m\tilde{B}(x,\mu)\bigr]}{\rd \log \mu}=
\label{eq:RGE_for_B_FT}
\tilde{\gamma}_B(x,\mu)=
2\Gamma^\rc[\as(\mu)] \tilde L(x,\mu) + \gamma_B^\text{nc}[\as(\mu)]\,.
\end{equation}
This RGE can be integrated to obtain the $\mu$-evolution of $\tilde B(x,\mu)$:
\begin{align}\label{eq:evolve_B_ft}
\tilde{B}(x, \mu)={}&e^{K(\mu, \mu_0)}(i e^{\gamma_E} x \mu_0)^{\omega(\mu, \mu_0)} \tilde{B}(x, \mu_0)
\,\equiv\tilde U_B(x,\mu,\mu_0)\tilde{B}(x, \mu_0)\,,\\
\omega(\mu, \mu_0)={}&- \!\int_{\as\left(\mu_0\right)}^{\as(\mu)} \frac{{\rm d} \alpha}{\alpha} \frac{\Gamma^\rc(\alpha)}{\hat{\beta}(\alpha)}\,,
\nn\\
K(\mu, \mu_0)={}&
\frac{1}{2}\Biggl[ \int_{\as\left(\mu_0\right)}^{\as(\mu)} \frac{{\rm d} \alpha}{\alpha} \frac{\Gamma^{\mathrm{c}}(\alpha)}{\hat{\beta}(\alpha)} \int_{\as(\mu_0)}^\alpha \frac{{\rm d} \alpha^{\prime}}{\alpha^\prime \hat{\beta}(\alpha^{\prime})}
-\int_{\as(\mu_0)}^{\as(\mu)} \frac{{\rm d} \alpha}{\alpha} \frac{\gamma_B^\mathrm{nc}(\alpha)}{\hat{\beta}(\alpha)}\Biggr]
,\nonumber
\end{align}
where $\tilde U_B(x,\mu_0,\mu)=\tilde U_B^{-1}(x,\mu,\mu_0)$ since the evolution kernel obeys a first-order differential equation whose solution is unique once a boundary condition has been specified.
An exact solution for $\omega(\mu,\mu_0)$ at N$^n$LL for any $n$ can be found in \rcite{Lepenik:2019jjk}. Fourier inverting \eq{evolve_B_ft} yields the evolution equation for the renormalized momentum-space jet function
\begin{align}\label{eq:EvolConvol}
B(\hat{s}, \mu)&=\int
\!\dd \hat{s}^{\prime}\, U_B(\hat{s}-\hat{s}^{\prime}, \mu, \mu_0)\, B(\hat{s}^{\prime}, \mu_0) \, \text{ with}
\\
U_B(\hat{s}, \mu, \mu_0)&=\frac{e^{K+\gamma_E\omega}}{\mu_0 \Gamma(-\omega)}\biggl[\frac{\mu_0^{1+\omega} \theta(\hat{s})}{\hat{s}^{1+\omega}}\biggr]_{+},\nonumber
\end{align}
where $K \equiv K(\mu,\mu_0)$ and $\omega \equiv \omega(\mu,\mu_0)$. Likewise, starting from \eq{RGE_for_B}, one can show that the renormalized forward-scattering matrix element evolves with the same RG kernel:
\begin{equation}
\mathcal{B}(\hat{s}, \mu)=\int
\!\rd \hat{s}^{\prime} \, U_B(\hat{s}^{\prime}, \mu, \mu_0) \, \mathcal{B}(\hat s - \hat{s}^{\prime}, \mu_0)\,.
\end{equation}

\subsection{Jet function for unstable quarks in a short-distance mass scheme}
\label{subsec:JF_unstable}
The jet function as defined in \eqs{def_calB}{def_B} exhibits an $\mathcal{O}(\Lambda_{\mathrm{QCD}})$ infrared renormalon ambiguity, see \rcite{Gracia:2021nut} for a full-fledged computation in the large-$\beta_0$ limit. The renormalon leads to an asymptotically divergent behavior of the perturbation series unless it is canceled by switching from the pole mass to a (low-scale) short-distance scheme such as the MSR mass~\cite{Hoang:2008yj,Hoang:2017suc}, see \rcite{Bachu:2020nqn} for a phenomenological analysis. The scheme change is implemented by shifting the argument of the jet function by the corresponding perturbation series \mbox{$\delta m\equiv m - m^{\mathrm{SD}} \sim \as \Gamt$} defining a suitable short-distance mass $m^{\mathrm{SD}}$ and consistently expanding in powers of the strong coupling~\cite{Fleming:2007qr}.
Concretely, it was argued in \rcites{Fleming:2007qr,Fleming:2007xt}, based on the study of the (b)HQET Lagrangian at leading power, \mbox{$\mathcal{L}_h = \bar h_v ( i\,v\cdot D\ - \delta m + i \Gamt/2 ) h_v$} with $D$ the ultracollinear covariant derivative, that the forward-scattering matrix element for a heavy quark with finite width $\Gamt$ and mass defined in a short-distance scheme
can be obtained from \eq{def_calB} by shifting the argument $\hat{s}\rightarrow \hat{s} -2 \delta m + i \Gamt$. Hence, the renormalized jet function defined in terms of $m^{\rm SD}$ and including the decay width reads
\begin{equation}
B(\hat s, \delta m, \Gamt, \mu) = \text{Im} \bigl[ \mathcal{B}(\hat{s} -2 \delta m + i \Gamt, \mu) \bigr]
=\exp\biggl(\!-2\delta m\frac{\partial}{\partial \hat s}\biggr)B(\hat s, 0,\Gamt, \mu) \,,
\end{equation}
where the exponential sitting out front indicates that $\delta m = \ord(\as)$ has to be treated as the parameter of a Taylor expansion around $B(\hat s, 0,\Gamt, \mu)$
to the given order in $\as$. The effect of the decay width on the momentum-space jet function is accounted for by convolving with a Breit-Wigner distribution~\cite{Fleming:2007xt}. As $B(\hat s, 0, 0, \mu) \equiv B(\hat s, \mu)$ we have
\begin{equation}
B(\hat s, \delta m, \Gamt, \mu) =
\exp\biggl(\!-2\delta m\,\frac{\partial}{\partial \hat s}\biggr)\!\!
\int
\frac{\mathd \hat s'}{\pi}\, B(\hat s'
, \mu) \, \dfrac{\Gamt}{(\hat s - \hat s')^2 + \Gamt^2}\,.
\label{eq:Ballincl}
\end{equation}
The momentum-space jet function with a non-zero $\Gamt$ obeys the same RG equation as its stable-quark counterpart, see \eqs{RGE_for_B}{EvolConvol}. A possible renormalization scale dependence of $\delta m$ is canceled (at leading power) by the corresponding dependence of $\hat{s}$ when consistently replacing $m \to m^{\mathrm{SD}}$ in \eq{shat} and making the dependence of the jet function on the jet invariant mass $M$ explicit.
Setting $\delta m=0$ results in the following functional form for the jet function:
\begin{align}\label{eq:coefGamma}
B(\hat s,0, \Gamt, \mu) ={}& \frac{\Gamt}{\pi\hat s_\Gamma^2}
\sum_{l=0}\biggl[ \dfrac{\as(\mu)}{4 \pi}\biggr]^l\sum_{k=0}^{2l}
\hat B_{lk}\biggl(\frac{\hat s}{\Gamt}\biggr)\log^k\biggl(\frac{\mu}{\hat s_\Gamma}\,\biggr)\,,\\
\hat s_\Gamma ={}& \sqrt{\hat s^2+\Gamt^2}\,,\nn
\end{align}
where the coefficients $\hat B_{lk}(\hat s / \Gamt)$ depend on powers of $\hat s / \Gamt$ and the combination
\begin{equation}
y\biggl( \frac{\hat s}{\Gamt} \biggr) \equiv \arctan\biggl(\frac{\hat s}{\Gamt} \biggr) + \frac{\pi}{2}\,,
\label{eq:def_y}
\end{equation}
see \eq{Bhatij}.
In the limit $\Gamt\to0$, which implies $y(\hat s / \Gamt)\to \pi$ and $\hat s_\Gamma\to\hat s$, the distributional structure of $B(\hat{s},\mu)$, see \eq{exp_calB}, re-emerges.

Combining \eqs{EvolConvol}{Ballincl} one can write the RG-evolved momentum-space jet function for $\delta m =0$ and $\Gamt \neq 0$ as a convolution of the corresponding jet function for stable quarks and a modified evolution kernel:
\begin{align}
B(\hat{s},0, \Gamt, \mu)&=\int
\dd \hat{s}^{\prime}\, U_B(\hat{s}-\hat{s}^{\prime}, \Gamt, \mu, \mu_0)\, B(\hat{s}^{\prime}, \mu_0)\,
\text{ with}
\\
U_B(\hat{s}, \Gamt, \mu, \mu_0)&=\frac{e^{K+\gamma_E\omega}}{\mu_0 \Gamma(-\omega)}\biggl(\frac{\mu_0}{\hat s_\Gamma}\biggr)^{\!\!1+\omega}
\frac{\Gamma (1 - {\omega}) \Gamma ({\omega})}{\pi \hat s_\Gamma}
\biggl\{\hat{s} \sin \biggl[{\omega} \,y\biggl(\frac{\hat s}{\Gamt}\biggr)\biggr] - \Gamt \cos \biggl[{\omega}\,y\biggl(\frac{\hat s}{\Gamt}\biggr)\biggr]\!\biggr\}.\nonumber
\end{align}
In Fourier space, including the effects of a finite width and switching the heavy quark mass scheme implies a multiplicative modification, which in turn translates into an additive modification of the exponent $\tilde b(x,\mu)$:
\begin{equation}
{\tilde B}(x,\delta m,\Gamt,\mu) = e^{2ix\delta m} e^{- \Gamt | x |} {\tilde B}(x,\mu)\,,
\label{eq:WB_FT}
\end{equation}
where now $x$ must be real if $\Gamma_{\!t}>0$.
To cancel the renormalon, the leftmost exponential must be expanded in powers of $\alpha_s$ and combined with ${\tilde B}(x,\mu)$ to a given order in the perturbative expansion.

The $\ord(\LQCD)$ renormalon of the position-space inclusive jet function cancels with the well-known renormalon of the pole mass
in the expression $\tilde B(x,\mu) e^{-2ix m}$ as can be inferred from the fact that this combination appears in the leading-power cross sections for processes with boosted heavy quarks, such as \eq{double_hemisphere_invariant}, which are (infrared and collinear safe) observables and hence unambiguous. This observation was confirmed by the explicit large-$\beta_0$ computation of \rcite{Gracia:2021nut}.
Therefore, the perturbation series of $\log[\tilde{B}(x,\mu)] - 2ix \delta m^{\rm SD}$ should also be renormalon-free. Taking $n$ times the derivative with respect to $\log(ix)$ results in
\begin{equation}
\frac{1}{2ix}\frac{\mathd^n}{\mathd \log^n(ix)}
\log\bigl[ m \tilde{B}(x, \mu_\delta)\bigr]- \delta m^{\rm SD} \equiv P_n\,,
\end{equation}
with $P_n$ representing a renormalon-free series in $\as$.
Different short-distance mass schemes are obtained by requiring $P_n=0$ for specific values of $x$, while maintaining that $\delta m^{\rm SD}$ is real. To this end $x$ is parametrized as $i x =\xi/R$ with $\xi > 0$ a dimensionless parameter and $R\gg\LQCD$ an infrared renormalization scale that plays the same role as the argument of the MSR mass in \rcites{Hoang:2008yj,Hoang:2017suc}. Finally, one can choose either $\mu_\delta=\mu$ or $\mu_\delta=R$. This generic class of schemes was first discussed in \rcite{Bachu:2020nqn} in the context of gap subtractions for the soft function. The corresponding mass shift takes the form:
\begin{equation}
\delta m^{\rm SD} = \frac{R}{2\xi} \sum^\infty_{l=\max(1,n-1)} \biggl[ \dfrac{\as(\mu_\delta)}{4 \pi}\biggr]^l \, \sum^{l+1-n}_{k=0} \frac{(k+n)!}{k!} \,\tilde b_{l,k+n}\log^k\biggl(\frac{\xi e^{\gamma_E}\mu_\delta}{R}\biggr)\,,
\end{equation}
where
the coefficients $\tilde b_{ik}$ are defined \eq{B_exp_series}. Since at $\mathcal{O}(\alpha_s^k)$ the highest power of $\log(ix)$ in $\log\bigl[ m \tilde{B}\bigr]$ is $k+1$, the series $\delta m^{\rm SD}$ for $n\geq 2$ starts at $\mathcal{O}(\as^{n-1})$. Larger values of $n$ therefore also imply that the asymptotic behavior of $\delta m^{\rm SD}$ sets in at higher orders. For $n\geq2$ the infinite series is formally independent of $\mu_\delta$ (and so is the mass $m^{\rm SD}$), therefore the two choices of $\mu_\delta$ yield identical schemes.

In \rcite{Jain:2008gb},
the $\mu$- and $R$-dependent ``jet-mass'' (low-scale) scheme was introduced by choosing $n=1$, $\xi=e^{-\gamma_E}$ and $\mu_\delta=\mu$:\footnote{The jet-mass $m^J$ should not be confused with the jet (invariant) mass $M$. Alternatively, the single-scale-dependent MSR mass can be used to remove the $\ord(\LQCD)$ renormalon as already mentioned in \sec{intro}.}
\begin{align}
\delta m^J(\mu,R) = m - m^J(\mu,R)
={}&
\frac{Re^{\gamma_E}}{2} \biggl\{\frac{\mathd}{\mathd \log(ix)}\log\bigl[ m \tilde{B}(x, \mu)\bigr] \biggr\}_{i x e^{\gamma_E}=1 / R} \label{eq:jet_mass_scheme}\\
={}& \frac{Re^{\gamma_E}}{2} \sum^\infty_{l=1} \biggl[ \dfrac{\as(\mu)}{4 \pi}\biggr]^l \sum^{l}_{k=0} \tilde m_{lk}\log^k\biggl(\frac{\mu}{R}\biggr)\,,
\text{ with}\nonumber\\
\tilde m_{lk} ={}& (k+1) \tilde b_{l,k+1}\,. \nonumber
\end{align}
The jet-mass is subject to the following (logarithmic) $\mu$-RGE~\cite{Jain:2008gb},
\begin{equation}\label{eq:muAnDim}
\mu\frac{\dd}{\dd \mu} m^J(\mu,R)=-Re^{\gamma_E}\Gamma^\rc[\as(\mu)]\,,
\end{equation}
as can be verified by inserting \eq{jet_mass_scheme} on the LHS, swapping the order in which the derivatives are taken, and using \eq{RGE_for_B_FT}.
The solution of this RGE is
\begin{equation}
m^J(\mu,R) = m^J(\mu_0,R) + \frac{R}{2}e^{\gamma_E} \omega(\mu_0,\mu)\,,
\end{equation}
with $\omega(\mu,\mu_0)$ defined in \eq{evolve_B_ft}.
In addition, the jet-mass obeys a (linear) R-evolution equation (R-RGE) derived from the $R$ independence of the pole mass:
\begin{align}\label{eq:RAnDim}
R\frac{\dd}{\dd R} m^J(R,R)={}& -\! R \gamma_R^J[\alpha_s(R)]=-R\sum_{l=1}^\infty \biggl[ \dfrac{\as(R)}{4 \pi}\biggr]^{l}\gamma_{l-1}^{R,J}\,,\, \text{ with}\\
\gamma_l^{R,J}={}& \frac{e^{\gamma_E}}{2} \Biggl[\tilde b_{l+1,1}-2\sum_{j=0}^{l-1}(l-j)\beta_{j}\tilde b_{l-j,1}\Biggr]\,.\nonumber
\end{align}
The function $\gamma_R^J$ is dubbed ``R-anomalous dimension''. One can always give up the $\mu$ dependence of the jet mass scheme using $\mu_\delta=R$, just like in the MSR mass or the non-derivative scheme introduced in \eq{jet_mass_ND} below.
A perturbative solution to the R-RGE can be found in \rcite{Hoang:2017suc}, while an algorithm to obtain an exact solution is provided in the appendix of \rcite{Lepenik:2019jjk}. To evolve $m^J$ from $(\mu_0,R_0)$ to $(\mu_1,R_1)$ one proceeds in steps, evolving first in $\mu$ to $(R_0,R_0)$, then to $(R_1,R_1)$ with R-evolution, and finally $\mu$-evolving to $(\mu_1,R_1)$. All in all we thus have
\begin{equation}
m^J(\mu_1,R_1) = m^J(\mu_0,R_0) + \frac{R_0}{2} e^{\gamma_E} \omega(\mu_0,R_0) + \Delta^J(R_0,R_1) + \frac{R_1}{2} e^{\gamma_E} \omega(R_1,\mu_1)\,,
\end{equation}
and $m^J(R,R)=m^J(R_0,R_0) + \Delta^J(R_0,R)$ where the last term resums logarithms of $R_1/R_0$ through R-evolution in a renormalon-free fashion.

\begin{figure*}[t!]
\subfigure[]
{\includegraphics[width=0.45\textwidth]{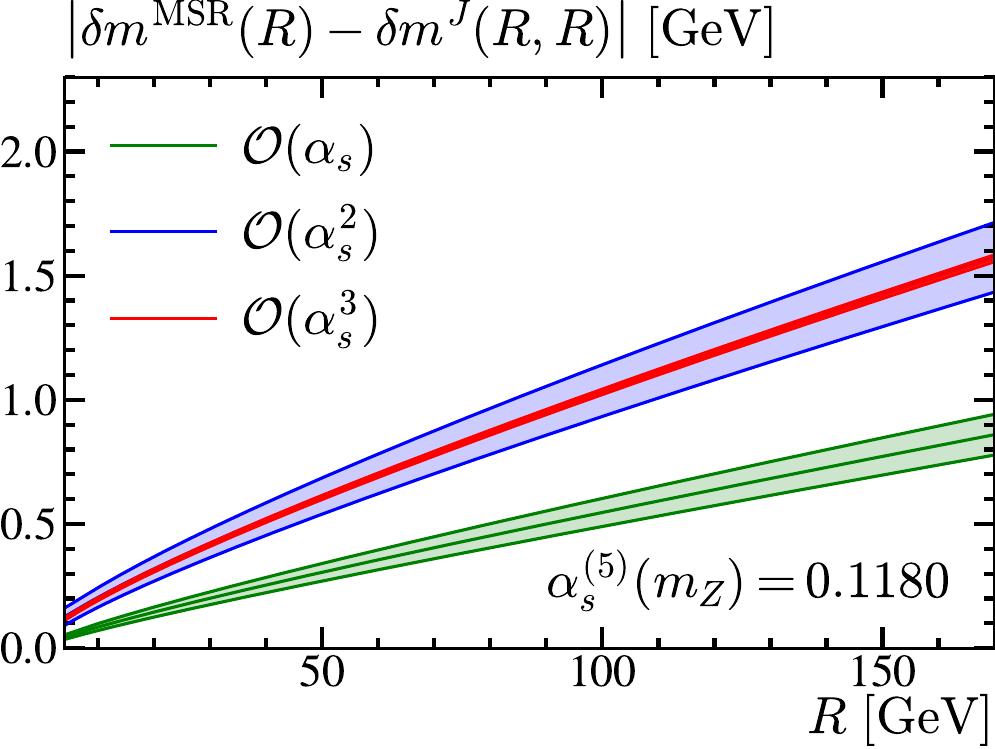}\vspace*{-1cm}}
~~ \subfigure[]
{\includegraphics[width=0.45\textwidth]{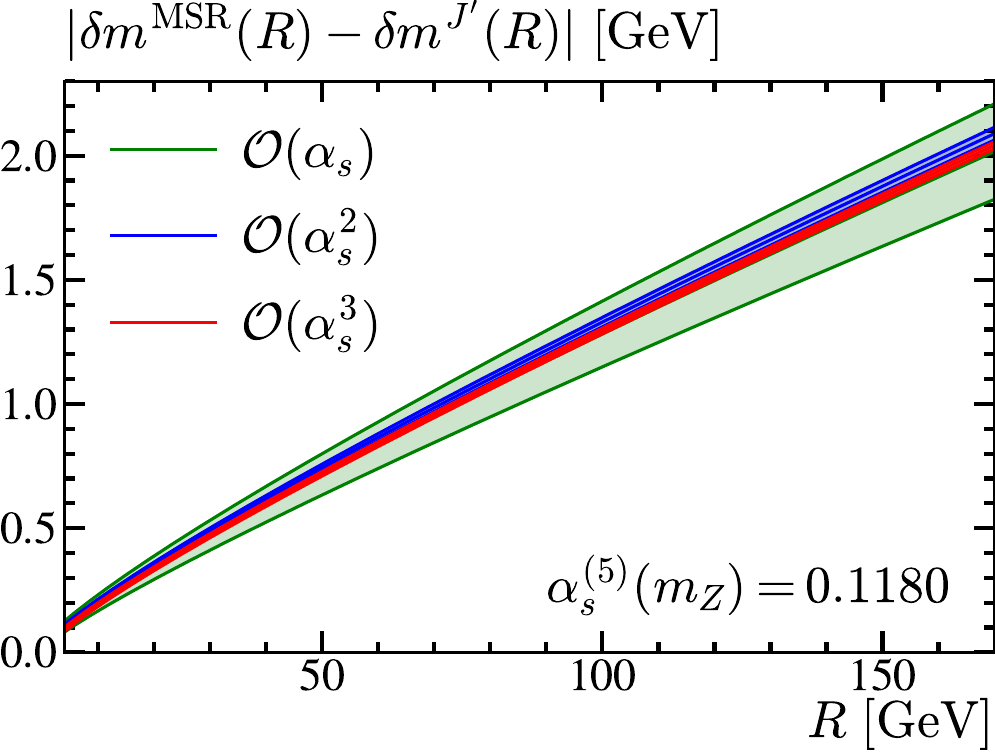}\vspace*{-1cm}}
\caption{Difference between the heavy quark MSR and jet-masses in the derivative~(a) and non-derivative~(b) schemes in absolute value. In the plots we use $n_\ell=5$ for the number of active light flavors and $\alpha_s^{(n_\ell=5)}(m_Z)=0.1180$, which is evolved to $\mu$ with five-loop accuracy using {\sc Revolver}~\cite{Hoang:2021fhn}. For the derivative scheme, which happens to be smaller than the MSR mass, we choose $\mu=R$. For both masses we expand $\alpha_s(R)$ in terms of $\alpha_s(\mu)$, and generate uncertainty bands varying $\mu$ in the range $R/2$ to $2R$. \label{fig:masses}}
\end{figure*}
In \rcite{Gracia:2021nut} an alternative $\mu$-independent short-distance (low-scale) scheme for the mass of the boosted heavy quark
using the parameter choice $n=0$, $\mu_\delta=R$ and $\xi=e^{-\gamma_E}$, was introduced.
We refer to the mass parameter in this scheme as the ``non-derivative jet-mass'' $m^{J^\prime}$. It is related to the pole mass by
\begin{align}\label{eq:jet_mass_ND}
\delta m^{J^\prime}\!(R) ={}& \frac{Re^{\gamma_E}}{2} \log\Bigl[ m \tilde{B}\Bigl(\frac{1}{i R e^{\gamma_E}}, R\Bigr)\Bigr]
= \frac{Re^{\gamma_E}}{2} \sum^\infty_{l=1} \biggl[ \dfrac{\as(\mu)}{4 \pi}\biggr]^l \sum^{l-1}_{k=0} \tilde m'_{lk}\log^k\biggl(\frac{\mu}{R}\biggr)\,,\\
\tilde m'_{l0} ={} & \tilde b_{l0} \,,\qquad\quad\ \, \tilde m'_{lk} = \frac{2}{k}\sum_{j=k}^{l-1} j\beta_{l-j-1} \tilde m'_{j,k-1}\,
\text{ for } k\ge 1\,,\, \nn\\
\Rightarrow\quad
\tilde m'_{21} ={}& 2 \beta_0\tilde b_{10} \,,\qquad \tilde m'_{31} =4\beta_0 \tilde b_{20} + 2 \beta_1\tilde b_{10} \,,\qquad
\tilde m'_{32} = 4 \beta_0^2\tilde b_{1 0} \,. \nn
\end{align}
The mass $m^{J^\prime}$ also obeys an R-evolution equation. The corresponding R-anomalous dimension is obtained from \eq{RAnDim} by replacing $\tilde b_{l1}\to \tilde b_{l0}$, which leads to an R-evolution term $\Delta^{J^\prime}$ analogous to $\Delta^J$.
The recursion relation to obtain $\tilde m'_{l,k\geq 1}$ was given for the first time in \rcite{Mateu:2017hlz}, and in the last line of \eq{jet_mass_ND} we explicitly give the results up to three-loop order.
Even though $m^{J^\prime}(R)$ does not depend on $\mu$, we indicate the residual renormalization-scale dependence of $\delta m^{J^\prime}(R)$ induced by truncation of the infinite sum over $l$ on the right-hand side of \eq{jet_mass_ND}.
The sum over $k$ is generated by re-expanding $\alpha_s(R)$ in powers of $\alpha_s(\mu)$.
This is necessary since to properly cancel the renormalon in the perturbation series of the jet function the subtraction term $\delta m^{J^\prime}$ must be expressed in terms of the strong coupling evaluated at the same scale $\mu$ as the jet function. Finally, since $\delta m^{J^\prime}$ does not involve any derivatives, the asymptotic behavior sets in faster than for the original jet-mass scheme. This may be advantageous in situations in which only few perturbative orders are known (for instance in quarkonium or B decays).
On the other hand, the derivative jet-mass scheme requires less perturbative ingredients than the non-derivative scheme. In particular, at $\ord(\as^4)$ the only missing piece in $\delta m^{J}$ is the four-loop non-cusp anomalous dimension of the jet function. In contrast, for $\delta m^{J^\prime}$ at the same order we would need also the finite piece of the jet function.

To preserve the power counting of the EFT in \eq{shat} we assume $R\sim\Gamt\sim \hat s$ in the MSR mass or either of the two jet-mass schemes.
The series in \eqs{jet_mass_scheme}{jet_mass_ND} define renormalon-free mass schemes that are explicitly local, can be computed perturbatively, and constitute valid alternatives to the MSR mass.
The one- and two-loop contributions to $\delta m^J$ were computed in \rcite{Jain:2008gb}, and with the known three-loop anomalous dimension, it is possible to obtain it at $\mathcal{O}(\as^3)$.
We present the three-loop correction for the two jet-mass schemes, i.e.\ $\delta m^J$ and $\delta m^{J^\prime}$
in \sec{new_results}. Using those results, we show in the two panels of \fig{masses} the absolute differences between the (top quark) MSR mass and the two jet-mass schemes, respectively, for $R\in[4,170]\,$GeV.
We estimate uncertainties by varying the strong-coupling renormalization scale $\mu$ by the conventional factor of two around $R$, see the caption of \fig{masses} for more details.
While the non-derivative jet-mass scheme appears to be well behaved starting from $\mathcal{O}(\alpha_s)$, exhibiting order-by-order convergence towards the MSR benchmark, the original definition in \eq{jet_mass_scheme} involving a derivative has its lowest order outside the two- and three-loop uncertainty bands. Hence, we conclude that the asymptotic behavior sets in faster in the non-derivative scheme, which can be used already at one-loop. At higher orders, both schemes are effective in removing the pole-mass renormalon. We use the MSR as a benchmark to compare both jet-mass schemes since it is not related to the jet function and is directly linked to the relation between the $\MSb$ and pole schemes. From our study one concludes that starting from two loops the three masses can be used.
After all, it is desirable to have several viable options for choosing a mass scheme in order to systematically assess the theoretical uncertainties of future top mass determinations from processes with boosted top jets.

\subsection{Relations among jet function coefficients}
\label{subsec:relations}
In this section we relate the various coefficients appearing in \eq{exp_calB} using properties presented in \app{plus_distributions}, and show how to generate the
coefficients of logarithms and $\cL^{k\geq0}$ distributions at a given loop order
from anomalous dimension and lower-loop ingredients.
The conversion between position and momentum space is performed using \eq{FT}, resulting in the relations
\begin{equation}\label{eq:rel_Fur}
\tilde B_{lk} = \frac{(- 1)^k}{k!} \sum_{j = k - 1}^{2 l - 1} B_{l j} |j| !\,\kappa_{j + 1 - k} \,,\qquad
B_{lk} = \frac{1}{|k|!} \sum_{j = k + 1}^{2 l} \tilde{B}_{l j} (- 1)^j j!\,\hat{\kappa}_{j - k - 1} \,,
\end{equation}
where the $\kappa_i$ and $\hat \kappa_i$ are given in \eq{GamRed}. The imaginary part of $\cB$ can be taken with the help of \eq{ImPart} and implies the following relations among coefficients:
\begin{align}\label{eq:rel_Im}
B_{lk} ={}& (- 1)^{k+1} \sum_{j = 0}^{l - \left\lceil \frac{k + 1}{2} \right\rceil} \binom{2 j + 2 + k}{2 j + 1}\dfrac{(- \pi^2)^j\max (1,k+1)}{2j+2+k}\mathcal{B}_{l, k + 2 j + 1} \,,\\
\cB_{lk} = \, & (- 1)^{k+1} \sum_{j = 0}^{l - \left\lceil \frac{k}{2} \right\rceil} \binom{2 j + k}{2 j} \frac{(- \pi^2)^j (2^{2 j} - 2)}{\max(1,2j+k)} \mathbb B_{2 j}\,B_{l, k + 2 j - 1} \,, \nonumber
\end{align}
where $\mathbb B_n$ denotes the $n$-th Bernoulli number and
$\lceil n \rceil$ is the closest integer larger or equal to $n$ (also known as the {\it ceiling} function).
The position-space jet function coefficients $\tilde{B}_{lk}$ and the coefficients $\tilde b_{lk}$ of the corresponding exponent in \eq{B_exp_series} are related by the following recursion relation valid for $l\ge 1$ and $0\leq k\leq 2l$,
\begin{equation}\label{eq:recExp}
\tilde B_{lk} = \frac{1}{l} \sum_{j = 1}^{l} \,j\! \sum_{n = \max [0, k +
2 (j-l)]}^{\min (j + 1, k)} \tilde b_{jn} \, \tilde B_{l - j, k - n} \,,
\end{equation}
where $\tilde B_{00}=1$. To invert this relation, we simply pull out the $j=l$ term in the outer sum, which forces $n=k$ in the inner sum, and isolate $\tilde b_{l n}$, finding again a recursive formula, now valid for $l\ge 1$ and $0\leq k\leq l+1$,
\begin{equation}\label{eq:recLog}
\tilde b_{l k} = \tilde B_{l k} - \frac{1}{l} \sum_{j = 1}^{l - 1}\, j\!\sum_{n = \max [0, k + 2 (j - l)]}^{\min (j + 1, k)} \tilde b_{jn} \, \tilde B_{l - j, k - n} \,,
\end{equation}
that serves to express $\tilde b_{l k}$ in terms of $\tilde B_{l k}$. It should be noted that not all $\tilde B_{l k}$ coefficients appear in \eq{recLog} as a consequence of \eqs{B_exp_series}{RGE_for_B_FT}.
Finally, the momentum-space jet-function coefficients for an unstable quark introduced in \eq{coefGamma} are related to those of the stable-quark forward-scattering matrix element $\cal B$ using \eq{ImGamma} as follows:
\begin{align}
\hat B_{lk}\biggl( \frac{\hat s}{\Gamt} \biggr) = \sum_{n = k}^{2 l} \binom{\,n\,}{k}\mathcal{B}_{l n} (- 1)^{\left\lceil\frac{n - k}{2} \right\rceil} \biggl[y\biggl( \frac{\hat s}{\Gamt} \biggr)\biggr]^{n - k} \biggl( \frac{\hat s}{\Gamt} \biggr)^{[(n - k)\,{\rm mod}\,2]}\,,
\label{eq:Bhatij}
\end{align}
where $[n\,{\rm mod}\,2]$ is the remainder of $n/2$, which is zero (one) for even (odd) numbers.
Relations between other coefficients are obtained applying the previous results sequentially.

In the rest of this section, we derive relations that can be used to express the logarithmic coefficients appearing in the renormalized quantities of \eq{exp_calB} in terms of (cusp, non-cusp and strong-coupling) anomalous dimension and lower-loop non-logarithmic coefficients. We start with the renormalized position-space jet function. Solving the RGE in \eq{RGE_for_B_FT} at leading order in $\as$ using the parametrization in \eq{B_exp_series} we find $\tilde{b}_{12}=\Gamma_{\!0}^{c}$, which due to exponentiation implies that the highest logarithmic term at each loop order of $\tilde B$ is determined only by the one-loop cusp anomalous dimension coefficient:

\begin{equation}
\tilde{B}_{l, 2 l} = \frac{1}{l!}(\Gamma_{\!0}^{c})^l\,.
\label{eq:Btilden2n}
\end{equation}
Iteratively solving \eq{RGE_for_B_FT} now fixes all other logarithmic terms, i.e.\ $\tilde B_{lk}$ with $l\geq1$ and $1\leq k\leq 2l-1$, in terms of the non-logarithmic $\tilde B_{n\le l-1,0}$ and anomalous dimension coefficients via the relation
\begin{equation}
\tilde{B}_{l k} = \frac{1}{k} \!\left[ (1-\delta_{k1})\!\!\!\!\sum^{l - 1}_{n = \left\lfloor \frac{k - 1}{2} \right\rfloor} \!\!\!\!2\tilde{B}_{n, k - 2} \Gamma^\rc_{\!l - n - 1} + \!\!\!\!\sum^{l - 1}_{n =\left\lfloor \frac{k}{2} \right\rfloor}\!\!\! \tilde{B}_{n , k - 1}(2n \beta_{l - n - 1}+ \gamma^B_{l - n - 1}) \right]\!.
\label{eq:recBtilde}
\end{equation}
The first sum in \eq{recBtilde} vanishes for $k=1$ since $\tilde B_{n,-1}=0$ for any $n$, but we nevertheless enforce this with a Kronecker symbol for clarity. Here $\lfloor n \rfloor$ is the closest integer smaller or equal to $n$ (also known as the {\it floor} function).

In a similar manner, we can solve the momentum-space RGE in \eq{RGE_for_B}, using \eq{BRed} for the convolution of two plus distributions. Alternatively, we can translate \eqs{Btilden2n}{recBtilde} via \eq{rel_Fur}, to obtain
\begin{align}\label{eq:BnR}
B_{l, 2 l-1} = \frac{2(\Gamma^\rc_{\!0})^l}{(l-1)!}\,,\qquad
B_{l, 2 l-2} = (1-2l)\frac{(\Gamma^\rc_{\!0})^{l - 1}}{(l - 1) !} \biggl[ \gamma_0^B + \frac{2}{3} (l -1) \beta_0 \biggr],
\end{align}
where $l\geq 1$, and
the recursive relation
\begin{align}\label{eq:BbarRest}
B_{lk} ={}& \frac{2 (1 - \delta_{k0})}{{\rm max}(1,k-1)}\! \! \sum^{l - 1}_{j = \left\lfloor \frac{k}{2} \right\rfloor} \!\!\!\!B_{j, k - 2} \Gamma^\rc_{\!l - j - 1} - \frac{1}{{\rm max}(1,k)}\!\!\sum^{l - 1}_{j = \left\lceil \frac{k}{2} \right\rceil}\!\! B_{j ,k - 1} (2 j \beta_{l - j - 1} \!+ \gamma^B_{l - j - 1})
\\
& + \frac{2(-1)^k}{k!} \sum^{2 l - 2}_{n= k+1} (n-1)! (-1)^n\zeta_{n - k + 1}\!\! \sum^{l - 1}_{j = \left\lceil \frac{n}{2}\right\rceil} \!\!B_{j,n-1} \Gamma^\rc_{\!l - j - 1}\,,\nn
\end{align}
for $0\leq k \leq 2l-3$ and $l\geq 2$. Here $\zeta_s=\sum_{j=1}^\infty j^{-s}$ is the Riemann zeta function and the first sum in \eq{BbarRest} does not contribute for $k=1$ since $B_{l,-2}=0$ for all $l$, but we again enforce this with a Kronecker symbol.

To obtain the logarithmic expansion coefficients of ${\cal B}(s,\mu)$ we iteratively solve its RGE in \eq{RGE_for_B} using of \eq{BcalRed} and find relations similar to the ones of the momentum-space jet function. Concretely, we have for the two leading logarithmic coefficients at each loop order
\begin{equation}
{\cal B}_{l , 2 l} = \frac{(\Gamma^\rc_{\!0})^l}{l!} \,,\qquad {\cal B}_{l , 2 l-1} = \frac{(\Gamma^\rc_{\!0})^{l - 1}}{(l - 1) !} \biggl[ \gamma_0^B + \frac{2}{3} (l -1) \beta_0 \biggr],
\end{equation}
where the first relation is valid for $l\geq 0$ while the second holds for $l\geq1$. The rest of coefficients with $l\geq2$ and $1\leq k \leq 2l-2$ are generated by
\begin{align}
{\cal B}_{l k} ={}& \frac{1}{k} \Biggl[2 (1 - \delta_{k 1})\! \! \! \sum^{l - 1}_{j = \left\lfloor \frac{k - 1}{2} \right\rfloor}{\cal B}_{j , k - 2}\, \Gamma^\rc_{\!l - j - 1} +\!\!\!\sum^{l - 1}_{j = \left\lfloor \frac{k}{2}\right\rfloor} {\cal B}_{j , k - 1} (2 j \beta_{l - j - 1} + \gamma^B_{l - j - 1}) \Biggr]\\
& +\! \frac{2(-1)^k}{k !} \sum^{2 l - 2}_{n = k}(-1)^n n! \,\zeta_{n - k + 2} \sum^{l - 1}_{j = \left\lceil \frac{n}{2} \right\rceil} {\cal B}_{j n}\, \Gamma^\rc_{\!l - j - 1} \,.\nn
\end{align}
Just like in \eqs{recBtilde}{BbarRest} we inserted here a Kronecker delta for clarity.

For completeness, we also give the corresponding relations for the coefficients of ($m$ times the) logarithm of the position-space jet function $\tilde b(x,\mu)$ in \eq{B_exp_series}, which can also be extracted from \eq{RGE_for_B_FT}. The coefficients of the single and double logarithm are determined by
\begin{equation}\label{eq:nonRec}
\tilde{b}_{l 1} = \gamma^B_{l-1} + 2\!\sum_{j = 1}^{l -1} j \beta_{l - j - 1}\, \tilde{b}_{j0}\,,\qquad
\tilde{b}_{l 2} =\Gamma^\rc_{\!l-1} + \sum_{j = 1}^{l -1} j \beta_{l - j - 1}\, \tilde{b}_{j1}\,.
\end{equation}
It should be noted that the single sums in both expressions vanish for $l=1$.
These relations can e.g.\ be employed to obtain the cusp and non-cusp anomalous dimension coefficients from $\tilde{b}_{l,k\leq2}$.
All other logarithmic coefficients are fixed by
\begin{equation}\label{eq:Rec}
\tilde{b}_{lk} = \frac{2}{k} \sum_{j =k - 2}^{l -1} \!\!\!j \beta_{l - j - 1} \tilde{b}_{j, k - 1} \,,
\end{equation}
valid for $l\geq2$ and $3\leq k\leq l+1$.

Finally, we note that because of \eqs{nonRec}{Rec} the logarithmic coefficients $\tilde m_{l,k\geq1} = (k+1)\, \tilde{b}_{l,k+1} $ of the series relating the jet- and the pole-mass schemes in \eq{jet_mass_scheme} obey the simple recursion relation
\begin{equation}\label{eq:JMrec}
\tilde{m}_{lk} = \frac{2}{k} \Biggl[ \delta_{k1} \Gamma^\rc_{\!l-1}
+\sum_{j = {\rm max}(1,k-1)}^{l - 1} \!\! j \,\beta_{l - j - 1} \,\tilde{m}_{j,k-1} \Biggr] ,
\end{equation}
valid for $l\geq1$ and $1\leq k\leq l$ and in agreement with \eq{muAnDim}.
Therefore, the $\tilde{m}_{lk}$ are fixed by $\tilde m_{j0}=\tilde b_{j1}$ (with $j\le l-1$), and the coefficients of the cusp anomalous dimension and the $\beta$-function. All the anomalous dimension coefficients required to apply the relations in this section up to three-loop order are collected in \app{expansion}.

\section{Computation of the jet function}
\label{sec:computation}

\begin{figure*}
\centering
\subfigure[$C_FC_A^2$]{\includegraphics[width=0.3\textwidth]{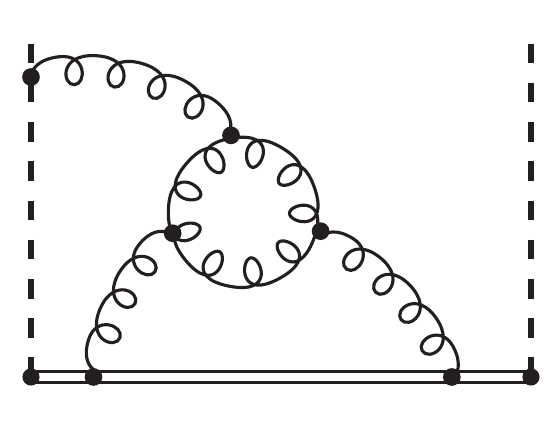}
\label{fig:FDEx5}}
\subfigure[$C_FT_F^2\nl^2$]{\includegraphics[width=0.3\textwidth]{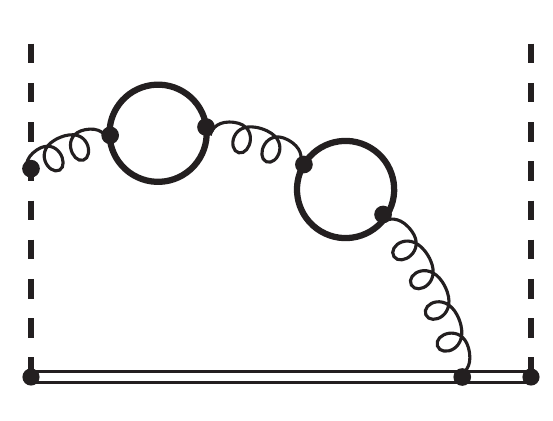}
\label{fig:FDEx2}}
\subfigure[$C_F^3$, $C_F^2C_A$, $C_FC_A^2$]{\includegraphics[width=0.3\textwidth]{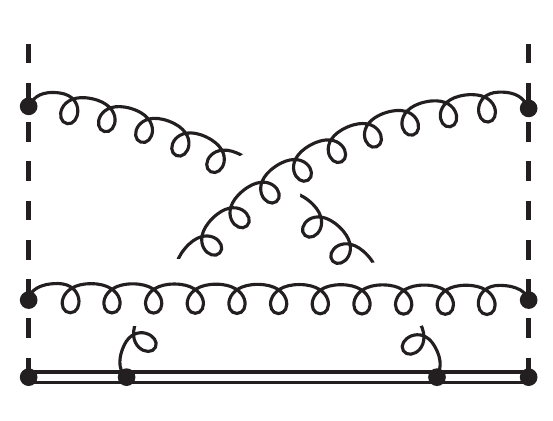}
\label{fig:FDEx3}}
\\[0.1cm] %
\subfigure[$C^2_F T_F \nl$, $C_FC_AT_F\nl$]{\includegraphics[width=0.3\textwidth]{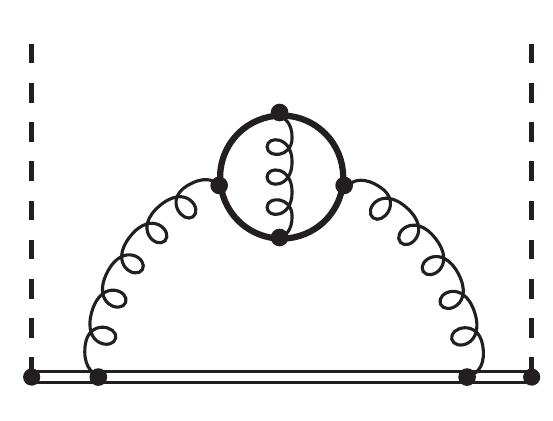}
\label{fig:FDEx1}}
\subfigure[$C^2_F T_F \nl$, $C_FC_AT_F\nl$]
{\includegraphics[width=0.3\textwidth]{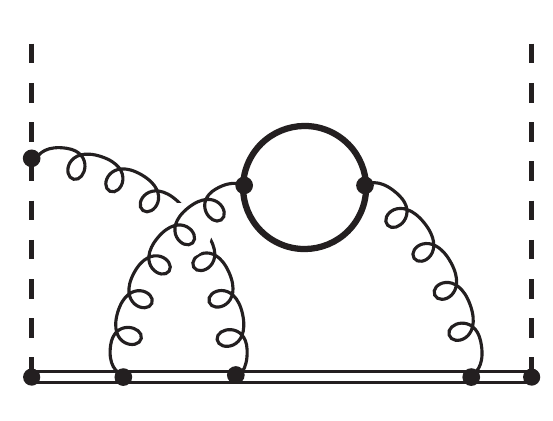}
\label{fig:FDEx4}}
\subfigure[$C^2_F T_F\nl$]
{\includegraphics[width=0.3\textwidth]{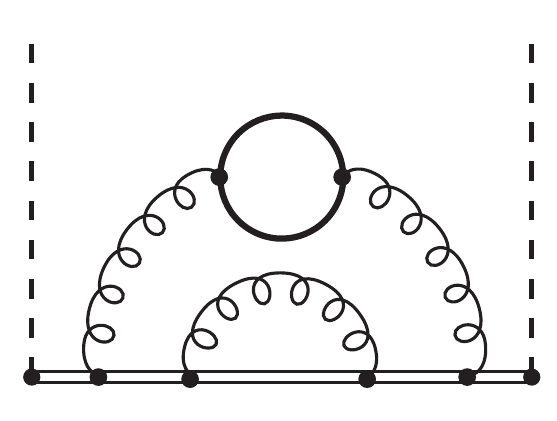}
\label{fig:FDEx6}}%
\caption{Sample of Feynman diagrams contributing to the jet function at three loops. Double, curly, dashed and solid lines represent heavy-quarks, gluons, light-like Wilson lines and light quarks, respectively. Each diagram contains the color factors indicated at the bottom of the panel.}%
\label{fig:FDEx}%
\end{figure*}

In this section we briefly describe our computation of the bHQET jet function by means of Feynman diagrams up to third loop order. To begin with, we note that in many aspects our calculation resembles that of the soft function for heavy-to-light quark decays (or equivalently the partonic version of the B-meson shape function) in \rcite{Bruser:2019yjk}. This soft function can, like the jet function in \eq{BWilsonCorr}, be expressed in terms of a vacuum correlator of light-like and time-like Wilson lines, albeit with different endpoints. We will therefore follow here a similar calculational strategy as \rcite{Bruser:2019yjk}. The general workflow is practically the same at each loop order, hence the following explanations are mostly generic. To illustrate the variety of Feynman diagrams involved in the computation of the three-loop jet function, we show a selection of them in \fig{FDEx}.

The bare jet-function matrix element in \eq{def_calB} depends only on the Lorentz invariant $\hat{s}=2\, v \cdot r$ as can be seen as follows. In our setup we have three Lorentz vectors $r$, $v$ and $\nb$, where $r$ appears in \eq{def_calB} as the Fourier-conjugate momentum of the position vector, and $v^2=1$ and $\nb^2=0$ are fixed. Since the jet function is Lorentz invariant, it can therefore only depend on the scalar products $r^2$, $r \cdot \nb$, $r \cdot v$, $\nb \cdot v$. Without loss of generality, we choose the momentum routing in all Feynman diagrams such that $r$ only flows through the heavy quark propagators in the same direction as $v$. Therefore, $r$ only appears in the scalar product $r \cdot v$ and the jet-function is independent of $r^2$ and $r \cdot \nb$. The (HQET-type) heavy quark propagators (which equal those of the time-like Wilson lines $W_v$) then take the form
\begin{align}
\imineq{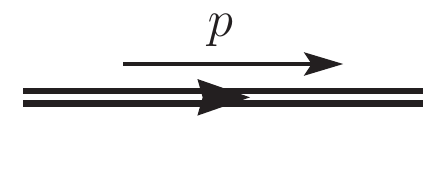}{8} = \frac{i}{v \cdot (p+r)+i\eta } = \frac{i}{v \cdot p + \hat{s}/2+i\eta } \,,
\label{eq:prop_bHQET}
\end{align}
where $p$ is a linear combination of loop momenta and $\hat{s}/2$ acts as an external offshellness.
The only remaining scalar product to discuss is therefore $\nb \cdot v \gg 1$.
An $\nb$ dependence of a Feynman diagram could only arise from gluons coupling to the light-like Wilson lines $W_n$ or $W_n^\dagger$.
The Wilson line $W_n$ defined in \eq{def_Wilson} is, however, invariant under rescaling $\nb \to \alpha\, \nb$
and so are, therefore, the individual Feynman diagrams.
Since upon loop integration the diagrams could depend on $\nb$ only via $\nb \cdot v$, this rescaling invariance implies that the jet function (matrix element) as well as single Feynman graphs are in fact independent of $\nb \cdot v$.
We can therefore set $\hat{s}=-1$, and $\nb \cdot v =1$ in our calculation and restore the $\hat{s}$ dependence of the resulting bare jet function matrix element by dimensional analysis.

Our computation of the bare jet function matrix element follows the general framework of many modern QCD multi-loop calculations. We start by generating the relevant Feynman diagrams with \texttt{qgraf}~\cite{Nogueira:1991ex}.\footnote{In our setup we replace the four-gluon vertex by a four-gluon interaction mediated by a non-propagating auxiliary particle following~\rcites{Draggiotis:1998gr, Duhr:2006iq, Weinzierl:2016bus}. In this way the color structure is disentangled from the Lorentz structure of each diagram and can be calculated separately facilitating the automated computation.}
Based on the denominator structure arising from the propagators each diagram is then assigned to an integral family (IF) using the \texttt{Mathematica} package \texttt{Looping}~\cite{Looping}. For this mapping procedure \texttt{Looping} implements an algorithm outlined in \rcite{Pak:2011xt} in order to eliminate the ambiguities due to loop momentum shifts, i.e.\ to identify loop integrands that only differ by the choice of loop momentum routing through the propagators. All planar Feynman diagrams are conveniently accommodated in one (maximal) planar IF (with at most two loop momenta flowing through a propagator). For the non-planar diagrams \texttt{Looping} automatically identifies a small number of suitable IFs. Next, the integrand of each diagram is obtained using \texttt{FORM}~\cite{Ruijl:2017dtg}: We start by applying the Feynman rules to obtain an algebraic expression, followed by the computation of the color factor using the \texttt{color.h} package~\cite{vanRitbergen:1998pn}. After that, Lorentz indices are contracted and all Dirac traces arising in diagrams with closed light quark loops are evaluated. All necessary input files for processing the individual diagrams with \texttt{FORM} are automatically generated by \texttt{Looping}.

Starting at two loops, diagrams can have linearly-dependent (light-like and time-like) Wilson-line propagator denominators. These dependencies are removed in \texttt{Looping} by applying the multivariate partial fractioning algorithm outlined in \rcite{Pak:2011xt}, which makes use of a Gr\"obner basis computation. The resulting scalar integrals are mapped to IFs,
which are defined through a complete set of linearly-independent propagator denominators, and are thus ready for automated integration-by-parts (IBP) reduction.
In order to simplify the expressions \texttt{Looping} automatically identifies all scaleless integrals and discards them in the first place. Furthermore, it exploits symmetries
(related to momentum shifts)
within each family to reduce the number of occurring integrals.

The IBP reduction to an (at best) minimal set of master integrals (MIs) is performed for each IF separately using \texttt{FIRE6}~\cite{Smirnov:2019qkx} in combination with \texttt{LiteRed}~\cite{Lee:2012cn, Lee:2013mka}. The MIs found in this way are then mapped across the IFs using \texttt{Looping} in order to identify relations among MIs from different IFs. After that, we are left with a total of 21 MI candidates. To check for additional linear relations between these MI candidates not found by our IBP setup, we use a strategy described in \rcite{Bruser:2018rad} in the context of the three-loop calculation of the inclusive (massless) quark jet function. Namely, we use dimensional recurrence relations~\cite{Tarasov:1996br, Tarasov:1997kx, Lee:2009dh, Lee:2010wea, Baikov:1996iu} implemented in \texttt{Looping} to express the $d$-dimensional MI candidates \mbox{$\vec{g}^{\,(d)}=(g_1^{(d)},\dots,g_{21}^{(d)})$} in terms of $(d\pm2)$-dimensional integrals. We then perform an IBP reduction of the latter to write the MI candidates in $d$ dimensions as a linear combination of MIs in $d\pm2$,
\begin{align}
\vec{g}^{\,(d)}= T_\mp^{(d \pm 2)} \vec{g}^{\,(d \pm 2)} \,,
\end{align}
where $T_\mp^{(d \pm 2)}$ are matrices with entries that depend rationally on $d$.
These matrices must satisfy the condition
\begin{align}
\vec{g}^{\,(d)}= T_\mp^{(d\pm2)} T_\pm^{(d)} \vec{g}^{\,(d)}\,,
\end{align}
from which additional nontrivial relations among the MI, if present, can directly be read off.
In our case we find one more relation reducing the number of independent MIs to 20.
The number of Feynman diagrams, nontrivial scalar integrals, and MIs relevant for our jet function calculation at one, two, and three loops is displayed in \tab{estimate_integrals}.

\begin{table}[t!]
\centering
\begin{tabular}{cccc}
\toprule
& $1$ loop & $2$ loops & $3$ loops \\ \midrule
Feynman diagrams & $4$ & ${\sim}50$ & ${\sim}1100$ \\
Scalar integrals & $3$ & ${\sim}70$ & ${\sim}5400$ \\
Master integrals & $1$ & $3$ & $20$ \\
\bottomrule
\end{tabular}
\caption{Number of Feynman diagrams, non-vanishing scalar integrals, and master integrals contributing to the jet function at one, two and three loops. The number of Feynman diagrams and scalar integrals may vary depending on the setup and serves as a rough complexity measure.} \label{tab:estimate_integrals}
\end{table}

Finally, we compute the MIs analytically as explained in some detail in \app{computationMI}. The results for all MIs required to three-loop order are presented in \app{calB_and_MI}. All diagrams are added up and the results for the MIs are inserted to obtain the bare jet function matrix element $\cB^\bare(\hat{s})$. As a stringent test of our setup we perform the calculation in a general covariant
gauge.
We have checked that the dependence on the gauge parameter
drops out in the bare jet function matrix element once expressed in terms of a minimal set of MIs. Taking the imaginary part by means of \eq{Imag} yields, according to \eq{def_B}, the bare jet function result. The latter is renormalized (upon expansion in $\eps$) as explained in \sec{JF_renormalization}, where we also argue that the renormalization and the translation between the matrix element $\cB$ and the jet function $B$ in fact commutes.

\section{Analytic results for the jet function and jet-mass scheme}
\label{sec:new_results}

In this section we summarize the results of our computations. The expressions we obtain for the bare one- and two-loop forward-scattering matrix elements agree with the previous results of \rcites{Fleming:2007xt,Jain:2008gb} and are given for arbitrary dimension $d$ (i.e.\ to all orders in $\varepsilon$) in \subsecs{1loop}{2loop}. The (position-space) three-loop jet function, which is the main result of this paper, is presented and briefly discussed in \subsec{3loop}. The corresponding expressions for the renormalized momentum-space jet function can be found in \app{JF}. We also include explicit results for the jet-mass scheme defined in \eq{jet_mass_scheme}. Recall that the non-logarithmic coefficients of the non-derivative jet-mass equal those of the logarithm of the position-space jet function as shown in \eq{jet_mass_ND},
where also the logarithmic coefficients are already given in terms of anomalous dimension and non-logarithmic coefficients.

\subsection{One loop}
\label{subsec:1loop}
The one-loop coefficient in the expansion of the bare forward-scattering matrix element in \eq{exp_calB} can be expressed in terms of the one-loop MI depicted in \fig{MI1L} as follows:
\begin{equation}
\mathcal{B}_{1}^\bare
=
\,
4\,\frac{d-2}{d-4} \,
\MI{1}{}\, C_F\,.
\end{equation}
The $d$-dimensional result for $\MI{1}{}$ is given in \app{one-loop_result}. The (logarithm of the) renormalized position-space jet function at one loop is, according to \eqs{exp_calB}{B_exp_series}, given by
\begin{equation}
\tilde B_{10}=\tilde b_{10}=\biggl(4 + \frac{\pi^2}{6}\biggr)C_F, \qquad
\tilde B_{11}=\tilde b_{11}=\tilde m_{10}=\gamma^B_0 \,,
\qquad
\tilde B_{12}=\tilde b_{12}=\frac{\tilde m_{11}}{2} =\Gamma^\rc_{\!0}\,,
\end{equation}
where at this order $\Gamma^\rc_{\!0}=\gamma^B_0=4 C_F$, rendering the two logarithmic coefficients equal. With $N_c=3$, the numerical values of the non-logarithmic coefficients in \eqsss{exp_calB}{B_exp_series}{jet_mass_scheme} are $\tilde B_{10} = \tilde b_{10}=7.52658$ and $\tilde m_{10}= 16/3$.

\subsection{Two loops}
\label{subsec:2loop}
For the two-loop coefficient of the bare jet-function matrix element we have
\begin{align}
\mathcal{B}_{2}^\bare
=&
\biggl[ \frac{32 (d-2)^2 (2 d-7) (2 d-5)}{(d-4)^3 (d-3)} \, \MI{2}{a} \biggr] C_F^2 \\ \nonumber
& +\biggl[\frac{8 (2 d-5) (3 d^4-34 d^3+153 d^2-302 d+192)}{(d-4)^3 (d-3)
(d-1)} \, \MI{2}{a}-\frac{8 (d^2-7 d+13)}{(d-4)^2} \, \MI{2}{b} \\ \nonumber
& -
\frac{8(2d-7)}{(d-4) (d-3)} \, \MI{2}{c}\biggr] C_FC_A +\biggl[ \frac{32 (d-3) (d-2) (2 d-5)}{(d-4)^2 (d-1)} \, \MI{2}{a}\biggr] C_F T_F n_\ell\,,
\end{align}
with the two-loop MIs depicted in \fig{MI2L} and explicitly given to all orders in $\eps$ in \app{two-loop_result}.
The two-loop coefficients in \eq{B_exp_series} are thus
\begin{align}
\tilde b_{20} ={}& \biggl(\frac{5918}{81}-\frac{101 \zeta_3}{9}+\frac{139 \pi ^2}{108}-\frac{17
\pi ^4}{180}\biggr) C_FC_A + \biggl(\frac{4 \zeta_3}{9}-\frac{2248}{81}-\frac{5 \pi ^2}{27}\biggr) C_FT_F n_\ell\,,
\label{eq:2L_JF}\\
\tilde b_{21} ={}& 2 \tilde b_{10} \beta_0 + \gamma^B_1\,,\qquad
\tilde b_{22} = \beta_0 \gamma^B_0 + \Gamma^\rc_{\!1}\,,\qquad
\tilde b_{23} = \frac{2}{3} \beta_0 \Gamma^\rc_{\!0}\,,\nn
\end{align}
where the logarithmic pieces are fixed by \eqs{nonRec}{Rec} and numerically we have $\tilde b_{20} = 252.299 - 19.3644 n_\ell$.
Notice that no $C_F^2$ is present in \eq{2L_JF} due to the non-Abelian exponentiation theorem. The two-loop coefficients for the jet-mass scheme defined in \eq{jet_mass_scheme} are therefore ($N_c = 3$)
\begin{align}
\tilde m_{20}
={}&
\biggl( \frac{2188}{27}-20 \zeta_3-\frac{4 \pi ^2}{3} \biggr) C_FC_A
-\frac{752}{27} C_FT_Fn_\ell
= 175.346 - 18.5679 n_\ell\,,\\
\tilde m_{21} ={}& 2 \bigl(\tilde m_{10} \beta_0 + \Gamma^\rc_{\!1}\bigr)\,,\qquad
\tilde m_{22} = 2 \beta_0 \Gamma^\rc_{\!0}\,,\nn
\end{align}
where the logarithmic coefficients can also be determined with \eq{JMrec}. The renormalized position-space jet function at two loops is, according to \eq{exp_calB}, given by the coefficients
\begin{align}
\tilde B_{20}
={}&
\biggl(\frac{4 \zeta_3}{9}-\frac{2248}{81}-\frac{5 \pi ^2}{27} \biggr) C_F T_F n_\ell
+
\biggl(\frac{5918}{81}-\frac{101 \zeta_3}{9}+\frac{139 \pi ^2}{108}-\frac{17 \pi ^4}{180} \biggr) C_F C_A
\\ \nn & +
\biggl( 8+\frac{2 \pi ^2}{3}+\frac{\pi ^4}{72} \biggr) C_F^2=280.624 - 19.3644 n_\ell\,,
\\
\nn \tilde B_{21}
={}&
\gamma_1^B + \bigl(2 \beta_0 + \gamma_0^B\bigr) \tilde B_{10}\,,
\qquad\qquad\qquad\qquad\tilde B_{23}
=
\Gamma^{\rm c}_{\!0}\biggl(\dfrac{2}{3} \beta_0 + \gamma^{B}_0\biggr)\,,
\\ \nn \tilde B_{22}
={}&
\Gamma_{\!1}^{\text
c} + \beta_0 \gamma_0^B + \dfrac{1}{2} \bigl(\gamma^B_0\bigr)^2 +
\Gamma_{\!0}^{\rm c} \tilde B_{10}\,,\qquad\qquad
\tilde B_{24}
=
\dfrac{1}{2}(\Gamma^{\rm c}_{\!0})^2\,.
\end{align}
The logarithmic terms can also be obtained recursively using \eqs{BnR}{BbarRest}.

\subsection{Three loops}
\label{subsec:3loop}
From our explicit calculation we obtain the following three-loop coefficients of the logarithm of the position-space jet function defined in \eq{B_exp_series}:\footnote{The expression of $\cB^\bare_3$ in terms of MIs for arbitrary $d$ can be obtained from the authors upon request.}
\begin{align}
\tilde b_{30}
={}&
\biggl( \frac{203 \pi ^2 \zeta_3}{27}-\frac{105398 \zeta_3}{243}+\frac{236 \zeta_3^2}{9} + \frac{902 \zeta_5}{9} + \frac{31952191}{26244} + \frac{93821 \pi
^2}{8748}
\label{eq:JF_RESULT}
\\ \nonumber & -\frac{3023 \pi ^4}{4860} + \frac{1031 \pi ^6}{10206} \biggr) C_FC_{A}^2
+
\biggl( \frac{3488 \zeta_3}{243}+\frac{846784}{6561}-\frac{8 \pi ^2}{243}+\frac{52 \pi ^4}{1215} \biggr) C_FT^2_F n_\ell ^2
\\ \nonumber
& + \biggl( \frac{10760 \zeta_3}{81}+\frac{8 \pi ^2 \zeta_3}{9}+\frac{224 \zeta_5}{9}-\frac{124717}{486}-\frac{55 \pi ^2}{54}+\frac{148 \pi ^4}{405}\biggr) C_F^2 T_F n_\ell\\ \nonumber
& + \biggl( \frac{1664 \zeta_3}{81}-\frac{76 \pi ^2 \zeta_3}{27}-\frac{88 \zeta_5}{3}-\frac{5273287}{6561}-\frac{12793 \pi
^2}{2187}-\frac{421 \pi ^4}{1215} \biggr) C_FC_A T_F n_\ell\,,\nn\\
\tilde b_{31}
={}& 4 \tilde b_{20} \beta_0 + 2\tilde b_{10} \beta_1 + \gamma^B_2\,, \qquad\qquad\quad
\tilde b_{32} = 4 \tilde b_{10} \beta_0^2 + \beta_1 \gamma^B_0 + 2 \beta_0 \gamma^B_1 + \Gamma^\rc_{\!2}\,,\nn\\
\tilde b_{33}
={}& \frac{2}{3}\beta_1 \Gamma^\rc_{\!0} + \frac{4}{3} \beta_0 ( \beta_0 \gamma^B_0 + \Gamma_{\!1}^c)\,,\qquad\quad\,
\tilde b_{34} = \frac{2}{3} \beta_0^2 \Gamma_{\!0}^c\,.\nn
\end{align}
Recall that the non-Abelian exponentiation theorem predicts the absence of the color factors $C_F^3$ and $C_F^2C_A$ in \eq{JF_RESULT} and puts constrains on the diagrams contributing to the $C_F^2T_F n_\ell$ term, see \subsec{JF_def}. We have explicitly checked that our three-loop result for $\cB^\bare$ complies with non-Abelian exponentiation already before $\eps$ expansion, i.e.\ in $d$ dimensions.

Our calculation reproduces the known three-loop anomalous dimension coefficients $\gamma_2^B$ and $\Gamma^\rc_{\!2}$, see \app{expansion}, from the $1/\varepsilon$ coefficient of $\tilde Z_B$ according to \eq{consistency_1}. In particular, it represents the first direct computation of the non-cusp anomalous dimension $\gamma_2^B$ of the bHQET jet function. Moreover, our result for $\tilde b$ obeys the RG constraints in \eqs{nonRec}{Rec}. We emphasize that \eq{Rec} is an additional consistency test for our result whereas \eq{nonRec} allows to obtain the anomalous dimension from the logarithm of the position-space jet function. We also reproduce the $C_FT_F^2n_\ell ^2$~term of $\tilde{b}_{30}$ obtained in \rcite{Gracia:2021nut}, and our analytic result of the full $\tilde{b}_{30}$ lies within the uncertainties of the estimate there ($N_c = 3$):
\begin{align}\label{eq:comparison}
\tilde b^{\rm exact}_{30}={}&50.054\, n_\ell^2-1899.8 \,n_\ell+12834\,,\\
\tilde b^{\text{\cite{Gracia:2021nut}}}_{30}=&\,50.054\, n_\ell^2-(1600\pm 550)\, n_\ell+11700\pm 3200 \,.\nn
\end{align}
The three-loop coefficients for the jet-mass scheme defined in \eq{jet_mass_scheme} thus read
\begin{align}
\tilde m_{30} ={}&\biggl(\frac{128 \zeta_3}{9}+\frac{97856}{729} \biggr) C_F T^2_F n_\ell^2
+ \biggl( \frac{1184 \zeta_3}{9}+\frac{16 \pi ^4}{45}-\frac{6266}{27} \biggr) C_F^2 T_F n_\ell \\
&+ \biggl(\frac{272 \zeta_3}{27}-\frac{76 \pi ^4}{135}+\frac{1400
\pi ^2}{243}-\frac{662588}{729} \biggr) C_FC_A T_F n_\ell \nn\\
& + \biggl(120 \zeta_5+\frac{112 \pi ^2 \zeta_3}{9}-\frac{11848 \zeta_3}{27}+\frac{209 \pi ^4}{135}-\frac{6292 \pi
^2}{243}+\frac{1039619}{729} \biggr)C_FC_A^2\,,\nn\\
\tilde m_{31} ={}&4 \tilde m_{20} \beta_0 + 2 \tilde m_{10} \beta_1 + 2 \Gamma_{\!2}^c\,,~\quad
\tilde m_{32} =\, 2 \beta_1 \Gamma_{\!0}^c + 4 \beta_0 ( \tilde m_{10} \beta_0 +\Gamma_{\!1}^c)\,,~\quad
\tilde m_{33} = \frac{8}{3} \beta_0^2 \Gamma_{\!0}^c\,.\nn
\end{align}
Again, \eq{JMrec} can be used to obtain the logarithmic coefficients $m_{3,k=1,2,3}$. Numerically we have $\tilde m_{30} = 12791.1 - 1824.47 n_\ell + 50.443 n_\ell^2$ (for $N_c=3$).
Finally, the three-loop non-logarithmic coefficient of the renormalized position-space jet function in \eq{exp_calB} is
\begin{align}
\tilde{B}_{30}
={}&
\biggr( \frac{3488 \zeta_3}{243}+\frac{846784}{6561}-\frac{8 \pi ^2}{243}+\frac{52 \pi ^4}{1215} \biggr) C_F T_F^2 n_\ell^2
\\ \nn
& + \biggr(\frac{236 \zeta_3^2}{9}-\frac{105398 \zeta_3}{243}+\frac{203 \pi ^2 \zeta_3}{27}+\frac{902 \zeta_5}{9}+\frac{31952191}{26244}+\frac{93821 \pi ^2}{8748}
\\ \nn &
-\frac{3023 \pi ^4}{4860}+\frac{1031 \pi ^6}{10206}\biggr) C_F C_A^2 +
\biggr( \frac{32}{3}+\frac{4 \pi ^2}{3}+\frac{\pi ^4}{18}+\frac{\pi ^6}{1296} \biggr) C_F^3
\\ \nn
& + \biggr( \frac{1664 \zeta_3}{81}-\frac{76 \pi ^2 \zeta_3}{27}-\frac{88 \zeta_5}{3}-\frac{5273287}{6561}-\frac{12793 \pi
^2}{2187}-\frac{421 \pi ^4}{1215} \biggr) C_F C_A T_F n_\ell
\\ \nn
&+ \biggr( -\frac{404 \zeta_3}{9}-\frac{101 \pi ^2 \zeta_3}{54}+\frac{23672}{81}+\frac{4210 \pi ^2}{243}-\frac{529 \pi
^4}{3240}-\frac{17 \pi ^6}{1080} \biggr) C_F^2 C_A
\\ \nn
&+ \biggr( \frac{10904 \zeta_3}{81}+\frac{26 \pi ^2 \zeta_3}{27}+\frac{224 \zeta_5}{9}-\frac{178669}{486}-\frac{3103 \pi
^2}{486}+\frac{271 \pi ^4}{810} \biggr) C_F^2 T_F n_\ell\,,
\end{align}
whereas the logarithmic coefficients read
\begin{align}
\tilde B_{31}
={}&
\tilde{B}_{10}\bigl(2 \beta_1+\gamma^B_1\bigr)+\tilde{B}_{20}\bigl(4
\beta_0+\gamma^B_0\bigr)+\gamma^B_2\,,
\\
\nn \tilde B_{32}
={}&
\tilde{B}_{10}\biggl[3 \beta_0 \gamma^B_0+4
\beta_0^2+\Gamma^{\rm c}_{\!1}+\frac{1}{2} \bigl(\gamma^B_0\bigr)^2\biggr]+\tilde{B}_{20} \Gamma^{\rm c}_{\!0}
+\beta_1 \gamma^B_0+\Gamma^{\rm c}_{\!2}+\bigl(2\beta_0+\gamma^B_0\bigr) \gamma^B_1,
\\
\nn \tilde B_{33}
={}&
\tilde{B}_{10}\biggl(\frac{8}{3} \beta_0+
\gamma^B_0\biggr)\Gamma^{\rm c}_{\!0}+\frac{2}{3} \beta_1 \Gamma^{\rm c}_0+\biggl(\frac{4}{3}
\beta_0+\gamma^B_0\biggr) \Gamma^{\rm c}_{\!1}+\Gamma^{\rm c}_{\!0}\gamma^B_1
+ \frac{4}{3}\beta_0^2 \gamma^B_0
\\
& +\biggl(\beta_0+\frac{1}{6}\gamma^B_0\biggr)\bigl(\gamma_0^B\bigr)^2\,,\nn
\\
\nn \tilde B_{34}
={}&
\frac{1}{2} (\Gamma^{\rm c}_{\!0})^2 \tilde{B}_{10}+ \frac{1}{3} \beta_0 \Gamma^{\rm c}_{\!0}\bigl(5\gamma^B_0+2 \beta_0\bigr)+\frac{1}{2} \Gamma^{\rm c}_{\!0}
\bigl(\gamma^B_0\bigr)^2+\Gamma^{\rm c}_{\!0} \Gamma^{\rm c}_{\!1}\,,
\\
\tilde B_{35}
={}&
(\Gamma^{\rm c}_{\!0})^2\biggl(\frac{2}{3} \beta_0 +\frac{1}{2} \gamma^B_0\biggr)\,,
\qquad \qquad
\tilde B_{36}
=
\frac{1}{6}(\Gamma^{\rm c}_{\!0})^3\,,\nn
\end{align}
in agreement with \eqs{Btilden2n}{recBtilde}.
Plugging in the numerical values for the various SU(3) color factors we find for the Dirac delta piece $\tilde{B}_{30} = 14804.2 - 2045.56 n_\ell + 50.0538 n_\ell^2$. The corresponding results for the momentum-space jet function can be found in \app{JF}.

\begin{figure*}[t!]
\subfigure[]
{\includegraphics[width=0.49\textwidth]{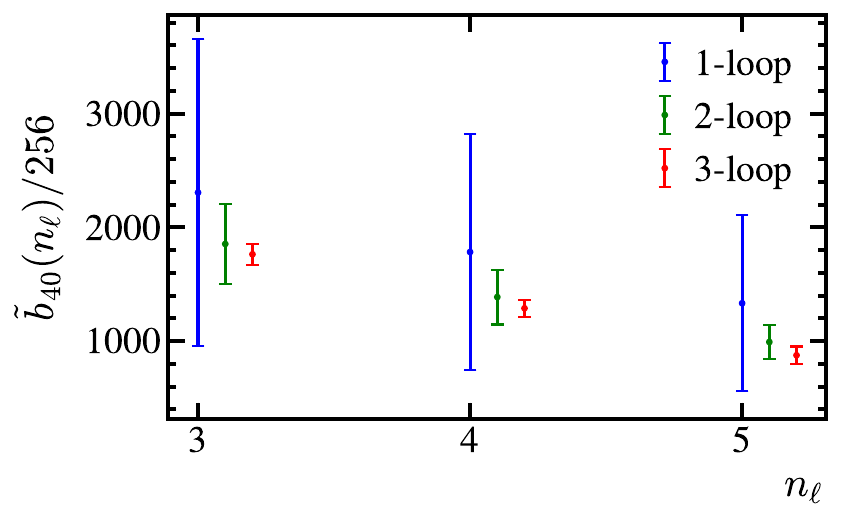}%
\label{fig:4loop}}
\subfigure[]
{\includegraphics[width=0.49\textwidth]{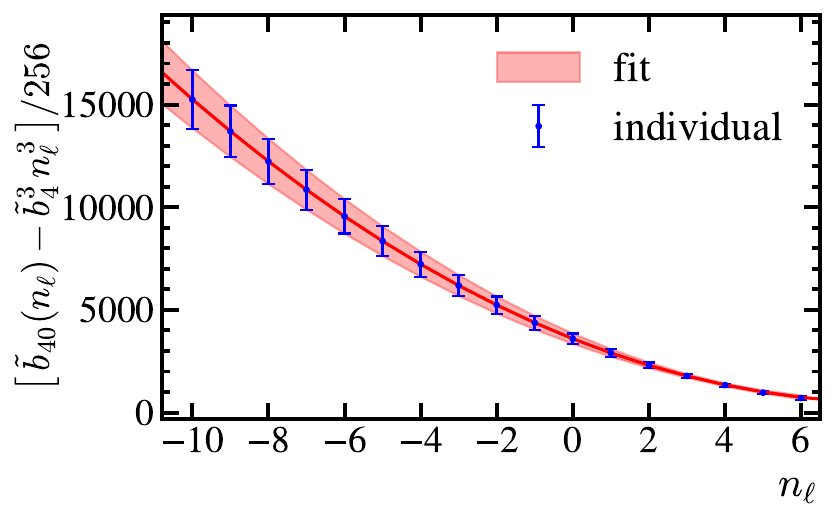}%
\label{fig:fit}}
\caption{Estimates of $\tilde{b}_{40}$ for various numbers of light flavors $n_\ell$, where the uncertainties have been computed varying the parameter $\lambda$. In the left panel we show the estimates when including lower-order information up to one (blue), two (green), and three (red) loops. In the right panel we use three-loop input to make predictions for a large set of $n_\ell$ values, estimating the uncertainty with $\lambda$ variation (in blue), and show as a red band the predictions obtained using the results for the various flavor coefficients, along with the corresponding uncertainties generated with error propagation. \label{fig:b4}}
\end{figure*}
\section{Estimate of the four-loop non-logarithmic jet function coefficient}
\label{sec:estimate}
In this section we follow the same strategy as that outlined in Sec.~10 of \rcite{Gracia:2021nut} to estimate the $\mathcal{O}(\alpha_s^4)$ jet function. Specifically, we predict $\tilde b_{40}(n_\ell)$, the non-logarithmic four-loop coefficient of the position-space jet function's exponent for different numbers of light flavors using the QCD beta function coefficients up to five loops~\cite{Tarasov:1980au,Larin:1993tp,vanRitbergen:1997va,Czakon:2004bu,Baikov:2016tgj,Chetyrkin:2017bjc,Luthe:2017ttg,Herzog:2017ohr}. These estimates may be useful as a cross check of future explicit four-loop computations, cf.\ \eq{comparison}. They are based on the dominance of the $u=1/2$ renormalon and use the R-evolution formalism applied to the non-derivative jet-mass scheme, whose coefficients, according to \eq{jet_mass_ND}, depend on $\tilde b_{l0}$ only. The logarithmic coefficients $\tilde b_{4,k\geq1}$ can be expressed in terms of the $\tilde b_{l\leq3,0}$ and anomalous dimension coefficients, see \eqs{nonRec}{Rec}. Since the four-loop non-cusp anomalous dimension is not known, no prediction can be made for $\tilde b_{41}$ at present. However, as the four-loop cusp anomalous dimension was computed in \rcites{vonManteuffel:2020vjv,Henn:2019swt,Henn:2019rmi,Bruser:2019auj,Moch:2018wjh,Moch:2017uml}, all ingredients for $\tilde b_{4,k\geq2}$ are known (see Appendix~\ref{app:expansion} and \rcite{Herzog:2017ohr})
and hence they are fixed by \eq{nonRec}.

To estimate $\tilde b_{40}$ we follow the same lines as \rcite{Gracia:2021nut}, to which we refer for details of the procedure. To start with, we set $N_c=3$ and write $\tilde b_{40}$ as a polynomial in $n_\ell$:
\begin{equation}\label{eq:flavDep}
\tilde b_{40}(n_\ell) = \sum_{i=0}^3 \tilde b_4^i \, n_\ell^i\,.
\end{equation}
\begin{table}[t!]
\centering
\begin{tabular}{ c ccc } \toprule $n_{\ell}$ & $\tilde{b}_{40}^{\text{1-loop}}$ & $\tilde{b}_{40}^{\text{2-loop}}$ & $\tilde{b}_{40}^{\text{3-loop}}$\\\midrule
3 & $590000 \pm 340000$ & $475000 \pm 90000$ & $451000 \pm 23000$\\
4 & $460000 \pm 270000$ & $355000 \pm 62000$ & $330000 \pm 19000$\\
5 & $340000 \pm 200000$ & $254000 \pm 38000$ & $224000 \pm 20000$\\\bottomrule
\end{tabular}
\caption{Estimate of the four-loop non-logarithmic coefficient of the logarithm of the position-space jet function based on renormalon dominance for various numbers of light flavors. The estimates in the second, third and fourth columns use as input one-, two- and three-loop results.\label{tab:4loop}}
\end{table}%
The highest-power coefficient is known from the large-$\beta_0$ computation carried out in \rcite{Gracia:2021nut}: \mbox{$\tilde{b}_4^3=-195.93$}. The expression resulting from the renormalon dominance argument
used to estimate $\tilde b_{l0}(n_\ell)$ using $m$-loop input
depends logarithmically on a free dimensionless parameter $\lambda$ (that appears due to fixed-order $\alpha_s(\lambda R)$ re-expansion in terms of $\alpha_s(R)$, see e.g.\ \rcite{Hoang:2017suc}) that is customarily varied between $1/2$ and $2$ to assign an uncertainty.
Moreover, it involves a sum over coefficients $S_k^\lambda$ which are linear combinations of $\tilde b_{n0}$ with $n\leq k+1$ such that if the sum included all $S_{k\leq l-1}^\lambda$ the prediction for $\tilde b_{l0}$ would be exact and \mbox{$\lambda$-independent} by consistency.

\begin{figure*}[t!]
\subfigure[]
{\includegraphics[width=0.31\textwidth]{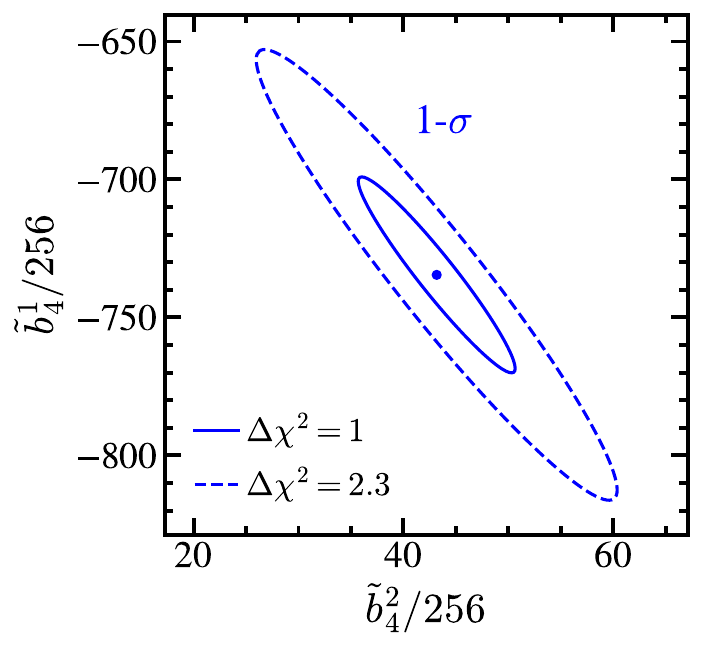}\vspace*{-1cm}}~
\subfigure[]
{\includegraphics[width=0.31\textwidth]{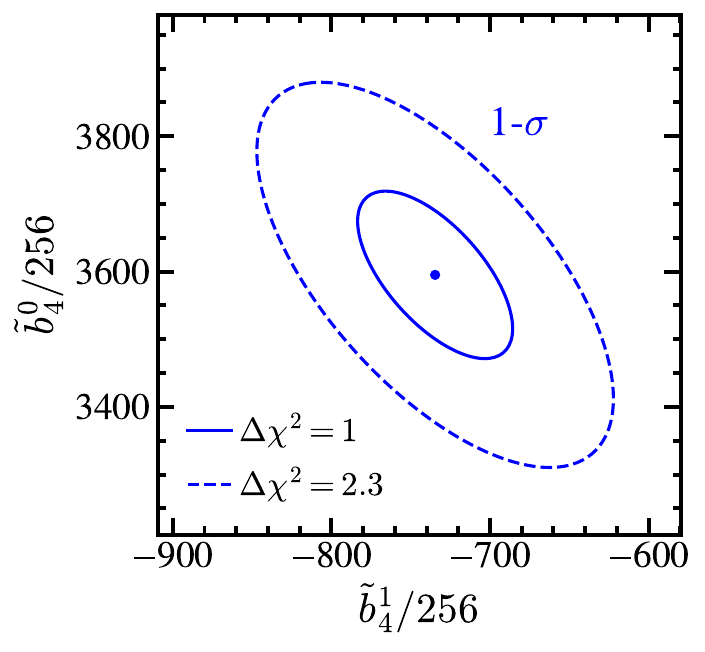}\vspace*{-1cm}}~
\subfigure[]
{\includegraphics[width=0.31\textwidth]{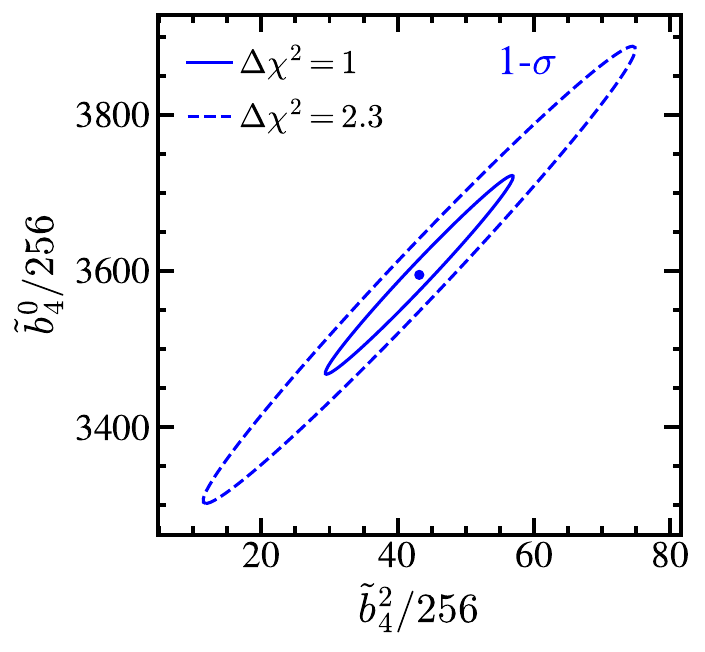}\vspace*{-1cm}}
\caption{Error ellipses for pairs of simultaneously determined $\tilde b_4^i$ coefficients. Solid ellipses correspond to one-$\sigma$ errors in each parameter ($39\%$ confidence level or $\Delta\chi^2=1$) while dashed ones are one-$\sigma$ in both dimensions ($68\%$ confidence level or $\Delta\chi^2=2.3$)\label{fig:ellipses}.}
\end{figure*}

To test the reliability of our uncertainty estimates from varying $\lambda$ we predict $\tilde b_{40}$ for $n_\ell=3,4,5$ using $S_k^\lambda$ up to one, two, and three loops, i.e.\ $k=0,1,2$, as input, and show the results in Table~\ref{tab:4loop} and \fig{4loop}. One can observe that the higher-order results are nicely contained in the lower-order uncertainty bands, which gives us confidence that our estimates are conservative. We obtain estimates for the flavor-dependent coefficients defined in \eq{flavDep} and their uncertainties as described in \rcite{Gracia:2021nut}, and to that end predict $\tilde b_{40}(n_\ell)$ for \mbox{$-10\leq n_\ell \leq 6$}, as shown in \fig{fit} using blue dots with error bars. Estimates for $n_\ell<0$ are employed rather than the physical $n_\ell>0$ since for a large and positive number of active flavors the property of confinement is lost. Moreover, flipping the sign of $\beta_0$ transforms the IR renormalon into a UV one, invalidating the procedure used to estimate higher-order coefficients.
The results are collected in Table~\ref{tab:flav}, and the correlation matrix $C$
among the various $\tilde b_4^i$ ordered from highest ($i=2$) to lowest ($i=0$) power of $n_\ell$ obtained from $\lambda$ variation (by simply analyzing an evenly-spaced grid) reads following \rcite{Gracia:2021nut}:
\begin{equation}\label{eq:corr}
C = \left(\begin{array}{ccc} 1 & - 0.980 & 0.994\\
- 0.980 & 1 & - 0.963\\
0.994 & - 0.963 & 1\end{array}\right),
\end{equation}
where the negative correlation coefficients have been decreased by a common scaling factor such that the uncertainties obtained for $n_\ell=3,4,5$ using error propagation on \eq{flavDep} are equal or larger than those obtained varying $\lambda$ in the direct determination for fixed $n_\ell$. Two-dimensional projections of the three-dimensional error ellipsoid are given in \fig{ellipses}. As shown in the red band of \fig{fit}, after having decreased the negative correlation, error propagation yields uncertainties for the various reconstructed $\tilde b_{40}$ which faithfully reproduce the ones obtained individually.

\begin{table}[t!]
\centering
\begin{tabular}{ c ccc c } \toprule order & $\tilde{b}^0_4$ & $\tilde{b}_4^1$ & $\tilde{b}_4^2$ & $\tilde{b}_4^3$\\\midrule
1 & $1110000\pm 640000$ & $-200000 \pm 120000$ & $9800 \pm 6500$ & $-195.93$\\
2 & $950000 \pm 200000$ & $-194000 \pm 51000$ & $11700 \pm 3700$ & $-195.93$\\
3 & $920000 \pm 66000$ & $-188000 \pm 18000$ & $11000 \pm 1100$ & $-195.93$\\\bottomrule
\end{tabular}
\caption{Numerical estimates for $\tilde b_4^{i\leq2}$ defined for the non-logarithmic coefficient $\tilde b_{40}(n_\ell)$ of the (logarithm of the) four-loop position-space jet function in \eq{flavDep} using perturbative input up to one-, two-, and three-loop order. For completeness, we also show the known value for $\tilde b_4^3$.\label{tab:flav}}
\end{table}

\section{Conclusions}
\label{sec:conclusion}
We have computed the three-loop inclusive jet function for boosted heavy quarks. This function is a universal building block in the factorization theorems for many collider observables involving heavy (top) quark final-state jets. Prominent examples are the doubly differential hemisphere mass, thrust, heavy-jet mass and C-jettiness
cross sections in $e^+e^-$ collisions, which play e.g.\ an important role for future precision determinations of the top mass.
The jet function is defined in the framework of bHQET, an effective field theory obtained from the collinear sector of SCET by integrating out the massive collinear modes associated with the heavy quark.

Our result
represents the last missing piece to obtain N$^3$LL$^\prime$-accurate self-normalized \mbox{2-jettiness} (or thrust) distributions in the resonance region of boosted top pair events at lepton colliders, and thus enables N$^4$LL resummation as soon as the pertinent anomalous dimensions become available. Once the N$^3$LL$^\prime$ hard-matching coefficient between SCET and bHQET is known, N$^3$LL$^\prime$ precision will also be attained for the corresponding unnormalized cross sections.
Moreover, using our new jet function result we have derived the relation between the pole mass and a family of renormalon-free short-distance ``jet-mass'' schemes for heavy quarks at $\mathcal{O}(\as^3)$, and the respective R-anomalous dimensions at the same order. We have compared two prototype jet-mass schemes to the MSR mass and found that they may eventually (once known to four loops) serve as viable alternatives to the latter. Finally, we have estimated the four-loop non-logarithmic part of the jet function based on the renormalon dominance assumption, which turned out to work well at the three-loop level. This may serve as a numerical check of a future four-loop calculation.

Our results will play a crucial role in calibrating the top quark mass parameter coded in parton-shower Monte Carlos against a well-defined quantum field theory mass, and will help elucidating the different roles played by the soft-function and top-quark--mass renormalons, which are equally severe. Hence, an update to the analysis carried out in \rcite{Bachu:2020nqn} is in order, and should help to gain further confidence that using the MSR mass scheme (or a jet-mass alternative) is suitable for a reliable determination of the top quark mass parameter.
A possible extension of our jet function calculation would be to allow for massive lighter quarks, most prominently the bottom.
While obtaining the full lighter-quark mass dependence at three loops is quite challenging, computing an expansion in small masses using the approach of \rcite{Bris:2024bcq} seems a promising strategy. One could also push our computation to the next perturbative order and derive, at least, the four-loop non-cusp bHQET jet function anomalous dimension. A first step in this direction has been already taken since we have obtained all three-loop master integrals one order higher in the $\eps$ expansion than necessary for our finite renormalized result.

\begin{acknowledgments}
This work was supported in part by the Spanish MECD grant PID2022-141910NB-I00, the JCyL grant SA091P24 under program EDU/841/2024, the EU STRONG-2020 project under Program No.\ H2020-INFRAIA-2018-1, Grant Agreement No.\ 824093. A.\,M.\,C.\ is supported by an FPU scholarship funded by the Spanish MICIU under grant no.\ FPU22/02506. R.\,B., M.\,S. and V.\,M.\ are grateful to the Erwin-Schr\"odinger International Institute for Mathematics and Physics for partial support during the Programme ``Quantum Field Theory at the Frontiers of the Strong Interactions'', July 31 - September 1, 2023. A.\,M.\,C.\ thanks the University of Freiburg for hospitality while parts of this work were completed.
The diagrams in this paper were drawn using \texttt{JaxoDraw}~\cite{Binosi:2008ig}.
\end{acknowledgments}

\appendix

\section{Renormalized momentum-space jet function}
\label{app:JF}
\allowdisplaybreaks

The coefficients of the renormalized momentum-space jet function in \eq{exp_calB} are given up to three loops by
\begin{align}
B_{1,-1}
={}&
\biggl( 4-\frac{\pi ^2}{2} \biggr) C_F\,,
\qquad\qquad\qquad B_{10}=-\gamma_0^{B}\,,
\qquad\qquad\qquad
B_{11}=2\Gamma_{\!0}^{\rm c}\,,\\[0.25cm]
\nonumber B_{2,-1}
={}&
\biggl(8-\frac{10 \pi ^2}{3}+\frac{13 \pi ^4}{360} -32 \zeta_3 \biggr)C_F^2
+
\biggl(\frac{68 \zeta_3}{9}-\frac{2248}{81}+\frac{59 \pi ^2}{27} \biggr)C_FT_Fn_\ell
\\ \nn & +
\biggl(\frac{5918}{81}-\frac{661 \pi ^2}{108}+\frac{23 \pi ^4}{180} -\frac{277 \zeta_3}{9}\biggr)C_FC_A\,,
\\
B_{20}
={}&
-\!B_{1,-1} \bigl(2 \beta_0+\gamma^B_0\bigr)+\Gamma^{\rm c}_0\biggl(\frac{\pi ^2}{3}
\gamma^B_0+4 \zeta_3\Gamma^{\rm c}_0\biggr) -\gamma^B_1\,,
\qquad\,\nn
B_{22}
=
-\bigl(2 \beta_0+3 \gamma^B_0\bigr)\Gamma^{\rm c}_0\,,\nn
\\
B_{21}
={}&
2 B_{1,-1} \Gamma^{\rm c}_0 + \gamma^B_0\bigl(2 \beta_0+\gamma^B_0\bigr)-\frac{2\pi ^2}{3}
(\Gamma^{\rm c}_0)^2+2 \Gamma^{\rm c}_1\,,
\qquad\quad~\,
B_{23}
=
2 (\Gamma^{\rm c}_0)^2\,,\nn\\[0.25cm]
\nn
B_{3,-1}=\,
&
\biggl( \frac{9536 \zeta_3}{27}-4 \pi ^2 \zeta_3-\frac{88 \zeta_5}{3}-\frac{5273287}{6561}+\frac{172715 \pi ^2}{2187}-\frac{2869 \pi
^4}{1215} \biggr) C_F C_A T_F n_\ell \\
&
+\biggl(\frac{846784}{6561} -\frac{8800 \zeta_3}{243}-\frac{2920 \pi
^2}{243}+\frac{148 \pi ^4}{1215} \biggr) C_F T_F^2 n_\ell^2 \nn
+\biggl(\frac{361 \pi ^2 \zeta_3}{9}-\frac{225686 \zeta_3}{243}
\\ & +\frac{236
\zeta_3^2}{9}+\frac{902 \zeta_5}{9}+\frac{31952191}{26244}-\frac{1076863 \pi ^2}{8748}+\frac{23881
\pi ^4}{4860}-\frac{3161 \pi ^6}{51030}\biggr) C_F C_A^2 \nn
\\
&
+ \biggl(\frac{3013 \zeta_3 \pi ^2}{18}-\frac{30428 \zeta_3}{27}\!+ \! 160 \zeta_3^2-\frac{2816 \zeta_5}{3}\!+\!\frac{23672}{81}-\frac{3106 \pi
^2}{27}\!+\!\frac{10163 \pi ^4}{3240}-\frac{\pi ^6}{216} \biggr)
\nn \\& \times\! C_F^2C_A
+ \biggl( \frac{42200 \zeta_3}{81}-\frac{506 \pi ^2 \zeta_3}{9}+\frac{3296 \zeta_5}{9}-\frac{178669}{486}+\frac{991 \pi ^2}{18}-\frac{269 \pi ^4}{810} \biggr) C_F^2 T_F n_\ell \nn \\
& +\biggl(\frac{304 \pi ^2 \zeta_3}{3}-\frac{448 \zeta_3}{3}+\frac{1280 \zeta_3^2}{3}-768 \zeta_5+\frac{32}{3}-\frac{28 \pi ^2}{3}+\frac{41 \pi
^4}{90}-\frac{13777 \pi ^6}{45360}\biggr)C_F^3\,, \nn
\\
B_{30}
={}&
B_{1,-1} \biggl[\frac{\pi ^2}{3}\Gamma_{\!0}^{\rm c}\bigl(2 \beta_0+
\gamma_0^B\bigr)-2 \beta_1+4 (\Gamma_{\!0}^{\rm c})^2 \zeta_3
-\gamma_1^B\biggr]-B_{2,-1} (4 \beta_0+\gamma_0^B)\nn
\\
& +\Gamma_{\!0}^{\rm c}
\biggl[2\zeta_3 \gamma_0^B \bigl(2 \beta_0 + \gamma_0^B\bigr)+8 \Gamma_{\!1}^{\rm c} \zeta_3+\frac{\pi ^2}{3} \gamma_1^B\biggr]+\frac{\pi ^4}{45}(\Gamma_{\!0}^{\rm c})^2
(4 \beta_0+ \gamma_0^B)\nn
\\
& +\frac{\pi
^2}{3} \Gamma_{\!1}^{\rm c} \gamma_0^B+8(\Gamma_{\!0}^{\rm c})^3 \biggl(3 \zeta_5-\frac{\pi ^2 \zeta_3}{3}\biggr)-\gamma_2^B\,,\nn
\\[3cm]
B_{31}
={}&
B_{1,-1} \biggl[2 \beta_0 \bigl(3 \gamma_0^B+4 \beta_0\bigr)-\frac{2\pi ^2}{3}
(\Gamma_{\!0}^{\rm c})^2+2 \Gamma_{\!1}^{\rm c}+ \bigl(\gamma^B_0\bigr)^2\biggr]+2 B_{2,-1} \Gamma_{\!0}^{\rm c} +2\gamma_0^B (\beta_1+\gamma_1^B)\nn
\\
&\! -\!8\zeta_3(\Gamma_{\!0}^{\rm c})^2 (3 \beta_0 +2 \gamma_0^B )-2 \pi ^2\Gamma_{\!0}^{\rm c}
\biggl[ \gamma_0^B\!\biggl(\beta_0 +\frac{1}{3} \gamma^B_0\biggr)\! +\frac{2}{3}
\Gamma_{\!1}^{\rm c}+\frac{\pi^2}{45}
(\Gamma_{\!0}^{\rm c})^2\biggr]\!+4 \beta_0
\gamma_1^B
+2 \Gamma_{\!2}^{\rm c}\,,\nn
\\
B_{32}
={}&
\!-\!\Gamma_{\!0}^{\rm c} B_{1,-1} \bigl(8 \beta_0+3 \gamma_0^B\bigr)+2 \pi ^2(\Gamma_{\!0}^{\rm c})^2 \bigl(
\beta_0+ \gamma_0^B\bigr)-\Gamma_{\!0}^{\rm c} \bigl(2 \beta_1+3
\gamma_1^B\bigr)\nn
\\
& \!-\!\Gamma_{\!1}^{\rm c} \bigl(4 \beta_0+3 \gamma_0^B\bigr)-3
\beta_0 \bigl(\gamma^B_0\bigr)^2-4 \beta_0^2 \gamma_0^B+20\zeta_3
(\Gamma_{\!0}^{\rm c})^3 -\frac{ 1}{2}\bigl(\gamma^B_0\bigr)^3\,,\nn
\\
B_{33}
={}&
2 (\Gamma_{\!0}^{\rm c})^2 B_{1,-1}+\Gamma_{\!0}^{\rm c} \biggl[\frac{4}{3} \beta_0\bigr( 5\gamma_0^B+2\beta_0\bigr)+4 \Gamma_{\!1}^{\rm c}+2
\bigl(\gamma^B_0\bigr)^2\biggr]-\frac{4\pi ^2}{3} (\Gamma_{\!0}^{\rm c})^3\,,\nn
\\
B_{34}
={}&
\!-\!5(\Gamma_{\!0}^{\rm c})^2 \biggl(\frac{2 \beta_0}{3}+\frac{\gamma_0^B}{2}\biggr)\,,
\qquad \qquad
B_{35}
=
(\Gamma_{\!0}^{\rm c})^3\,,\nn
\end{align}

\section{Cumulative jet function}
\label{sec:cum}The cumulative version of the stable ($\Gamt = 0$) jet function (sometimes needed for integrated cross sections) is defined by%
\footnote{The cumulative jet function for an unstable quark has its lower integration limit at $\hat s=-\infty$.}
\begin{equation}
B^{\rm cum}({\hat s}_c) = \int_0^{\hat s_c} \! \dd \hat s \, B(\hat s)\,,
\end{equation}
which is a real-valued ordinary function with support for positive argument and has dimensions of an inverse mass. The perturbative expansion of the cumulative jet function can be expressed in terms of the momentum-space coefficients $B_{jk}$ as
\begin{align}\label{eq:Bcum}
m B^{\rm cum}(\hat s_c) &\,= \sum^\infty_{l=0} \biggl[ \dfrac{\as(\mu)}{4 \pi}\biggr]^l \!
\sum^{2l}_{k=0} \mathscr{L}^k (\hat s_c, \mu) B^c_{lk}\,,
\\
\mathscr{L}^k (\hat s_c, \mu)&\equiv \theta(\hat s_c) \log^{k}\!\biggl(\frac{\hat s_c}{\mu}\biggr)\,, \nn
\end{align}
where $B^c_{jk}=B_{j,k-1}/{\rm max}(1,k)$.
The renormalized cumulative jet function obeys the same RGE as $B(\hat s, \mu)$, see \eq{RGE_for_B}. Again, we can derive recursion relations similar to \eqs{BnR}{BbarRest} (and even more handy) directly from this RGE:
\begin{equation}
B^c_{l , 2 l} = \frac{(\Gamma^\rc_{\!0})^l}{l!} \,,\qquad B^c_{l , 2 l-1} = - \frac{(\Gamma^\rc_{\!0})^{l - 1}}{(l - 1) !} \biggl[ \gamma_0^B + \frac{2}{3} (l -1) \beta_0 \biggr],
\end{equation}
where the first expression is valid for $l\geq 0$ and the second for $l\geq1$. Once these two sets of coefficients are known, the rest follow from the recursive relation
\begin{align}
B^c_{l k} = & \frac{1}{k} \Biggl[2 (1 - \delta_{k 1})\! \! \! \sum^{l - 1}_{j = \left\lfloor \frac{k - 1}{2} \right\rfloor}B^c_{j , k - 2} \Gamma^\rc_{\!l - j - 1} -\!\!\!\sum^{l - 1}_{j = \left\lfloor \frac{k}{2}\right\rfloor} B^c_{j , k - 1} (2 j \beta_{l - j - 1} + \gamma^B_{l - j- 1}) \Biggr]\\
& -\! \frac{2(-1)^k}{k !} \sum^{2 l - 2}_{n = k} n! (- 1)^{n} \zeta_{n- k + 2} \sum^{l - 1}_{j = \left\lceil \frac{n}{2} \right\rceil} B^c_{j n} \Gamma^\rc_{\!l - j - 1}\, ,\nn
\end{align}
valid for $l\geq2$ and $1\leq k \leq 2l-2$.

\section{Anomalous dimensions}
\label{app:expansion}
Here we collect the perturbative coefficients of the anomalous dimensions $\hat{\beta}$, $\Gamma^{\rm c}$, and $\gamma_B$ that are relevant for our three-loop calculation, as defined in \eqs{alpBare}{expansion_gamma}.

For the jet function computation we only need the one- and two-loop $\beta$-function coefficients,
\begin{equation}
\beta_0 = \frac{11 C_A}{3} - \frac{4 T_F n_\ell}{3}\,,\qquad
\beta_1 = \frac{34 C_A^2}{3}-\frac{20 C_A T_F n_\ell}{3}-4 C_F T_F n_\ell\,.
\label{eq:values_beta}
\end{equation}
The known higher-order coefficients enter our estimate in \sec{estimate} and can be found in \rcites{Herzog:2017ohr,Luthe:2017ttg}.

The (light-like) cusp anomalous dimension coefficients read up to three loops~\cite{Korchemsky:1987wg, Moch:2004pa}
\begin{align}
\label{eq:anom_dim_cusp}
\Gamma_{\!0}^{\rm c}={}& 4C_F\,,
\qquad
\Gamma^{\rm c}_{\!1}=\biggl(\dfrac{268}{9}-\dfrac{4\pi^2}{3}\biggr)C_FC_A-\dfrac{80}{9}C_F T_F n_\ell\,,\\
\Gamma_{\!2}^{\rm c} ={}&\biggl(\frac{490}{3}-\frac{536 \pi^2}{27}+\frac{44 \pi^4}{45}+\frac{88 \zeta_3}{3}\biggr) C_F C_A^2
+\biggl(\frac{160 \pi^2}{27} - \frac{1672}{27} -\frac{224 \zeta_3}{3}\biggr) C_F C_A T_F n_\ell \nonumber\\
&+\biggl(64 \zeta_3-\frac{220}{3}\biggr) C_F^2 T_F n_\ell-\frac{64}{27} C_F T_F^2n_\ell^2\,. \nonumber
\end{align}

The non-cusp anomalous dimension of the bHQET jet function was computed directly at one and two loops in \rcites{Fleming:2007xt,Jain:2008gb}, respectively. The three-loop coefficient can be derived based on RG consistency of \eq{double_hemisphere_invariant} using results given in \rcites{Hoang:2015vua,Bruser:2019yjk,Becher:2008cf}. We have
\begin{align}
\label{eq:anom_dim_no_cusp}
\gamma_0^B ={}&4 C_F, \quad \gamma_1^B=\biggl(\frac{1396}{27}-\frac{23 \pi^2}{9}-20 \zeta_3\biggr) C_F C_A+\biggl(\frac{4 \pi^2}{9}-\frac{464}{27}\biggr) C_F T_F n_\ell\,, \\
\gamma_2^B={}& \biggl(\frac{112 \pi
^2 \zeta_3}{9}-\frac{2468 \zeta_3}{9}+120 \zeta_5+\frac{192347}{729}-\frac{11797 \pi
^2}{243}+\frac{44 \pi ^4}{15}\biggr)C_F C_A^2\nonumber \\
&+
\biggl(\frac{4268 \pi
^2}{243}-\frac{1520 \zeta_3}{27}-\frac{42908}{729}-\frac{16 \pi ^4}{15}\biggr)C_F C_A T_F n_\ell
\nonumber \\&
+ \biggl(\frac{448 \zeta_3}{27}-\frac{10048}{729}-\frac{80 \pi
^2}{81}\biggr)C_F T_F^2 n_\ell^2\nonumber\\&+\biggl(\frac{1184
\zeta_3}{9}-\frac{5402}{27}+\frac{4 \pi ^2}{3}+\frac{16 \pi
^4}{45}\biggr)C_F^2 T_F n_\ell \,.\nonumber
\end{align}

\section{Useful relations for plus distributions and logarithms}
\label{app:plus_distributions}
\begin{table}[t!]
\centering
\begin{tabularx}{0.85\textwidth}{m{0.3cm}*{5}{X} m{2cm} m{2cm}}
\toprule
$i$ & \myalign{c}{$0$} & \myalign{c}{$1$} & \myalign{c}{$2$} & \myalign{c}{$3$} & \myalign{c}{$4$} & \myalign{c}{$5$} & \myalign{c}{$6$} \\
\midrule
$\kappa_{i}$ & $1$ & $0$ & $\phantom{-}\frac{\pi ^2}{12}$ & $-\frac{\zeta_3}{3}$ & $\frac{\pi ^4}{160}$ & \myalign{r}{$-\frac{\zeta_5}{5}-\frac{\pi ^2 \zeta_3}{36}$} & $\frac{\zeta_3^2}{18}+\frac{61 \pi^6}{120960}$ \\
$\hat \kappa_{i}$ & $1$ & $0$ & $-\frac{\pi ^2}{12}$ & $\phantom{-}\frac{\zeta_3}{3}$ & $\frac{\pi ^4}{1440}$ & \myalign{r}{$\phantom{-}\frac{\zeta_5}{5}-\frac{\pi ^2 \zeta_3}{36}$} & $\frac{\zeta_3^2}{18}-\frac{\pi^6}{24192}$ \\
\bottomrule
\end{tabularx}
\caption{Expressions of the first $\kappa_{i}$ and $\hat \kappa_{i}$ coefficients defined in \eq{GamRed}.} \label{tab:Gamma}
\end{table}
In this appendix we summarize useful relations to manipulate expressions with logarithms and plus distributions.
The definition of $L$, $\tilde L$ and $\cal L$ is given in \eq{plus_log}, and that of $\mathscr{L}$ in \eq{Bcum}. The plus distribution in \eq{plus_log} is defined by
\begin{equation}\label{eq:plusDef}
\int_0^A \!\rd x\, f (x) \biggl[ \theta (x) \frac{\log^n (x)}{x} \biggr]_+ \equiv
\int_0^1 \! \rd x\, \frac{f (x) - f (0)}{x} \log^n (x) +
\int_1^A \!\rd x\, f (x) \frac{\log^n (x)}{x}\,.
\end{equation}
To derive the relations in this appendix we frequently use the distributional identity
\begin{equation}\label{eq:expPlus}
\theta(x)\, x^{- 1 + \epsilon} = \frac{1}{\epsilon} \delta (x-\eta) + \sum^{\infty}_{n = 0}
\frac{\epsilon^n}{n!} \left[\theta (x) \frac{\log^n (x)}{x} \right]_+ = \sum^{\infty}_{n=-1} \frac{\epsilon^n}{|n|!} \mathcal{L}^{n}.
\end{equation}
Plus and Dirac delta distributions also arise from taking derivatives:
\begin{equation}
\frac{\rm d}{{\rm d}\hat s} \bigl[\mathscr{L}^n (\hat s)\bigr]
= {\rm max}(1,n) \,\mathcal{L}^{n-1}(\hat s)\,.
\end{equation}
To relate the momentum- and position-space jet functions we define two sets of coefficients dubbed $\kappa_{i}$ and $\hat \kappa_{i}$ that (for $i\ge2$) can be obtained through the following recursive relations:
\begin{equation}\label{eq:GamRed}
\kappa_i = \frac{1}{i} \Biggl[ (- 1)^i \zeta_{i} - \sum_{j =
1}^{i - 3} (- 1)^j \zeta_{j + 1} \kappa_{i - 1 - j} \Biggr],\quad
\hat{\kappa}_{i} = -\frac{1}{i} \Biggl[ (- 1)^i \zeta_{i} - \sum_{j = 1}^{i - 3} (- 1)^j \zeta_{j + 1} \hat{\kappa}_{i - 1 - j} \Biggr],
\end{equation}
with $\kappa_{0}=\hat \kappa_{0}=1$ and $\kappa_{1}=\hat \kappa_{1}=0$. The sums in both expressions only contribute for $i\ge 4$. The two sets of coefficients satisfy the constraint
\begin{equation}
\sum_{j = 0}^i \kappa_{i - j}\, \hat \kappa_{j} = 0\,,
\end{equation}
for $j \geq 1$. The values of the first few coefficients in both sets are given in Table~\ref{tab:Gamma}.
To convolve two plus distributions or a plus distribution with a regular logarithm raised to some integer power, we conveniently define another pair of two-dimensional sets denoted by $\xi_{ij}$ and $\hat \xi_{ij}$.
These are related to the $\kappa_i$ and $\hat{\kappa}_i$ by
\begin{align}
\hat{\xi}_{ij} ={}& \sum_{k = 0}^{i+j} \hat{\kappa}_{k} \sum_{\ell = \max (0, k - j)}^{\min (k,i)} \binom{\,k\,}{\ell} \,\kappa_{i - \ell}\, \kappa_{j + \ell - k}\,, \label{eq:XiRed} \\
\xi_{ij} ={}& (-1)^j \sum_{k = 0}^{i+j} \kappa_{k} \sum_{\ell = \max (0, k - j)}^{\min (k, i)}\binom{\,k\,}{\ell} (-1)^{\ell} \, \kappa_{i - \ell} \, \hat{\kappa}_{\!j + \ell - k}\,,\nn
\end{align}
where $\binom{\,k\,}{\ell} = k!/[\ell!\,(k-\ell)!]$ is the binomial coefficient.
Expressions for the first few coefficients of the $\xi_{ij}$ and $\hat \xi_{ij}$ can be found in \tab{Xi}.

For the sake of compactness of the expressions in the remainder of this appendix, we will refrain from writing out the $\mu$ dependence of $\cal L$, $\mathscr{L}$, $\tilde L$, and $L$ explicitly.
\begin{table}[t]
\centering
\begin{tabular}{c|cccc}
$\hat \xi_{ij}$ & $0$ & $1$ & $2$ & $3$ \\ \hline
$0$ &$\;1\;$ & $0$ & $0$ & $0$ \\
$1$ & $\;0\;$ & $-\frac{\pi ^2}{6}$ & $\zeta_3$ & $-\frac{\pi ^4}{90}$ \\
$2$ & $\;0\;$ & $\zeta_3$ & $-\frac{\pi ^4}{360}$ & $2 \zeta_5-\frac{\pi ^2 \zeta_3}{6}$ \\
$3$ & $\;0\;$ & $-\frac{\pi ^4}{90}$ & $2 \zeta_5-\frac{\pi ^2 \zeta_3}{6}$ & $\zeta_3^2-\frac{23 \pi ^6}{15120}$ \\
\end{tabular}
\hfill
\begin{tabular}{c|cccc}
$\xi_{ij}$
& $0$ & $1$ & $2$ & $3$ \\ \hline
$0$ & $\;1\;$ & $0$ & $0$ & $0$ \\
$1$ & $\;0\;$ & $\frac{\pi ^2}{6}$ & $\zeta_3$ & $\frac{\pi ^4}{90}$ \\
$2$ & $\;\frac{\pi ^2}{6}\;$ & $\zeta_3$ & $\frac{11 \pi ^4}{360}$ & $\frac{\pi ^2 \zeta_3}{6}+2 \zeta_5$ \\
$3$& $\;0\;$ & $\frac{7 \pi ^4}{180}$ & $\frac{\pi ^2 \zeta_3}{3}+2 \zeta_5$ & $\zeta_3^2+\frac{\pi ^6}{112}$ \\
\end{tabular}
\caption{Expressions of the first $\hat \xi_{ij}$ and $\xi_{ij}$ coefficients defined in \eq{XiRed}. In both tables, the first row (column) labels the first (second) index $i$ ($j$). The coefficients $\hat \xi_{ij}$ are symmetric under $i \leftrightarrow j$, while $\xi_{ij}$ have no symmetry.} \label{tab:Xi}
\end{table}

\subsubsection*{Fourier transform of distributions}

In order to get $\tilde B$ from $B$ and vice versa, it is convenient to write a closed form for the Fourier transform of plus distributions and its inverse:\footnote{The result in the first line is obtained using analytic continuation as explained below \eq{FT_invFT_B}.}
\begin{align}\label{eq:FT}
\int\! \rd \hat s \, e^{- i \hat s x} \mathcal{L}^n(\hat s)
={}&
|n|! \sum_{j = 0}^{n + 1} \frac{(- 1)^j}{j!} \kappa_{n + 1 - j}
\tilde L^j (x)\, ,\\
\frac{1}{2 \pi} \int \!\rd x\, e^{i \hat s x} \tilde L^n (x-i\eta) ={}&(- 1)^n n!
\sum_{j = -1}^{n - 1} \frac{\hat{\kappa}_{n - j - 1}}{|j|!} \mathcal{L}^j (\hat s)\, .\nonumber
\end{align}

\subsubsection*{Plus distributions through imaginary parts}

When taking the imaginary part of $\cal B$ to obtain $B$, plus distributions naturally arise.
Using the identity
\begin{equation}\label{eq:Imag}
{\rm Im}\Biggl[\!\biggl(-\frac{\hat s}{\mu} - i\eta\biggr)^{\!\!-1+\xi}\,\Biggr]
=\theta(\hat s)\biggl(\frac{\hat s}{\mu}\biggr)^{\!\!-1+\xi } \sin (\pi \xi)
= \biggl(\frac{\hat s}{\mu}\biggr)^{\!\!-1+\xi } \frac{\theta(\hat s)\pi}{\Gamma(\xi)\Gamma(1-\xi)}\,,
\end{equation}
we can take the imaginary part of the bare forward-scattering matrix element in \eq{def_calB} before expanding in $\varepsilon$. Expressing each side as a series in powers of $\xi$ with the help of \eq{expPlus}, the following relation can be inferred:\footnote{One can also obtain this result by taking a derivative with respect to $\hat s$ of the simpler relation
\begin{equation}
\frac{1}{\pi }{\rm Im}\biggl[\log^{n+1}\biggl(\frac{\mu }{-\hat s-i \eta }\biggr)\biggr]
= (- 1)^{n} \sum_{j = 0}^{\lfloor \frac{n}{2}\rfloor}
\biggl( \begin{array}{c} n + 1\\ 2 j + 1\end{array} \biggr)(-\pi^2)^j
\mathscr{L} ^{n-2 j} (\hat s)\,.
\end{equation}}
\begin{equation}\label{eq:ImPart}
{\rm Im} \left[ L^n(\hat s+i \eta)
\right] = (- 1)^{n} \sum_{j = 0}^{\lfloor \frac{n}{2}\rfloor}
\binom{n+1}{2j+1} \frac{{\rm max}(n-2j,1)}{n+1} (-\pi^{2})^j \mathcal{L}^{n - 2 j - 1}(\hat{s}) \,.
\end{equation}
For the transformation in the opposite direction according to \eq{from_calB_to_B} we use
\begin{equation}
\int \frac{\mathd \hat s'}{\pi} \frac{\mathcal{L}^{n} (\hat s')}{\hat s' - \hat s}
= (-1)^{n} \sum_{j = 0}^{\lceil \frac{n}{2}\rceil} \binom{n+1}{2j}\,
\dfrac{(-\pi^2)^j(2^{2j}-2)}{\max (1,n+1)}\,\mathbb B_{2 j}\, L^{n+1-2 j}\! (\hat s) \, ,
\end{equation}
where $\mathbb B_{n}$ is the n-th Bernoulli number.
If we allow for a finite width, the imaginary part can be taken using the following result:
\begin{align}\label{eq:ImGamma}
{\rm Im} \biggl[ \frac{1}{(- \hat s - i \Gamma)}\log^k \biggl( \frac{\mu}{- \hat s- i \Gamma} \biggr) \! \biggr] \!={}&
\frac{\Gamma}{\hat s_\Gamma^2} \sum_{j = 0}^k \binom{\,k\,}{\,j\,} (- 1)^{\left\lceil \frac{k- j}{2} \right\rceil}L_\Gamma^j \biggl[y\biggl(\frac{\hat s}{\Gamma}\biggr)\biggr]^{k - j} \Bigl( \frac{\hat s}{\Gamma} \Bigr)^{[(k - j)\,{\rm mod}\,2]},\nn \\
L_\Gamma ={}& \log \biggl(\frac{\mu}{\hat s_\Gamma} \biggr) \,.
\end{align}
As in \sec{theoretical_setup}, $[n\,{\rm mod}\,2]$ stands for the remainder of $n/2$ and $\hat s_\Gamma=\sqrt{\hat s^2 + \Gamma^2}$.

\subsubsection*{Convolution of a distribution and a logarithm}

Convolutions of plus distributions and (powers of) logarithms appear in the renormalization of the forward-scattering matrix element $\cB$, and evaluate to
\begin{equation}
\int\! \rd \hat{s}' \mathcal{L}^\ell(\hat s') L^n(\hat s - \hat s')
=
n! \, |\ell|!\! \!\sum_{k = 0}^{n + \ell + 1}\! (-1)^{n+k} L^k (\hat s) \! \! \!\sum_{j = \max (0, k - n)}^{\min (k, \ell + 1)} \! \! \frac{\xi_{\ell
+ 1 - j, n - k + j} }{(k - j) !j!}\,.
\label{eq:convpluslog}
\end{equation}
To convolve $\cB$ with its anomalous dimension we need a special case of \eq{convpluslog}:
\begin{equation}\label{eq:BcalRed}
\int \! \rd \hat{s}' \mathcal{L}^0(\hat s') L^n(\hat s - \hat s') = -L^{n+1} (-\hat s) + (-1)^n n!
\sum_{j =0}^{n - 1} \frac{(-1)^j}{j!} \zeta_{n - j +1}L^j (\hat s)\,,
\end{equation}
where for $n=0$ only the first term contributes. In order to compute the convolution of the cumulative jet function and the momentum-space anomalous dimension we need
\begin{equation}\label{eq:BcumRed}
\int\dd \hat{s}' \mathcal{L}^0(\hat s')
\mathscr{L}^n (\hat{s} - \hat{s}')
=\mathscr{L}^{n + 1} (\hat{s}) + (- 1)^{n} n!\sum_{j = 0}^{n - 1}\frac{(- 1)^{j}}{j!} \mathscr{L}^j (\hat{s}) \zeta_{n- j + 1} \,,
\end{equation}
where the sum does not contribute for $n=0$.

\subsubsection*{Convolution of two plus distributions}

Similarly, convolutions of two plus distributions occur when renormalizing the momentum-space jet function $B$ and obey the following relation:
\begin{align}
\int \!\rd \hat s' \, \mathcal{L}^n(\hat s') \mathcal{L}^\ell(\hat s - \hat s')
& = |n|!\, |\ell|!\!\!
\sum_{k = -1}^{n + \ell + 1}\!\mathcal{L}^{k}(\hat s)\!\!\!\! \sum_{j =
\max(0,k-n)}^{\min(k+1,\ell + 1)}\!\frac{{\rm max}(k+1,1)}{(k+1-j)!j!}
\,\hat{\xi}_{n + j - k, \ell + 1 - j}
\,. %
\label{eq:Brenorm}
\end{align}
Taking a derivative in \eq{BcumRed} with respect to $\hat s$ one obtains a particular case of \eq{Brenorm}:
\begin{equation}\label{eq:BRed}
\int\dd \hat{s}' \mathcal{L}^0(\hat s')\mathcal{L}^n(\hat s - \hat s') = \frac{(n + 2)\mathcal{L}^{n+1}(\hat s)}{{\rm max}(1,n + 1)}
+ (-1)^n|n|!\sum_{j = -1}^{n - 1} \frac{(- 1)^{j}}{|j|!} \mathcal{L}^{j}(\hat s) \zeta_{n - j + 1} \,,
\end{equation}
where the sum does not contribute for $n=-1$. This relation can be used to convolve the momentum-space jet function with its anomalous dimension.

\section{Renormalization factors}
\label{app:consistency}

The renormalization factor of the strong coupling,
\begin{equation}\label{eq:Zexpand}
Z_\alpha(\as)
=1 - \frac{\beta_0}{\varepsilon} \frac{\as}{4 \pi} + \biggl(\frac{\beta_0^2}{\varepsilon^2} - \frac{\beta_1}{2 \varepsilon} \biggr) \Bigl( \frac{\as}{4 \pi} \Bigr)^{\!2}+\mathcal{O}(\as^3)\,,
\end{equation}
formally admits a closed form valid to all orders in perturbation theory that can be obtained by integrating its $d$-dimensional RGE derived from \eq{alpBare} with the appropriate boundary condition
$Z_\alpha(0)=1$\,:
\begin{equation}\label{eq:Zalpha}
Z_{\alpha}(\as) = \exp\Bigl[-z\bigl(\hat{\beta}, \alpha_s\bigr)\Bigr]\,,\qquad
z(\gamma, \alpha_s) \equiv \int_0^{\alpha_s} \frac{{\rm d} x}{x}\frac{\gamma (x)}{\varepsilon + \hat{\beta} (x)} \,,
\end{equation}
where $\varepsilon$ regulates the lower integration endpoint, and hence cannot be set to zero.
In particular, one must not expand the integrand in $\eps$. Rather, upon expanding around small $\as$, which can be done conveniently before integrating, one obtains the perturbative expansion in \eq{Zexpand}. The ratio of $Z_\alpha(\alpha_1)/Z_\alpha(\alpha_2)$ can be written as the exponential of the integral in \eq{Zalpha} but with $\alpha_2$ and $\alpha_1$ as the upper and lower integration limits, respectively. Therefore, $\varepsilon$ can be set to zero, resulting in $Z_\alpha(\alpha_1)/Z_\alpha(\alpha_2) = \alpha_2/\alpha_1 + \ord(\eps)$
as can also be inferred from the $\mu$ independence of $\as^{\rm bare}$, see \eq{alpBare}.

Similarly, for the jet function renormalization factor one can write down a closed-form expression, which we derive in the following.
We start by writing $\log \tilde Z_B$ as a series in $1/\eps$\,:
\begin{equation}
\log \tilde Z_B
\equiv
\sum^\infty_{j=1} \dfrac{w_{j}(\as,\tilde L)}{\varepsilon^j}\,.
\label{eq:ZB_series_eps}
\end{equation}
Taking the derivative w.r.t.\ $\log\mu$
and using \eq{alpBare}, the Fourier transform of \eq{def_anomalous_dim}, and \eq{RGE_for_B_FT},
we obtain the RGE for $\tilde{Z}_B$,
\begin{equation}
\dfrac{\rd \log \tilde Z_B }{\rd \log \mu}=\dfrac{\partial \log \tilde Z_B}{\partial \log \mu}-2\as\bigl[\varepsilon+ \hat{\beta}(\alpha_s)\bigr]\dfrac{\partial \log\tilde Z_B}{\partial \as}
= -2\Gamma^\rc(\as) \tilde L(x,\mu) - \gamma_B^\text{nc}(\as)\,.
\label{eq:RGE_for_Z_FT}
\end{equation}
Inserting \eq{ZB_series_eps} into \eq{RGE_for_Z_FT}, and comparing the coefficients of $\eps^{-j}$,
we arrive at
\begin{subequations}
\begin{align}
\frac{\partial w_1(\as,\tilde L)}{\partial \as} &= \dfrac{1}{2 \as}\biggl[2
\Gamma^\rc (\as) \tilde L + \gamma_B^{\text{nc}} (\as)\biggr] \, ,
\label{eq:consistency_1}
\\
\frac{\partial w_{ j+1}(\as,\tilde L) }{\partial \as}
&=
\frac{1}{2\as} \frac{\partial w_{j}(\as,\tilde L) }{\partial \tilde L} - \hat{\beta} (\as) \frac{\partial
w_{ j} (\as,\tilde L) }{\partial \as} \quad (j\geq1)\,.
\label{eq:consistency_2}
\end{align}
\label{eq:consistency}%
\end{subequations}
Equation~\eqref{eq:consistency_1} provides an alternative way to derive the anomalous dimension from our calculation and implies that $w_{1}$ is linear in $\tilde L$. On the other hand, \eq{consistency_2} is a consistency relation that establishes by induction that all $w_{j\ge2}$ are also linear in $\tilde L$.
Hence $\log \tilde Z_B$ is linear in $\tilde{L}$ and can thus be written as
\begin{equation}
\log \tilde Z_B
=
\log \bigl[C_B (\as)\bigr] + z_B(\as) \tilde{L}\,,
\label{eq:ansatz_Z}
\end{equation}
where the $\mu$ dependence of $\tilde Z_B$ is both explicit through $\tilde L$ and implicit through $\alpha_s\equiv \alpha_s(\mu)$.
We can now exponentiate \eq{ansatz_Z} to obtain the ansatz
\begin{equation}
\tilde{Z}_B (x, \mu)
=\, C_B [\as(\mu)]
\, (i x \mu
e^{\gamma_E})^{z_B [\as(\mu)]}\,.
\label{eq:form_counterterm}
\end{equation}
By replacing this in \eq{RGE_for_Z_FT}, we can solve for the functions $C_B$ and $z_B$ and find
\begin{equation}
\label{eq:form_counterterm2}
C_B(\as)
=\,
\exp\! \biggl[ \frac{1}{2}
z(\gamma_B^{\text{nc}}+z_B,\as)
\biggr],
\quad \;
z_B (\as)
= z(\Gamma^\rc,\as)
\,,
\end{equation}
with $z(\gamma,\alpha_s)$ defined in \eq{Zalpha} and expanded in $\as$ in \eq{zexpansion}.
We stress again that the $\eps$ in $z(\Gamma^\rc,\as)$ regulates the integral at its lower limit and must not be set to zero before expanding in $\as$.
Performing the inverse Fourier transformation of \eq{form_counterterm} we also obtain a closed form for the momentum-space renormalization factor $Z_B$ and its inverse:%
\footnote{For the Fourier integral of $\tilde{Z}_B$ to converge we assume
$-1< \mathrm{Re}\bigl[z_B (\as)\bigr]<0$ and analytically continue in $\varepsilon$ the result to the region where ${\rm Re}\bigl[z_B (\as)\bigr]>0$.}
\begin{equation}
Z_B (\hat{s}, \mu)
=
\frac{\theta (\hat{s})C_B(\as)}{\hat{s}\, \Gamma [- z_B (\as)]} \biggl(
\frac{\hat{s}}{\mu e^{\gamma_E}} \biggr)^{\!\!- z_B (\as)}\! \!,
\quad
Z_B^{- 1} (\hat{s}, \mu)
=
\frac{\theta (\hat{s})C_B^{- 1}(\as)}{\hat{s}\, \Gamma [z_B
(\as)]} \biggl( \frac{\hat{s}}{\mu e^{\gamma_E}} \biggr)^{\!\!z_B
(\as)}\!\!.
\end{equation}
Interestingly, these resummed expressions are ordinary functions, and distributions are generated only upon expansion for small $\alpha_s$
using \eq{expPlus}.

In order to provide explicit expressions for $\log C_B$ and $z_B$ up to third loop order we need
\begin{equation}
z(\gamma,\alpha_s)=\frac{\alpha_s}{4\pi}\frac{\gamma_0}{\varepsilon} +
\frac{1}{2} \biggl(\frac{\alpha_s}{4\pi}\biggr)^{\!\!2} \biggl(\frac{\gamma_1}{\varepsilon }-\frac{\beta_0\gamma_0}{\varepsilon ^2}\biggr)
+ \frac{1}{3}\biggl(\frac{\alpha_s}{4\pi}\biggr)^{\!\!3} \biggl(\frac{\gamma_2}{\varepsilon}-\frac{\gamma_1\beta_0+\beta_1\gamma_0}{\varepsilon ^2}+\frac{\beta_0^2\gamma_0}{\varepsilon ^3}\biggr)+\mathcal{O}(\as^4)\,,
\label{eq:zexpansion}
\end{equation}
which can be derived by expanding the integrand in \eq{Zalpha} before integrating, and where the coefficients $\gamma_i$ are defined in analogy to \eq{expansion_gamma}. The coefficients $\beta_i$ are defined in
\eq{alpBare} and explicitly given in \eq{values_beta}. This can be
readily applied to $z_B (\as)$ using the rightmost expression in \eq{form_counterterm2}. For $C_B$ we find
\begin{align}
\log [C_B (\as)] ={}& \frac{1}{2} z(\gamma_B^{\rm nc},\alpha_s) +
\frac{\alpha_s}{4\pi}\frac{\Gamma_{\!0}^c}{2 \varepsilon ^2}
+\frac{1}{8}\biggl(\frac{\alpha_s}{4\pi}\biggr)^{\!\!2} \biggl(\frac{\Gamma^\rc_1}{ \varepsilon
^2}-\frac{3 \beta_0 \Gamma_{\!0}^c}{2 \varepsilon ^3}\biggr)\\
&+\frac{1}{36}\biggl(\frac{\alpha_s}{4\pi}\biggr)^{\!\!3} \biggl(\frac{11 \beta_0^2 \Gamma_{\!0}^c}{\varepsilon ^4}-\frac{5 \beta_0 \Gamma^\rc_1}{\varepsilon ^3}
-\frac{8\beta_1 \Gamma^\rc_0}{\varepsilon ^3}+\frac{2\Gamma^\rc_2}{\varepsilon ^2}\biggr)+\mathcal{O} (\as^4)\,. \nn
\end{align}

For completeness, let us also consider the evolution of the $Z$ factors for the jet function in position and momentum space.
Using \eq{form_counterterm} and taking the ratio of two renormalization factors $\tilde{Z}_B$ at $\mu$ and $\mu_0$ we can (after careful manipulation) set $\eps\to0$. In this way we obtain the RG-evolved expression
\begin{equation}
\tilde{Z}_B(x, \mu)=
\tilde{Z}_B(x, \mu_0) \, \tilde{U}_B(x,\mu_0,\mu)
\,,
\label{eq:evolve_Z_ft}
\end{equation}
where $\tilde{U}_B$ was defined in \eq{evolve_B_ft}. Note that, as required by consistency one has $\tilde Z_B (x,\mu)\tilde B(x, \mu) = \tilde Z_B (x,\mu_0)\tilde B(x, \mu_0) $.
For the momentum-space $Z$ factor we obtain accordingly
\begin{equation}
Z_B(\hat{s}, \mu)=\int
\!\dd \hat{s}^{\prime}\, U_B(\hat{s}-\hat{s}^{\prime}, \mu_0, \mu)\, Z_B(\hat{s}^{\prime}, \mu_0)\,,
\end{equation}
with the evolution kernel $U_B$ given in \eq{EvolConvol}.

\section{Summary of MI results}
\label{app:calB_and_MI}

\begin{figure}[!t]
\begin{center}
\includegraphics[width=3.3cm, height=1.5cm]{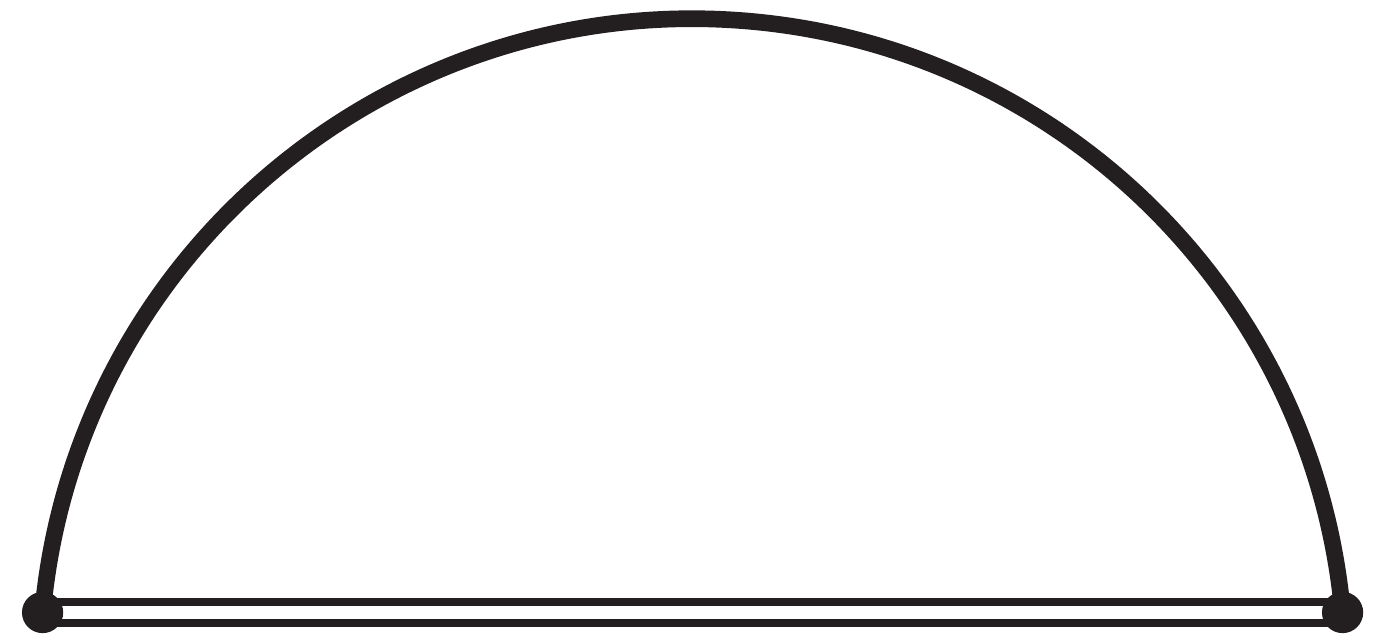}
\caption{Diagrammatic representation of $\MI{1}{}$. Double and solid lines represent heavy-quark and quadratic propagators, respectively.
\label{fig:MI1L}}
\end{center}
\end{figure}

\begin{figure*}[t]
\begin{center}
\subfigure[]
{\includegraphics[width=3.3cm, height=2cm]{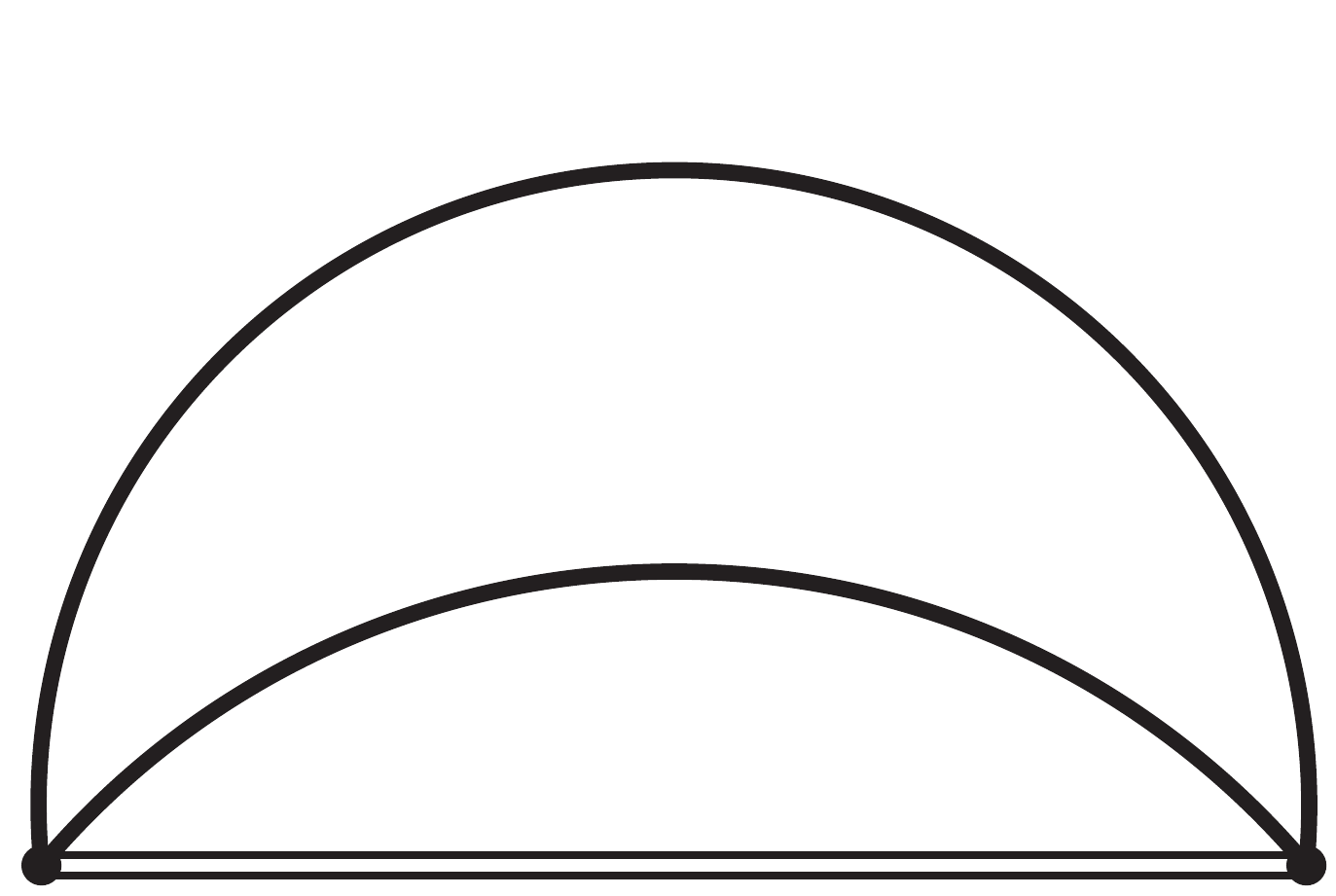}
\label{fig:MI2L1}}
\qquad
\subfigure[]{\includegraphics[width=3.3cm, height=2cm]{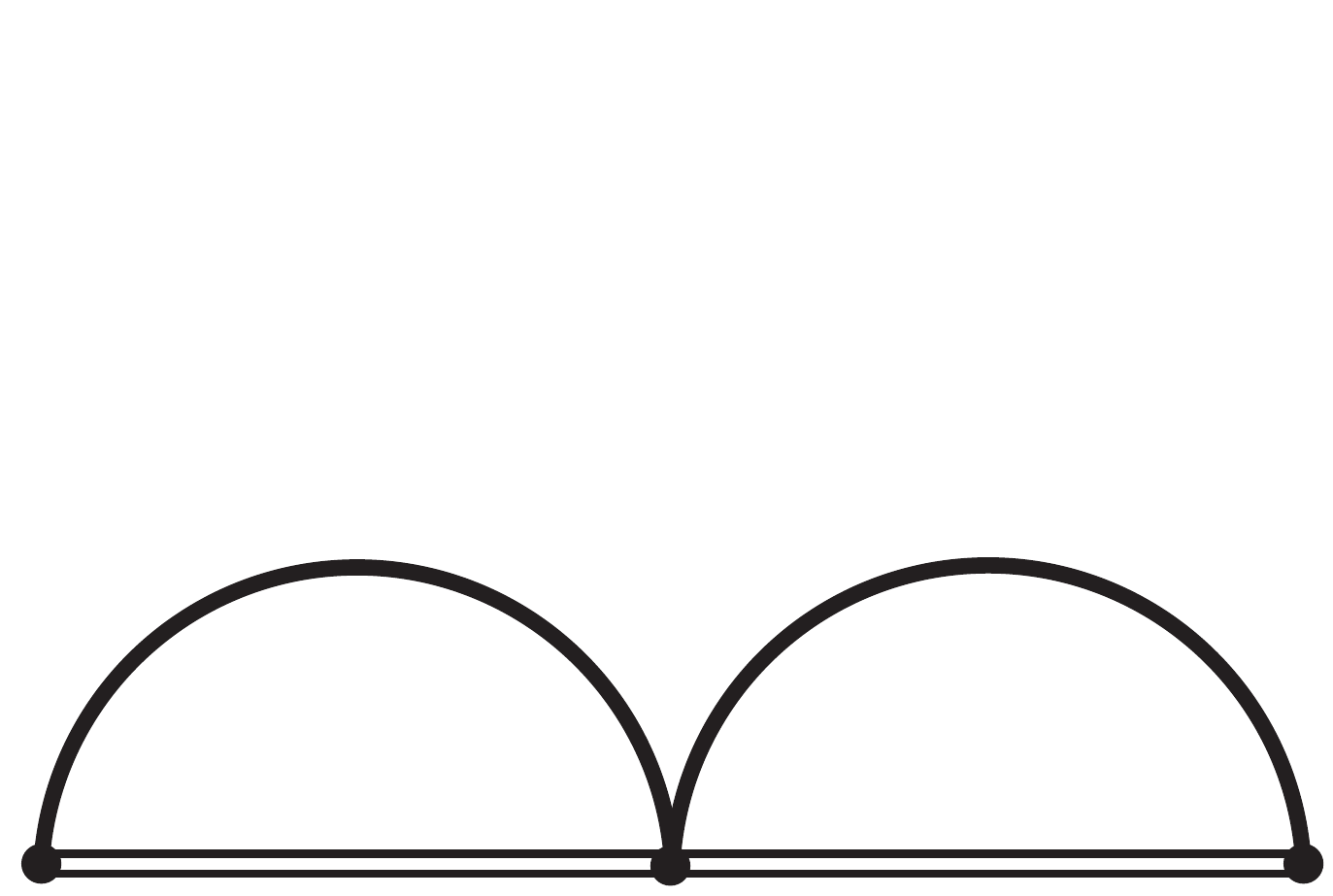}
\label{fig:MI2L2}}
\qquad
\subfigure[]{\includegraphics[width=3.3cm, height=2cm]{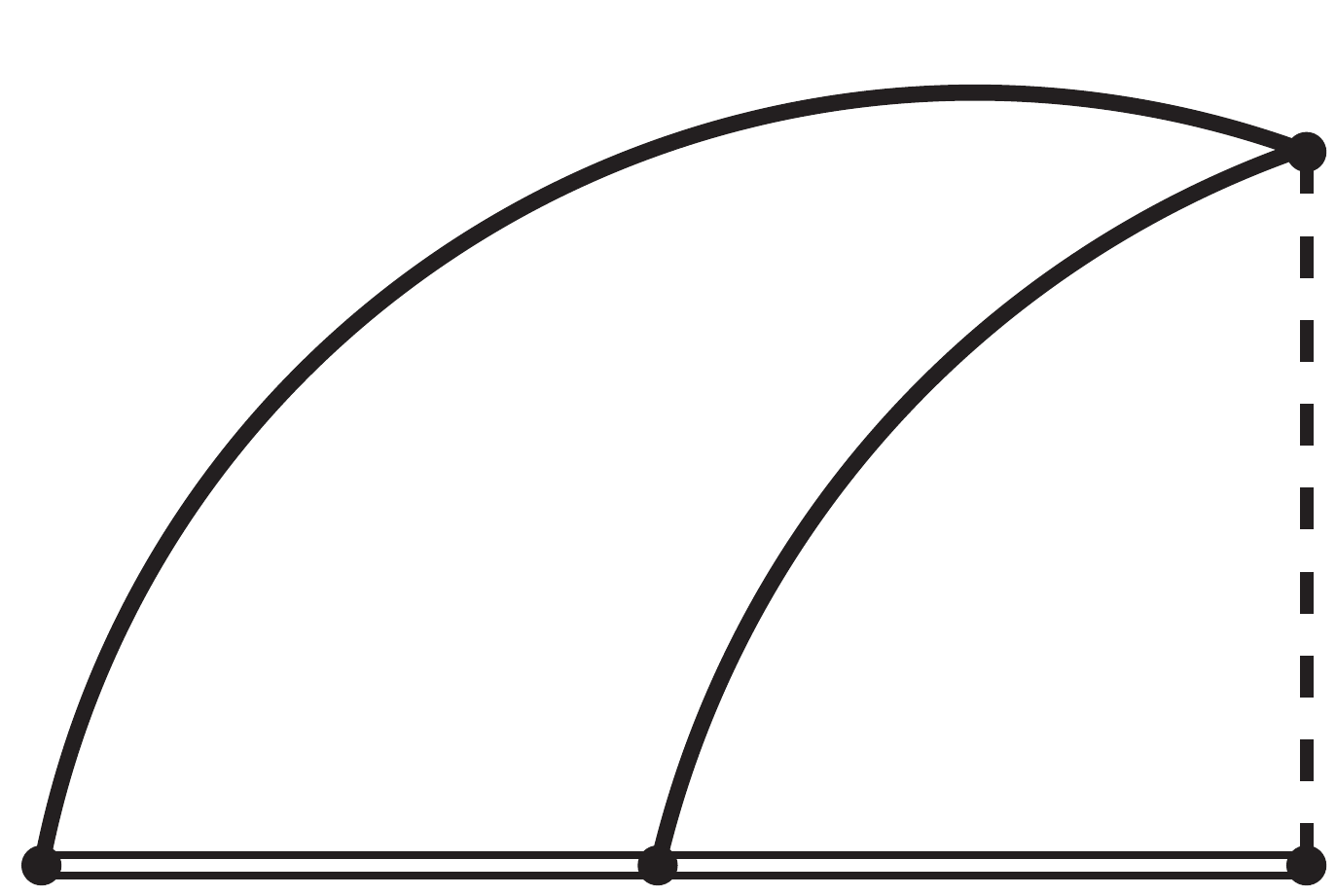}
\label{fig:MI2L3}}
\caption{Diagrammatic representation of the two-loop MIs.
Double, dashed and solid lines represent heavy-quark, light-like linear (eikonal), and quadratic propagators, respectively.}
\label{fig:MI2L}%
\end{center}
\end{figure*}

Here the analytic results for the one-, two-, and three-loop MIs shown in Figs.~\ref{fig:MI1L},~\ref{fig:MI2L}~and~\ref{fig:MI3L}, respectively, are given.
As stated in \sec{computation} we set w.l.o.g.\ $\hat{s}=-1$.
The expressions for arbitrary $d$ can be expanded with \texttt{HypExp} \cite{Huber:2005yg, Huber:2007dx} when necessary.
We checked all our MI results numerically using \texttt{FIESTA} \cite{Smirnov:2021rhf} and/or \texttt{PySecDec} \cite{Heinrich:2023til} up to $\mathcal{O}(\varepsilon^3)$. For conciseness, we suppress the causal $i\eta$ prescription for the propagators in this appendix.

Some of our results involve the hypergeometric functions
\begin{align}
\pFq{2}{1}{a_1,a_2}{b}{c}
=\;&\frac{\Gamma (b) }{\Gamma (a_1) \Gamma (b-a_1)} \int_0^1 \!\mathd z\,\frac{z^{a_1-1}
(1-z)^{b-a_1-1}}{(1- z c)^{a_2}}\,,
\label{eq:hypergeometric} \\ \nn
\pFq{3}{2}{a_1,a_2,a_3}{b_1,b_2}{c}
=\;&\frac{\Gamma (b_2) }{\Gamma (a_3) \Gamma (b_2-a_3)} \int_0^1 \!\mathd z\,\dfrac{z^{a_3-1}}{(1-z)^{a_3-b_2+1}} \pFq{2}{1}{a_1,a_2}{b_1}{z c} \,.
\end{align}
The former is symmetric in $a_1 \leftrightarrow a_2$ while the latter is invariant under $a_1 \leftrightarrow a_2 \leftrightarrow a_3$ and $b_1 \leftrightarrow b_2$.

\subsection{One-loop integral}
\label{app:one-loop_result}
The MI in \fig{MI1L} corresponds to the one-loop self-energy correction of the HQET propagator:
\begin{equation}
\MI{1}{}
=
\int \frac{\dd^d k}{i\pi^{d/2}} \frac{e^{\varepsilon \gamma_E} }{(-2v\cdot k+1)(-k^2)}
=
e^{\varepsilon \gamma_E}\,
\Gamma (3-d)\, \Gamma\bigl(\tfrac{d}{2}-1\bigr)\,.
\label{eq:MI1L}
\end{equation}

\subsection{Two-loop integrals }
\label{app:two-loop_result}

All three two-loop MIs in \fig{MI2L} can be mapped to the planar family
\begin{equation}
G(\nu_1, \ldots,\nu_7)\equiv e^{2\varepsilon \gamma_E} \int \frac{\dd^d k_1}{i\pi^{d/2}} \int \frac{\dd^d k_2}{i\pi^{d/2}} \, \frac{1}{D_1^{\nu_1} \cdots D_7^{\nu_7}} \,, \\
\end{equation}
with the (propagator) denominators
\begin{align}
\!\!\!\!\!\!D_1 =& - \!n \cdot k_1\,, & D_2 =& - \! n \cdot k_2\,, &D_3 =& - \!2 v \cdot k_1 + 1 \,, \\
\!\!\!\!\!\!D_4 =& - \!2 v \cdot k_2 + 1\,, & D_5 =& - \!k_1^2 \,, & D_6 =& - \! k_2^2\,,
&\!\!\!\!D_7 =& - \!(k_1 -k_2)^2.\nn
\end{align}
The relevant master integrals have been computed to all orders:
\begin{align}\label{eq:MI2L}
\MI{2}{a}=\,
& G(0,0,0,1,1,0,1)=
e^{2 \varepsilon \gamma_E} \,\Gamma^2 \bigl(\tfrac{d}{2}-1\bigr) \Gamma (5-2d)\,,
\\
\MI{2}{b}={}&
G(0,0,1,1,1,1,0)=
\bigl(\MI{1}{}\bigr)^2,
\nn\\
\MI{2}{c}={}&
G(0,1,1,1,1,0,1) =
\frac{e^{2 \varepsilon\gamma_E}}{2} \, \Gamma(7-2d) \Gamma^2\bigl(\tfrac{d}{2}-2\bigr)\,
\pFq{3}{2}{1,4-d,6-\frac{3d}{2}}{3-\frac{d}{2},5-d}{1}.\nn
\end{align}

\begin{figure*}[t]
\centering
\subfigure[]{\includegraphics[width=3.3cm, height=2cm]{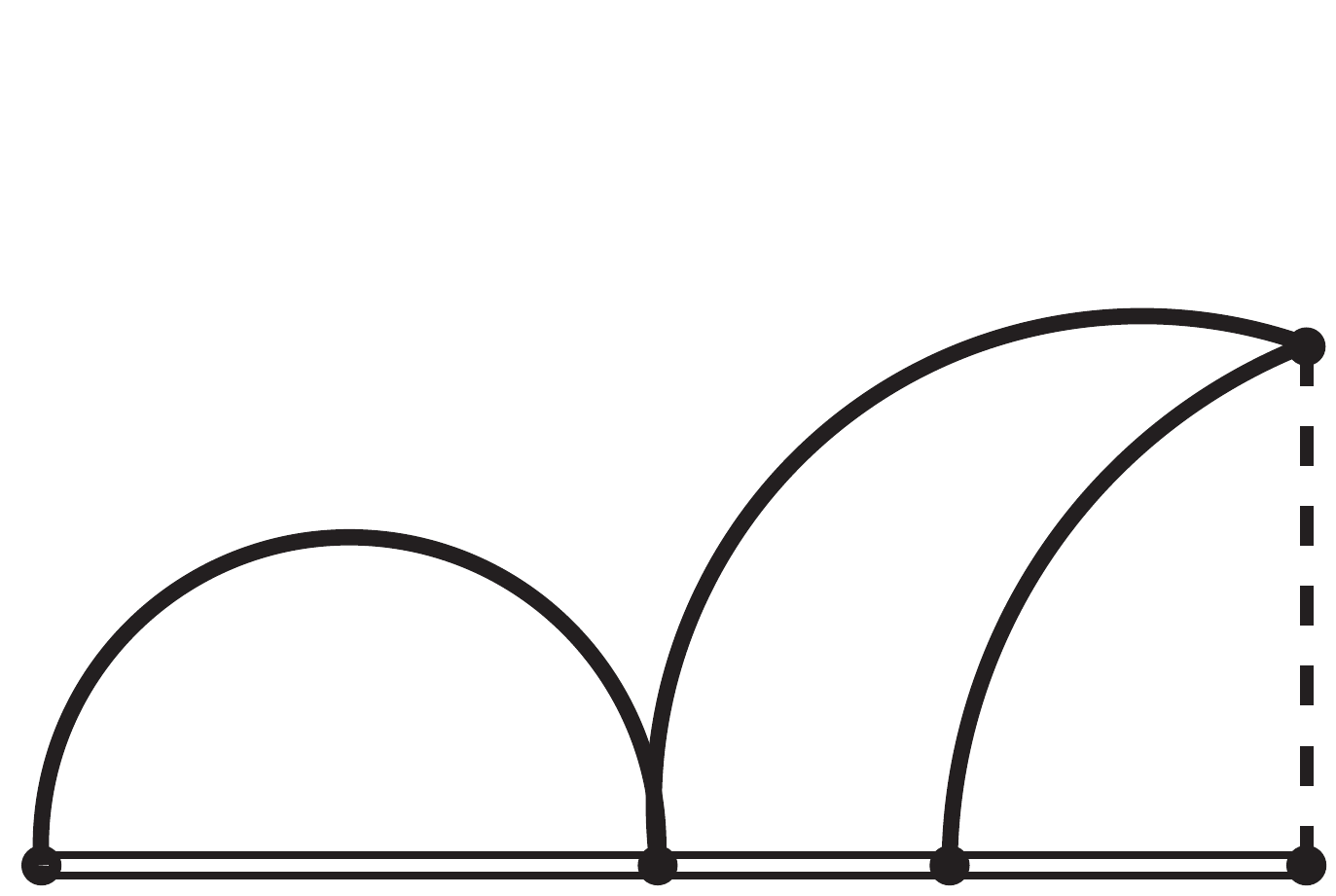}
\label{fig:MI3L17}} \hfill
\subfigure[]{\includegraphics[width=3.3cm, height=2cm]{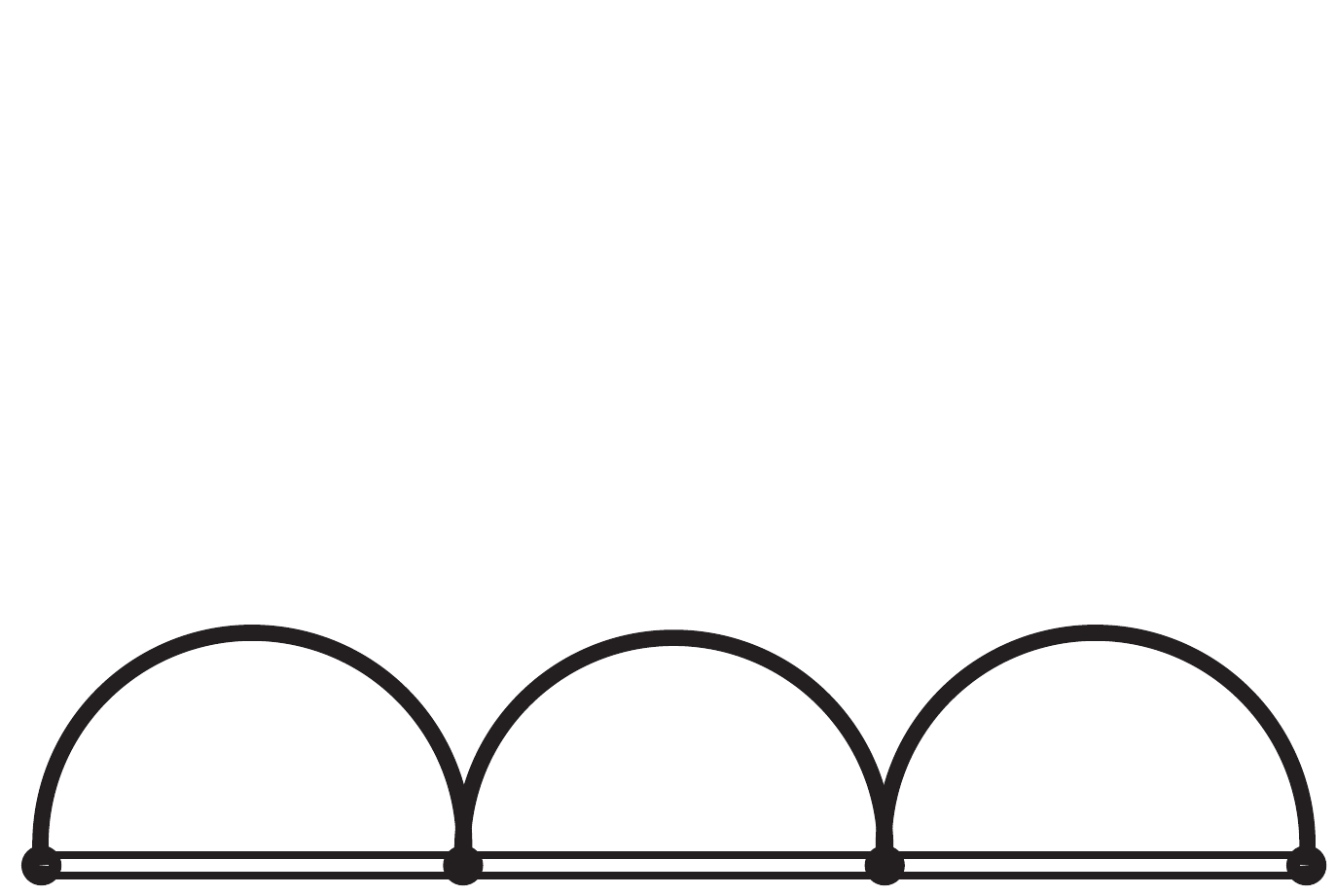}
\label{fig:MI3L07}} \hfill
\subfigure[]
{\includegraphics[width=3.3cm, height=2cm]{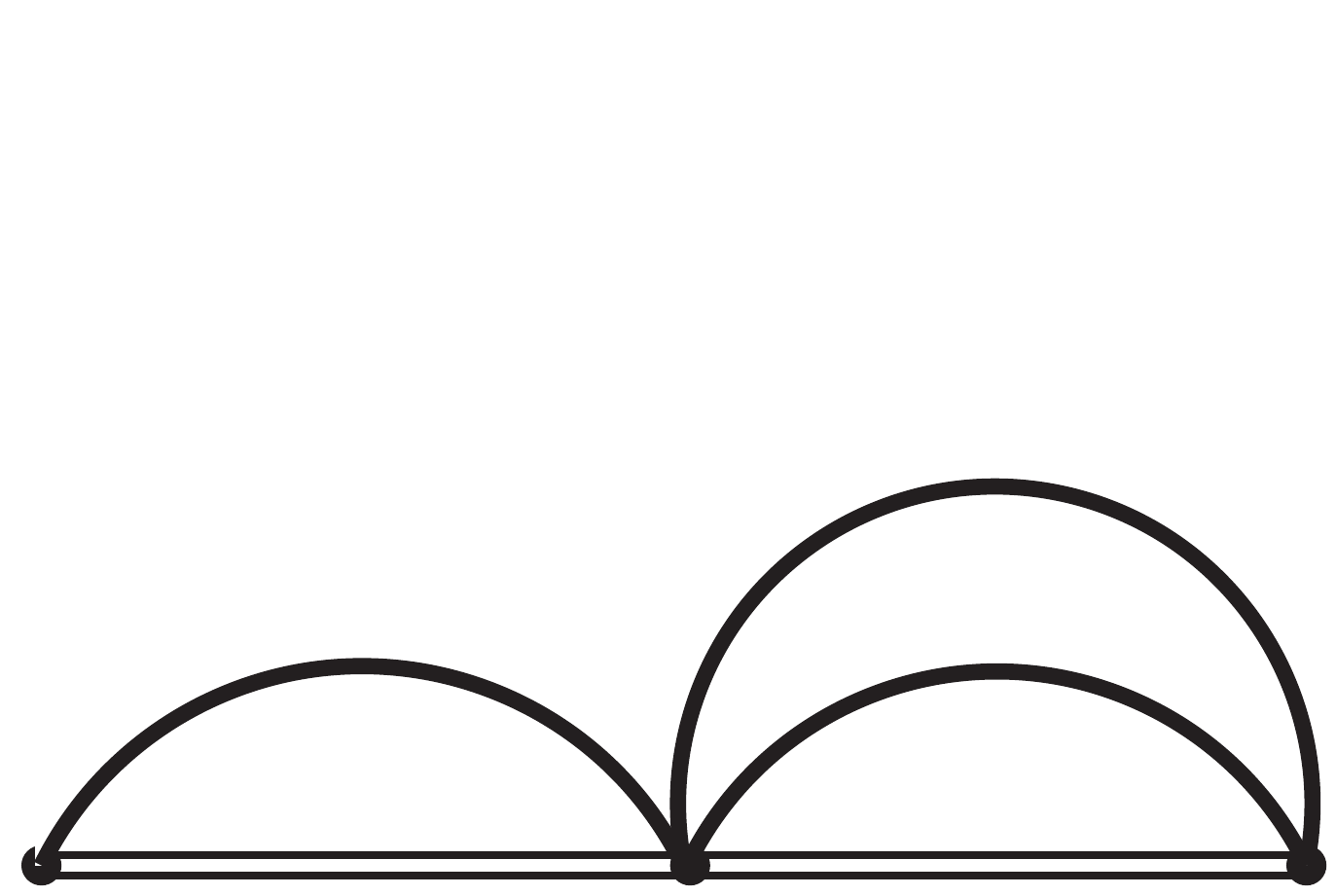}
\label{fig:MI3L05}} \hfill
\subfigure[]
{\includegraphics[width=3.3cm, height=2cm]{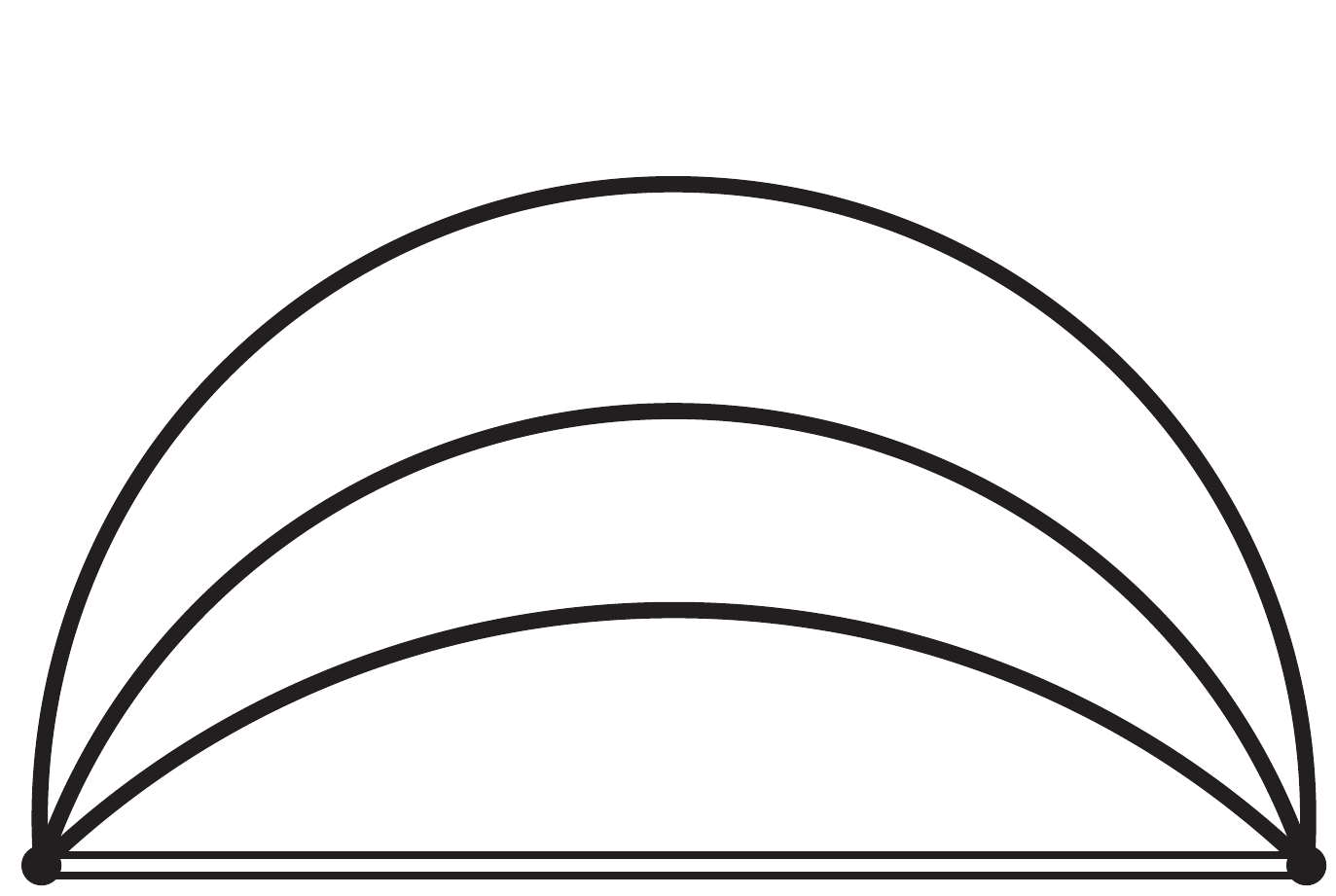}
\label{fig:MI3L01}}
\subfigure[]{\includegraphics[width=3.3cm, height=2cm]{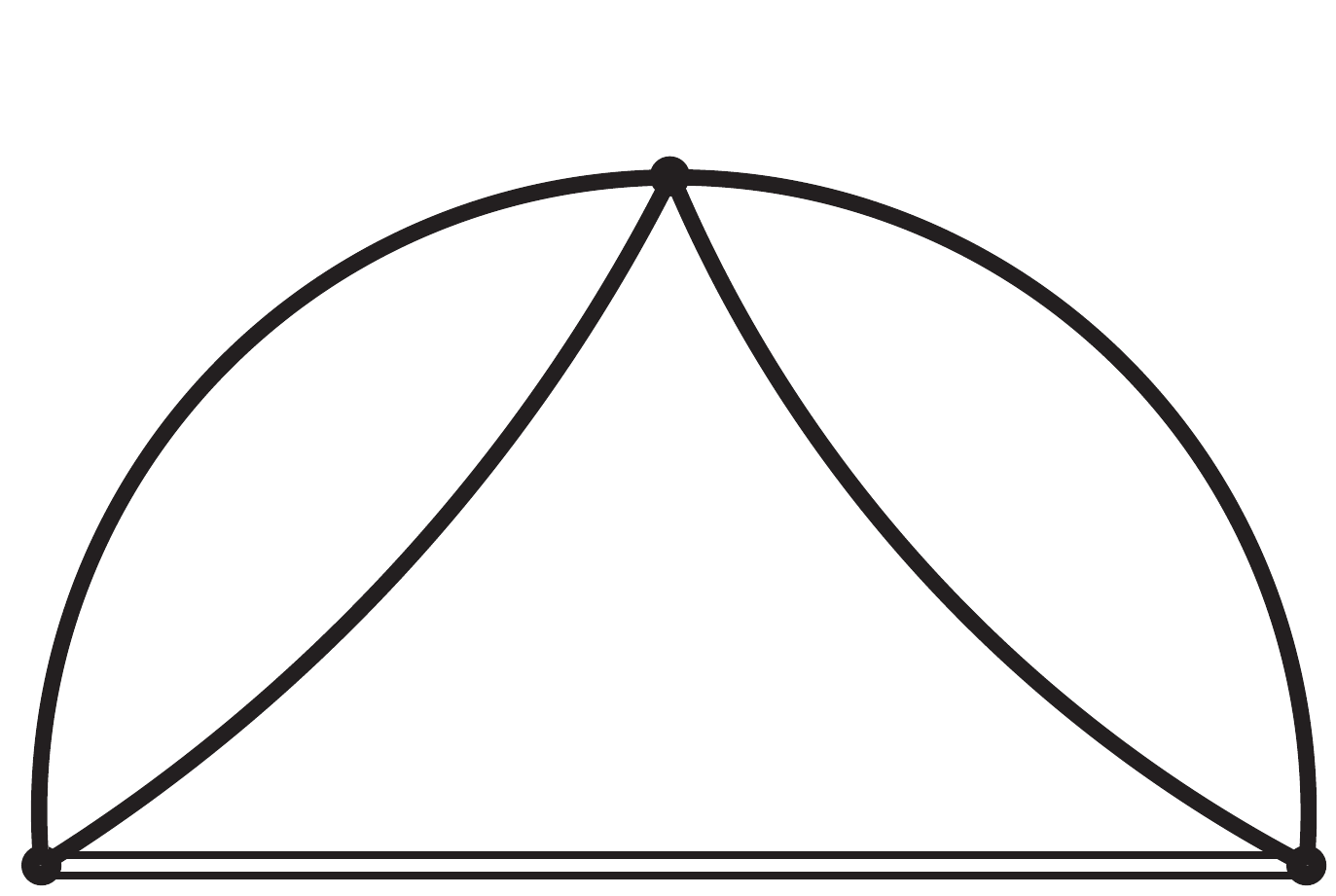}
\label{fig:MI3L02}} \hfill
\subfigure[]{\includegraphics[width=3.3cm, height=2cm]{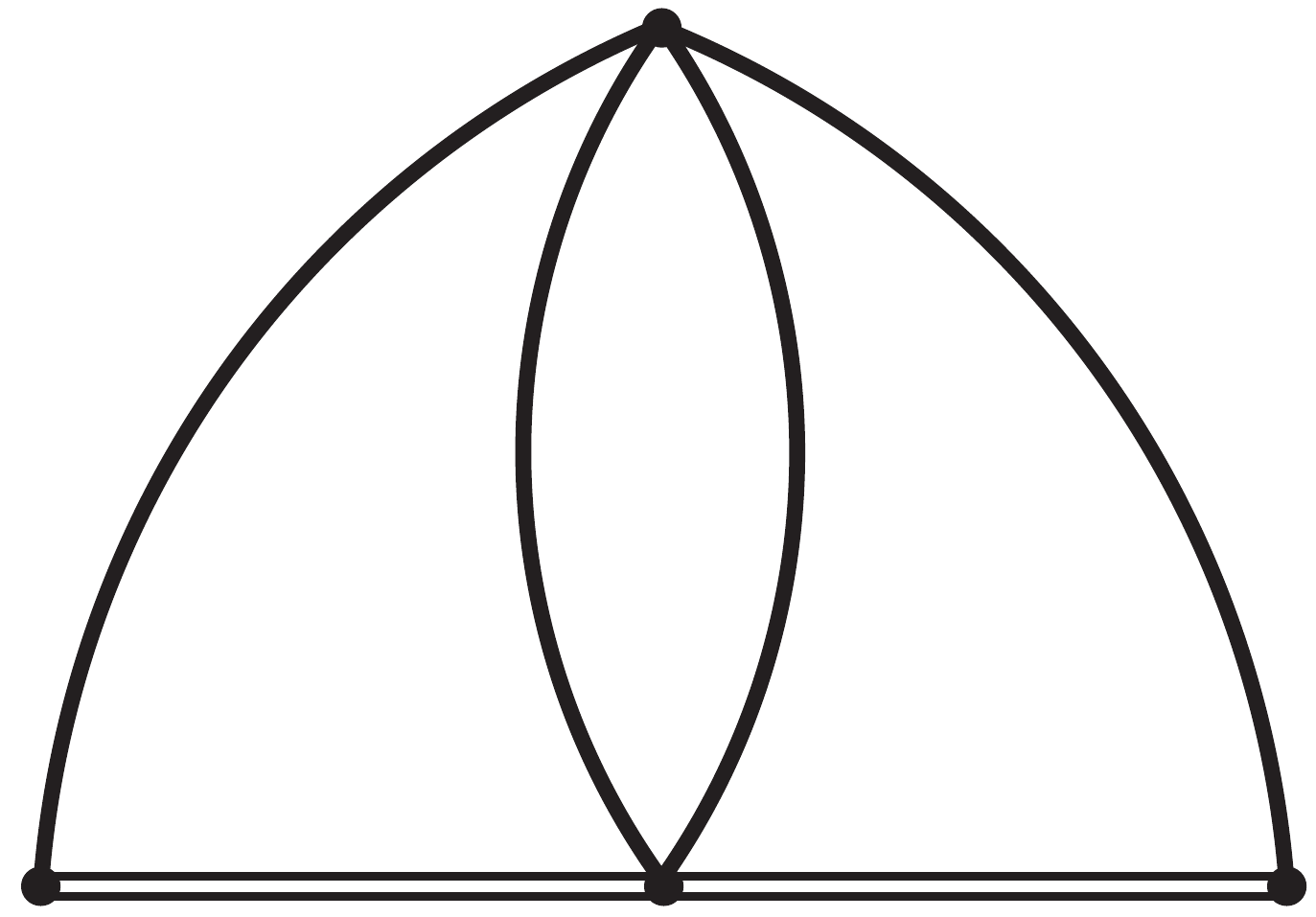}
\label{fig:MI3L03}} \hfill
\subfigure[]{\includegraphics[width=3.3cm, height=2cm]{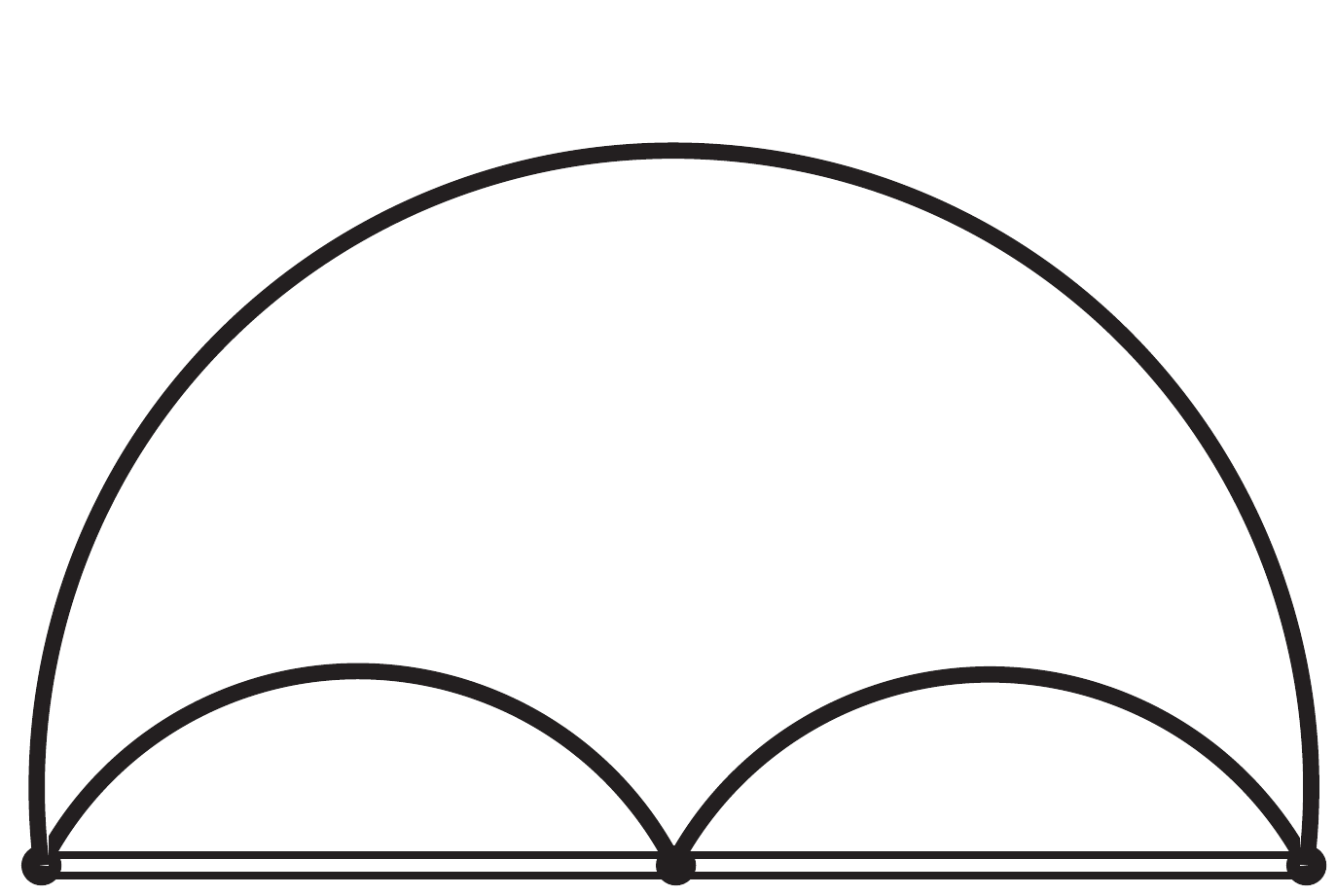}
\label{fig:MI3L04}} \hfill
\subfigure[]
{\includegraphics[width=3.3cm, height=2cm]{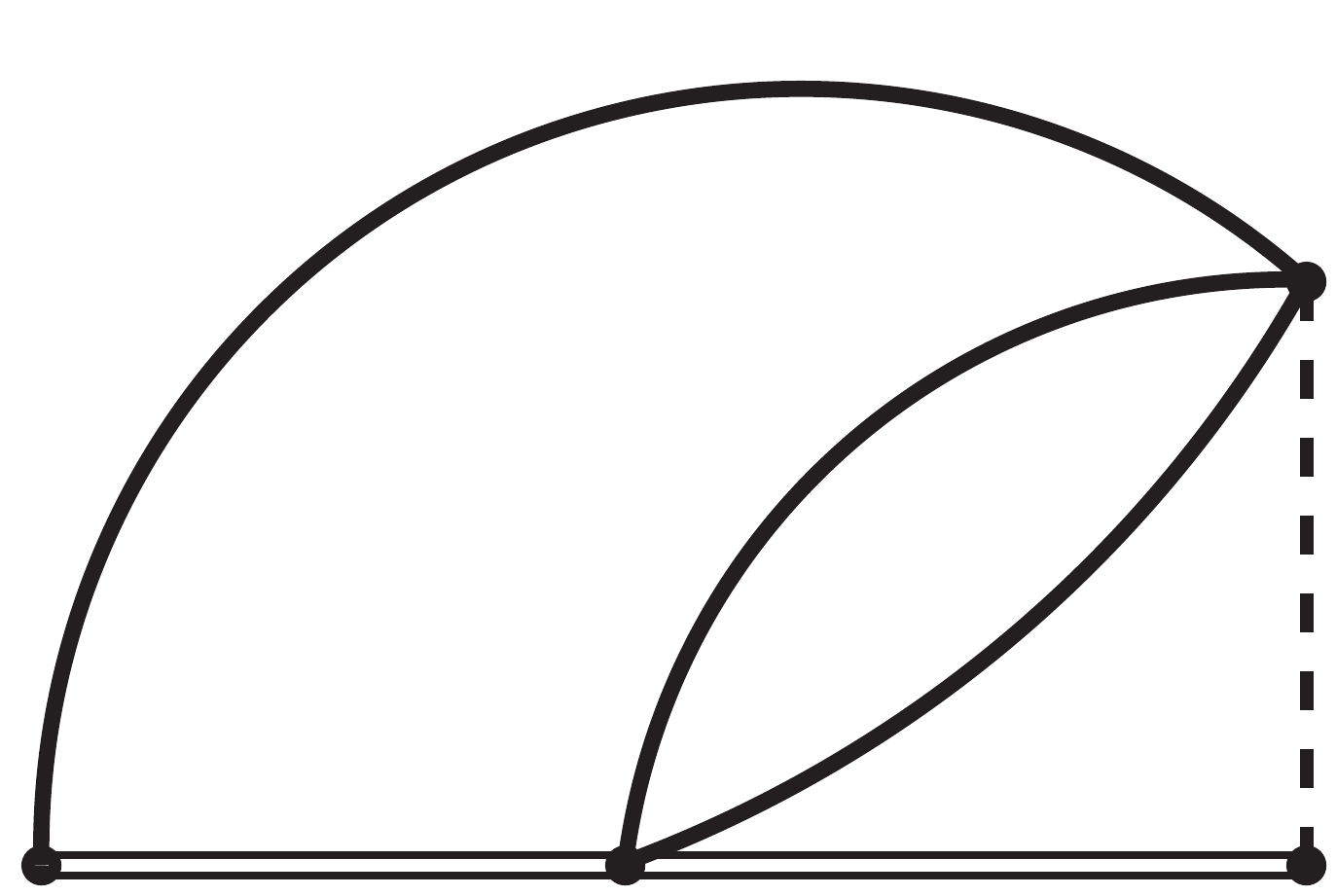}
\label{fig:MI3L09}}
\subfigure[]{\includegraphics[width=3.3cm, height=2cm]{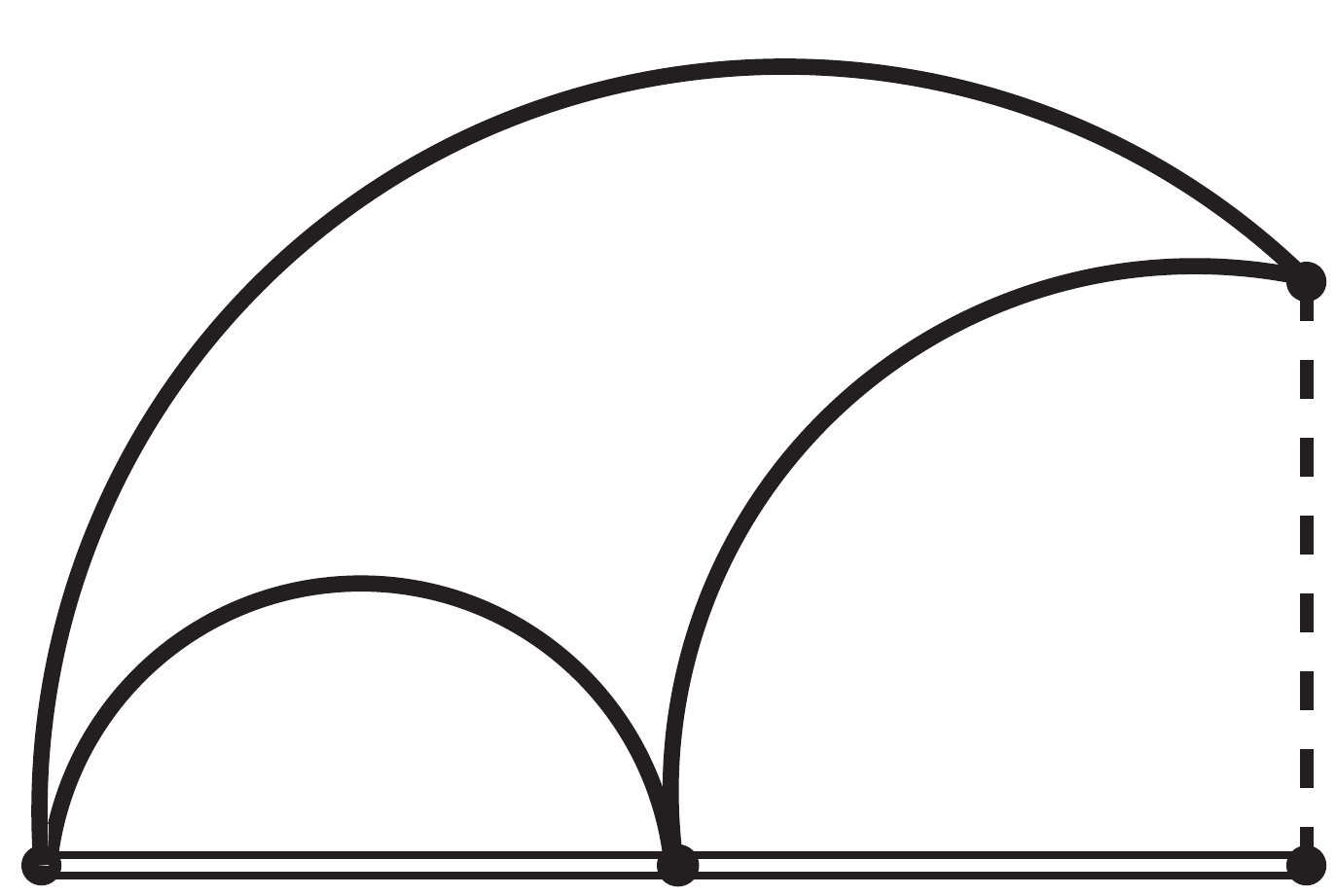}
\label{fig:MI3L10}} \hfill
\subfigure[]{\includegraphics[width=3.3cm, height=2cm]{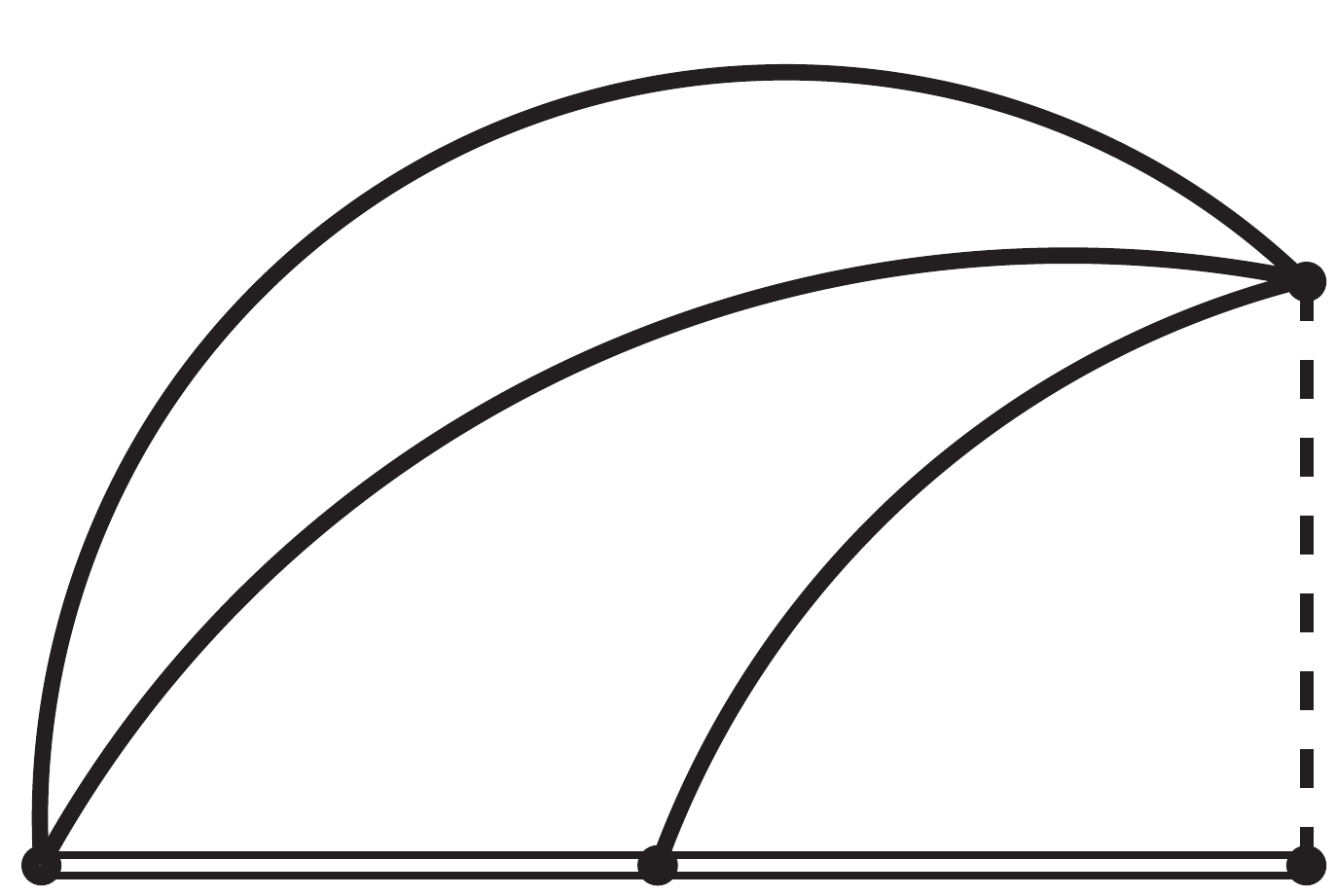}
\label{fig:MI3L11}} \hfill
\subfigure[]{\includegraphics[width=3.3cm, height=2cm]{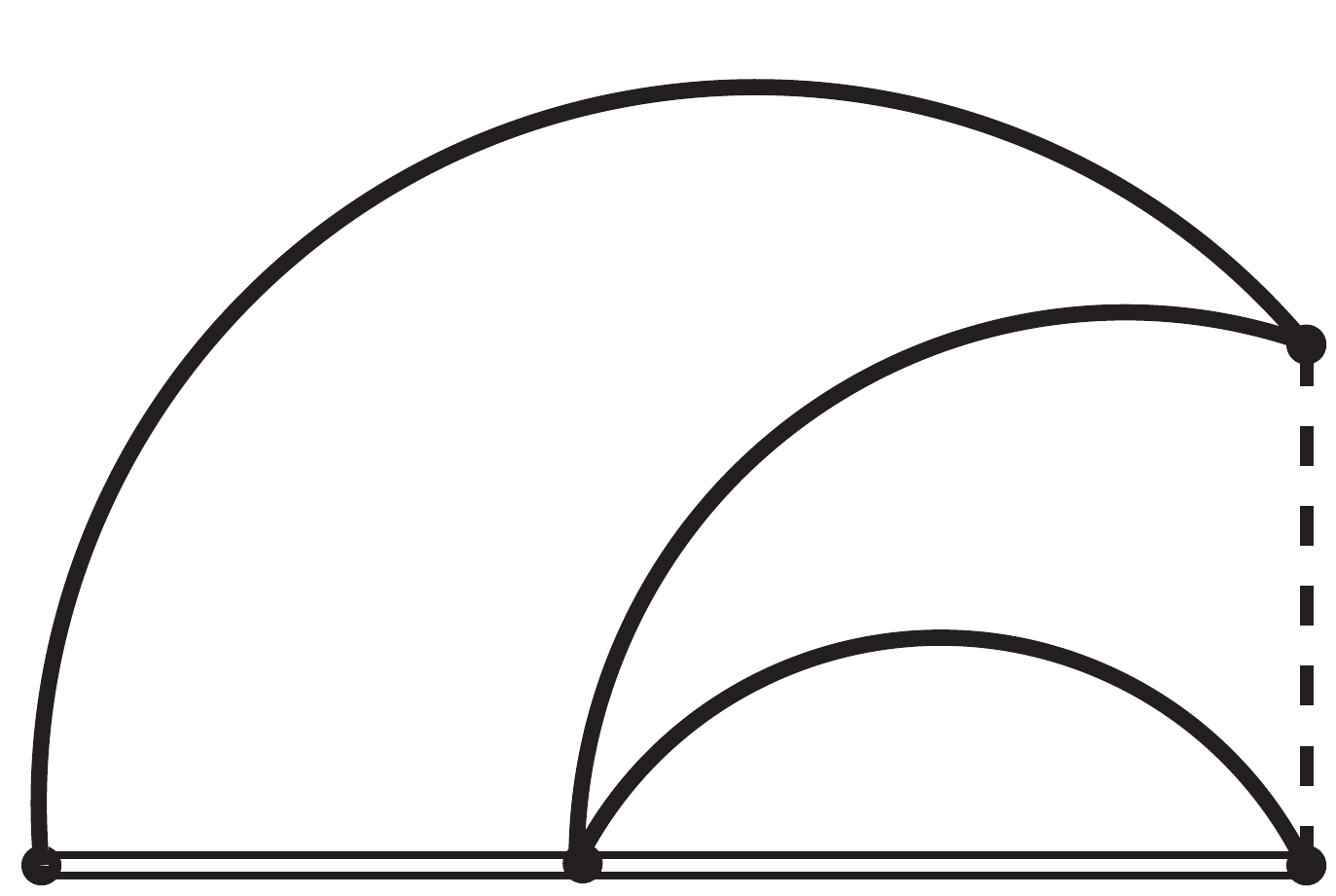}
\label{fig:MI3L14}} \hfill
\subfigure[]{\includegraphics[width=3.3cm, height=2cm]{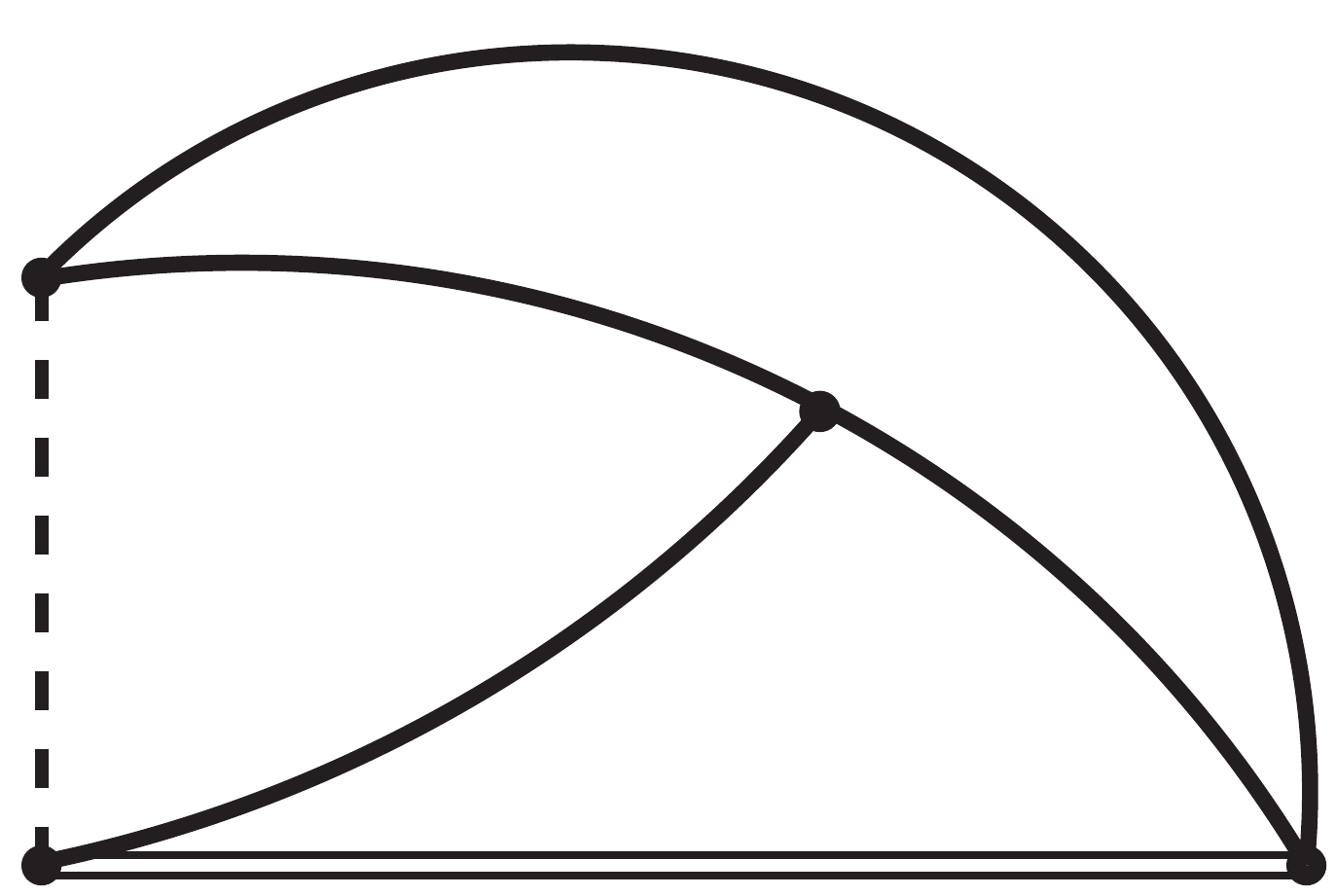}
\label{fig:MI3L08}}
\subfigure[]{\includegraphics[width=3.3cm, height=2cm]{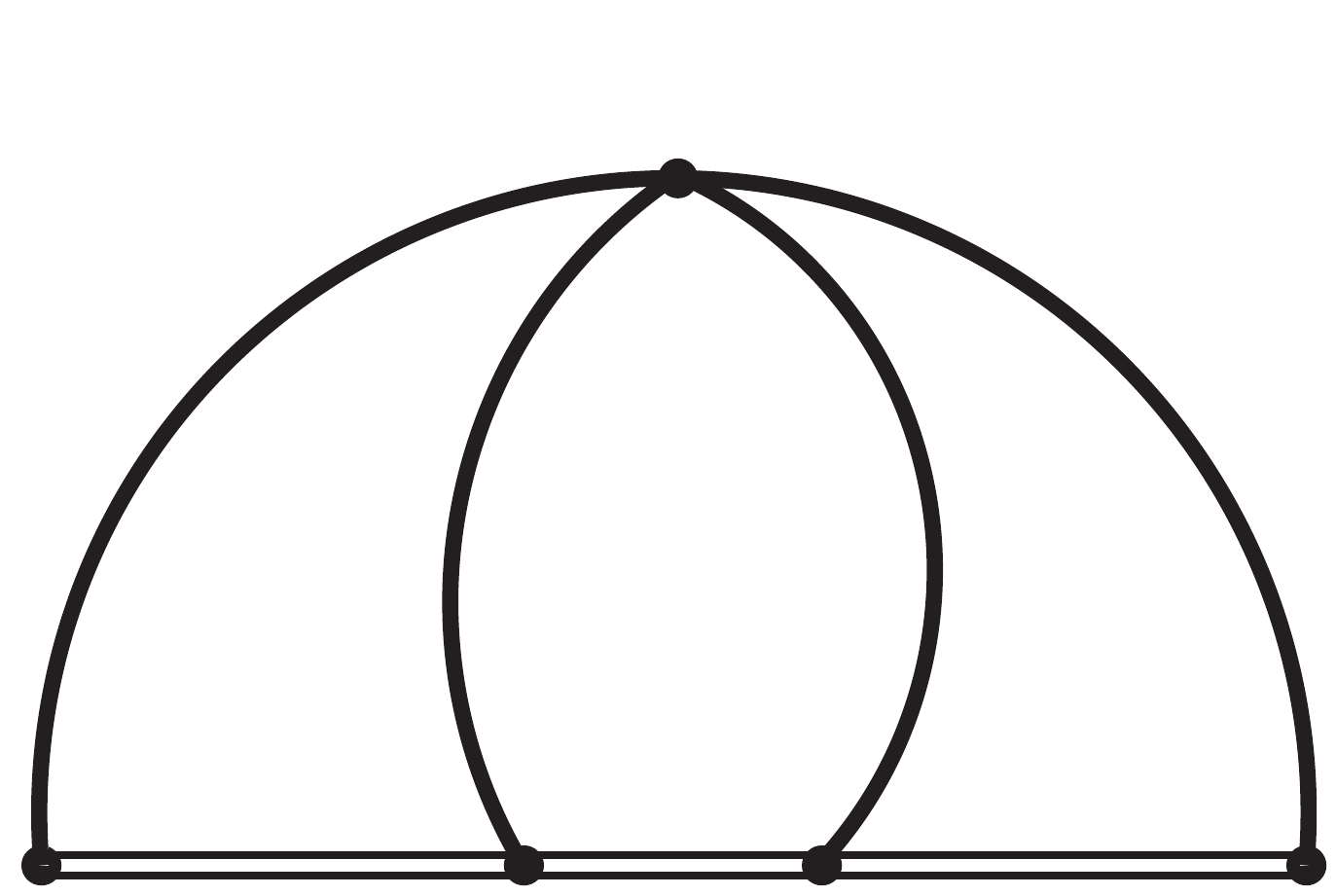}
\label{fig:MI3L06}} \hfill
\subfigure[]{\includegraphics[width=3.3cm, height=2cm]{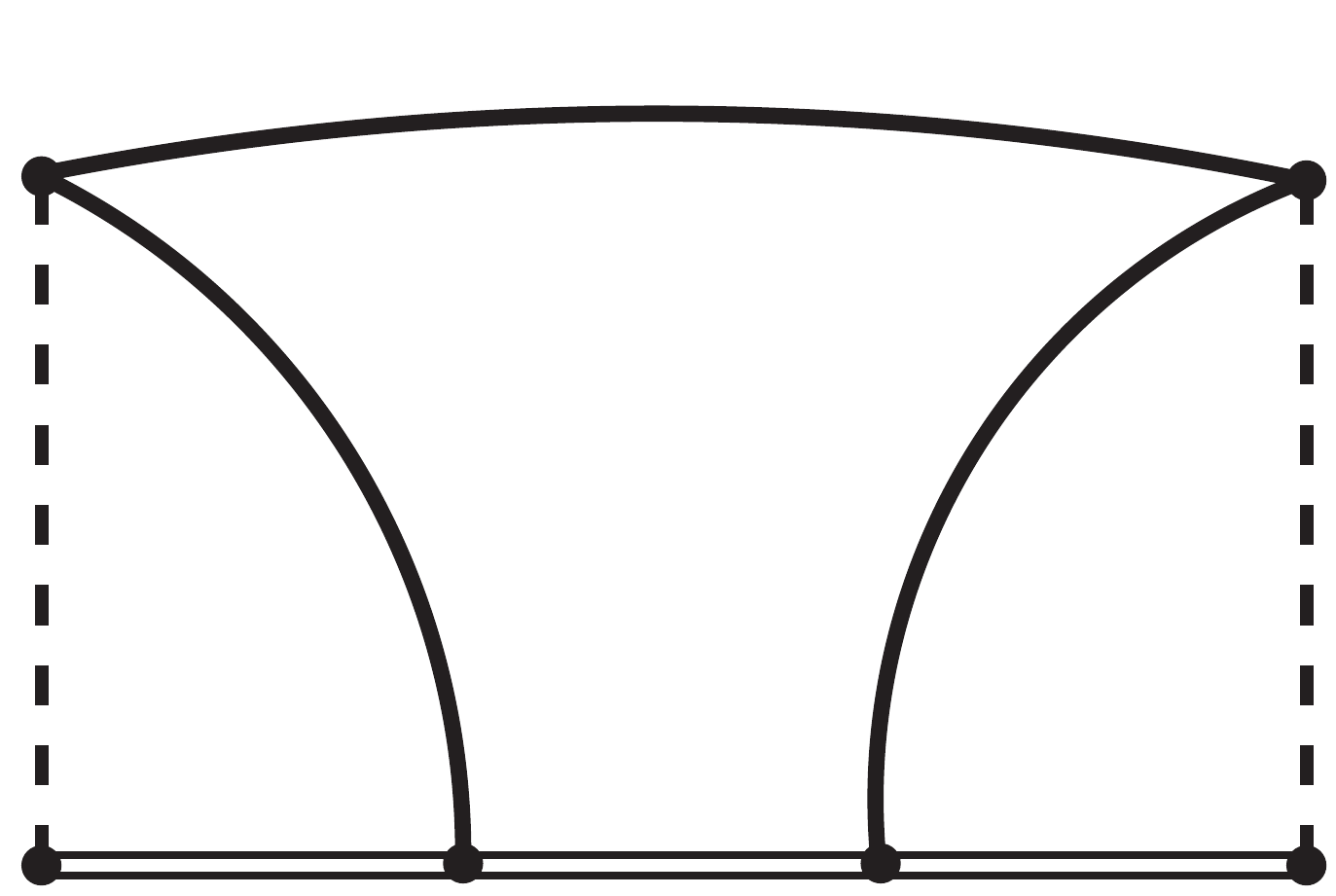}
\label{fig:MI3L19}} \hfill
\subfigure[]
{\includegraphics[width=3.3cm, height=2cm]{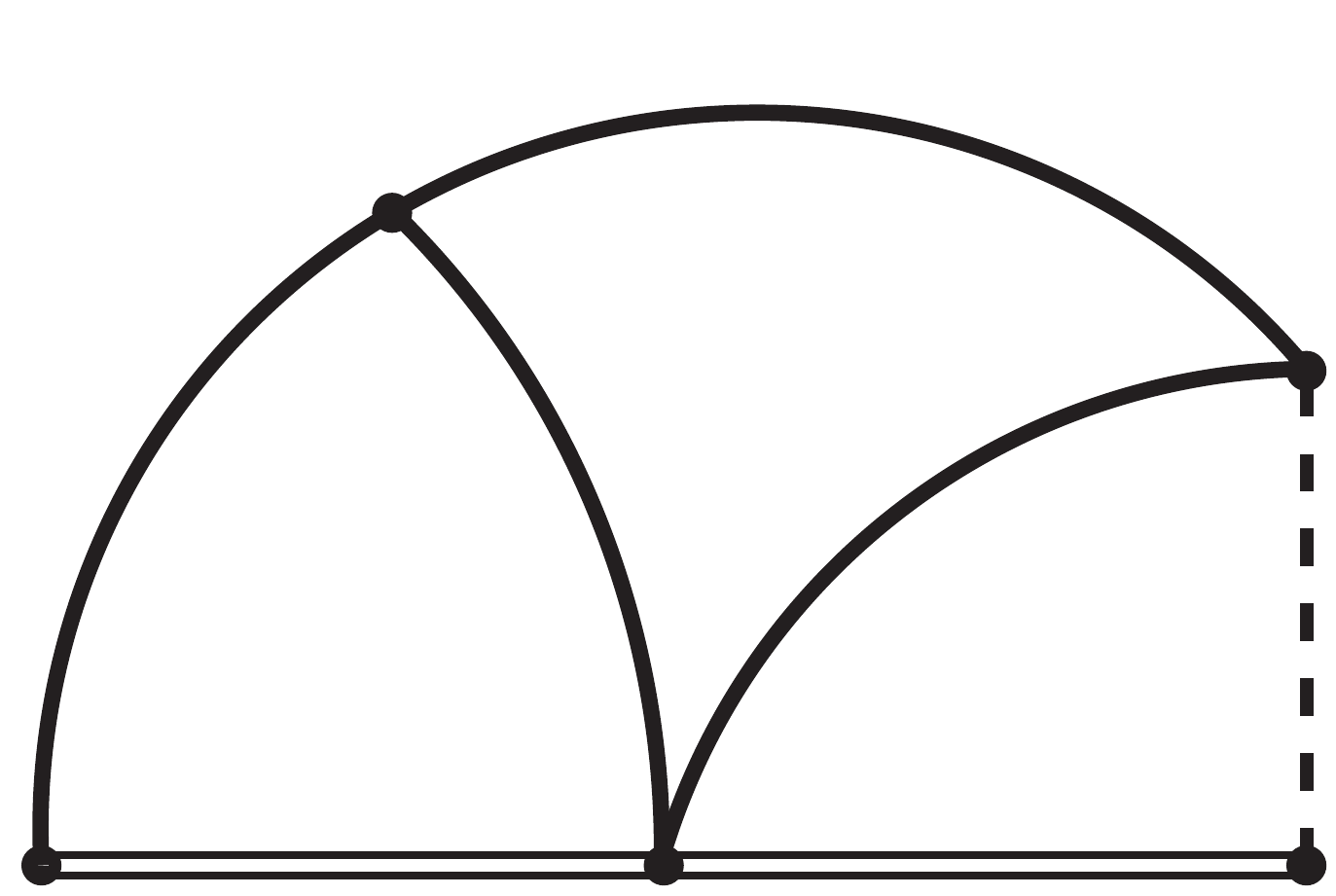}
\label{fig:MI3L12}} \hfill
\subfigure[]{\includegraphics[width=3.3cm, height=2cm]{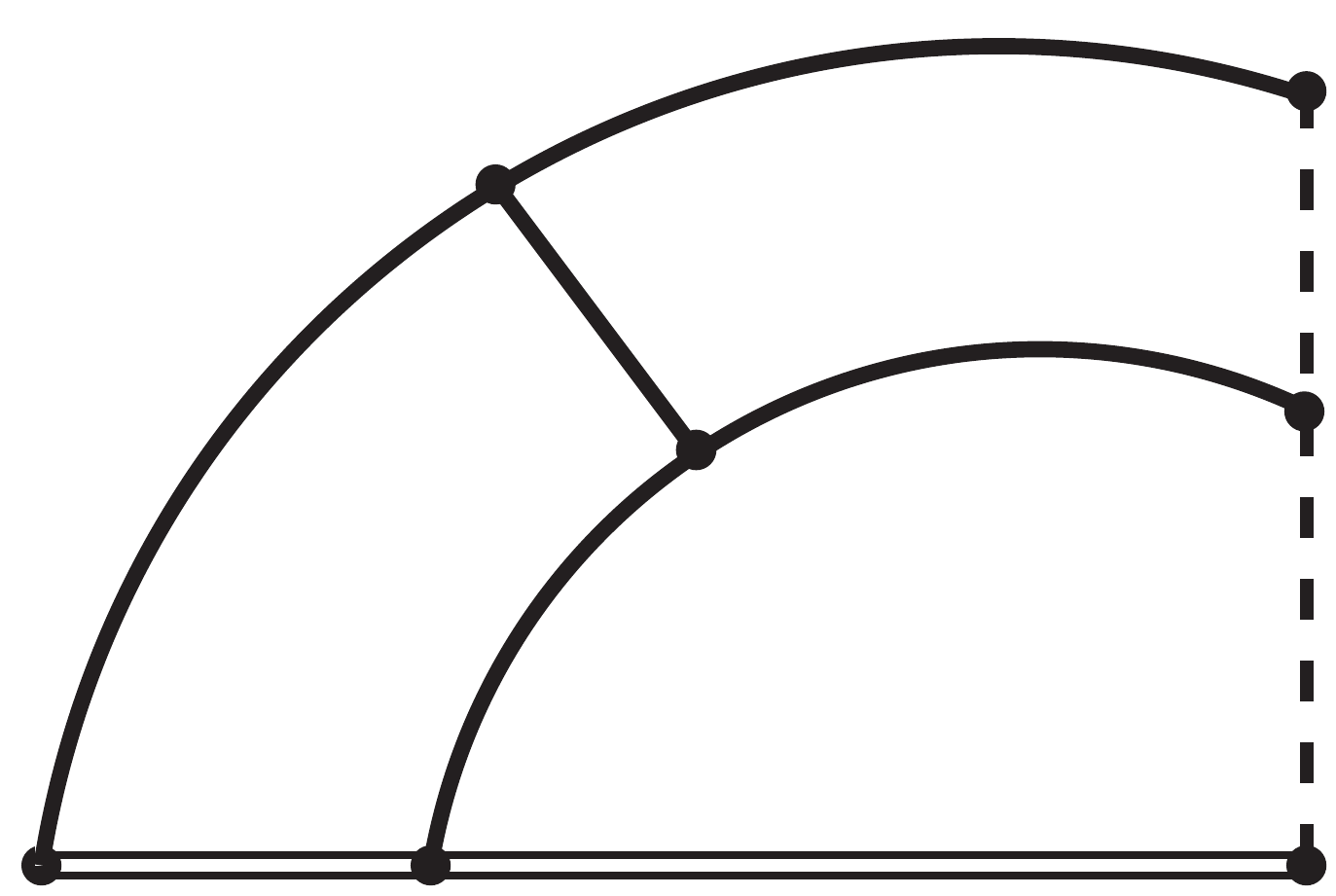}
\label{fig:MI3L13}}
\subfigure[]
{\includegraphics[width=3.3cm, height=2cm]{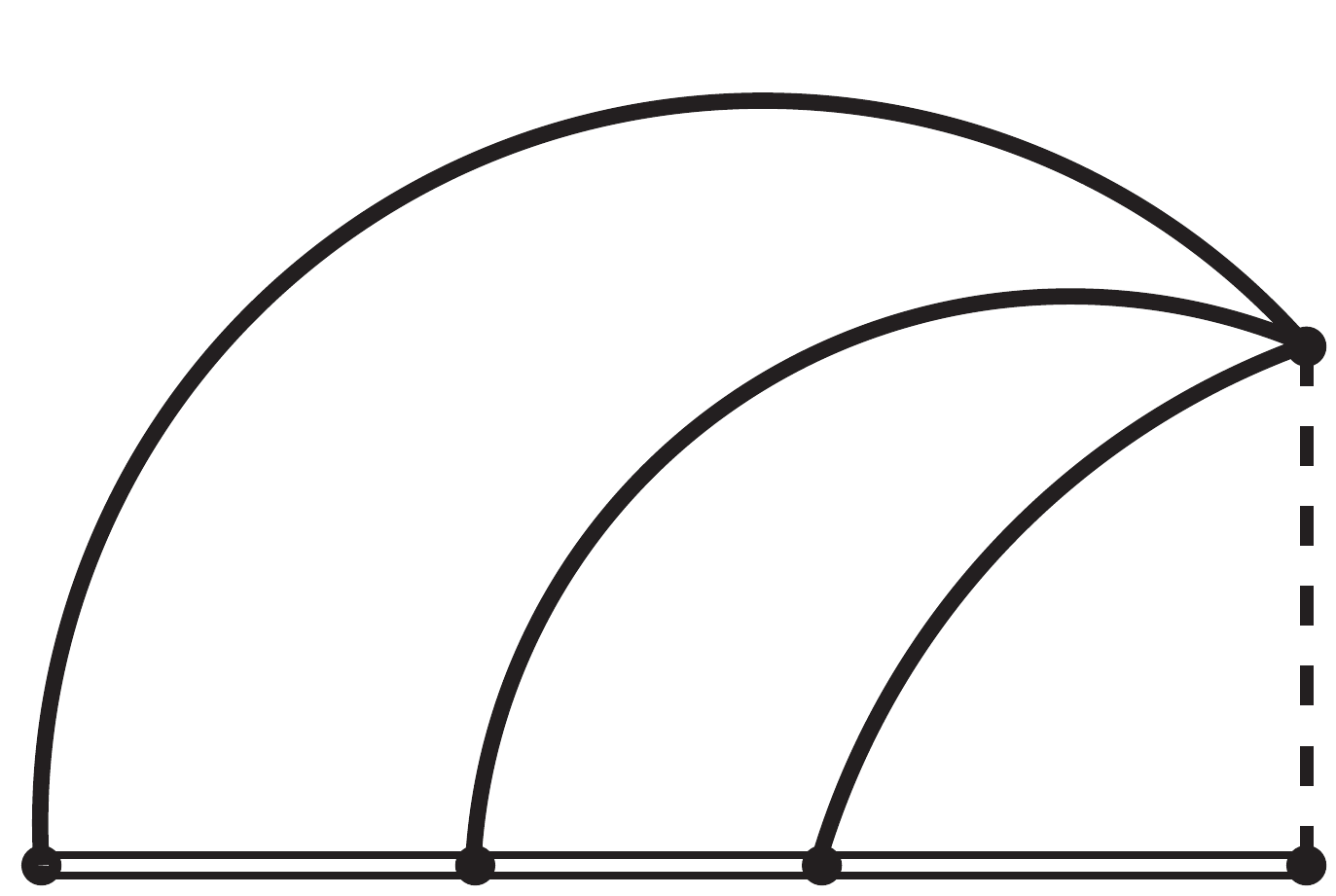}
\label{fig:MI3L16}} \hfill
\subfigure[]{\includegraphics[width=3.3cm, height=2cm]{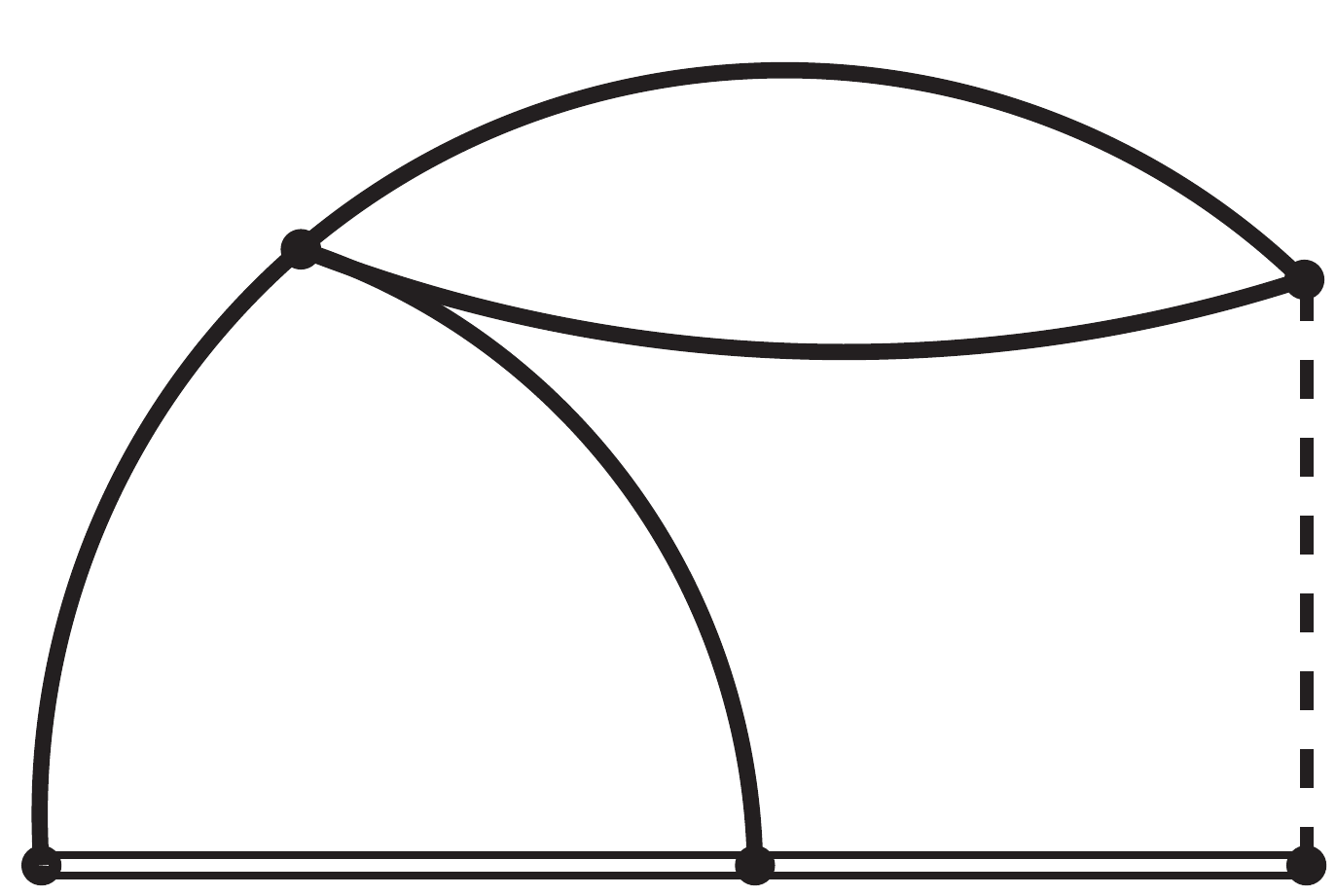}
\label{fig:MI3L18}} \hfill
\subfigure[]{\includegraphics[width=3.3cm, height=2cm]{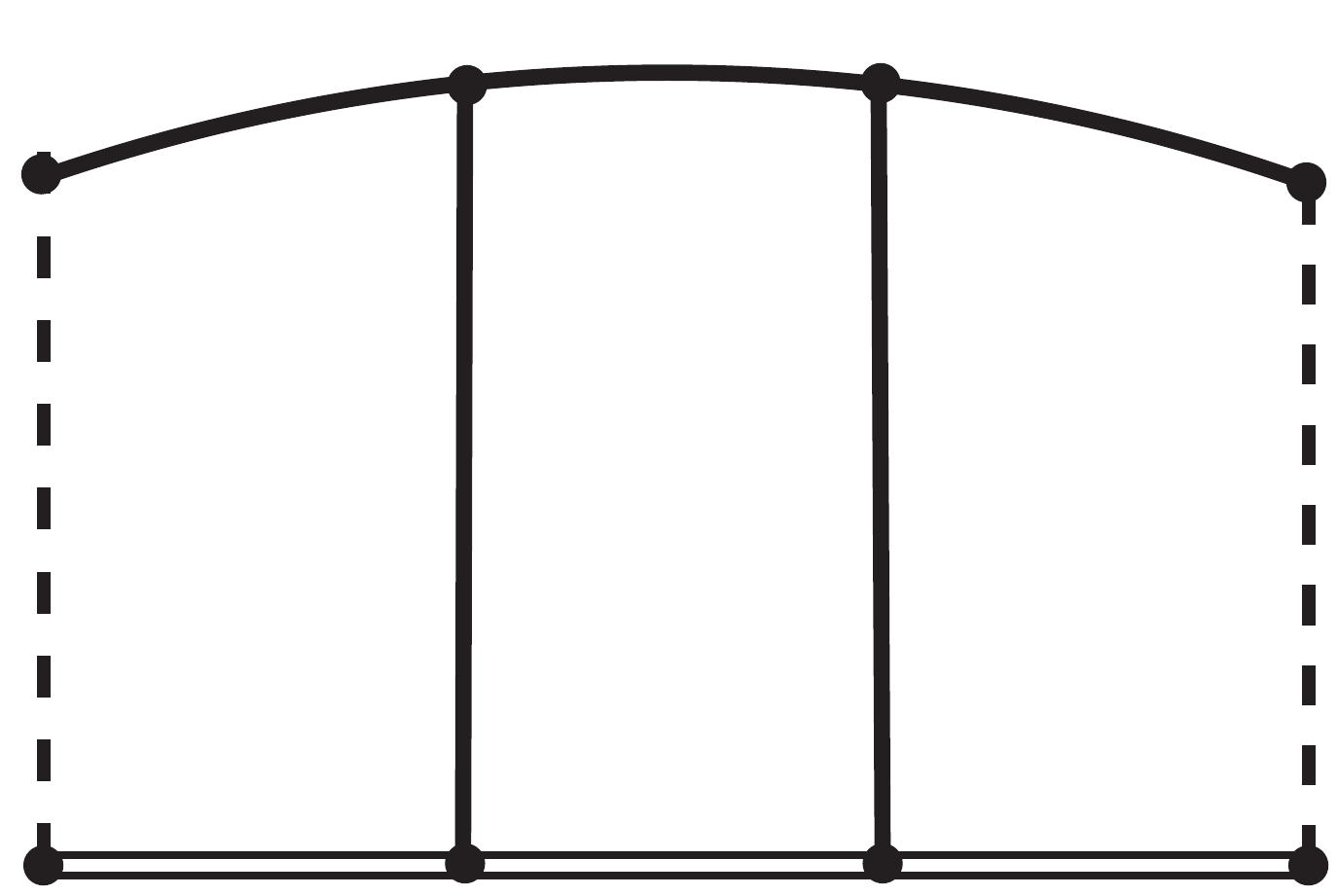}
\label{fig:MI3L20}} \hfill
\subfigure[]{\includegraphics[width=3.3cm, height=2cm]{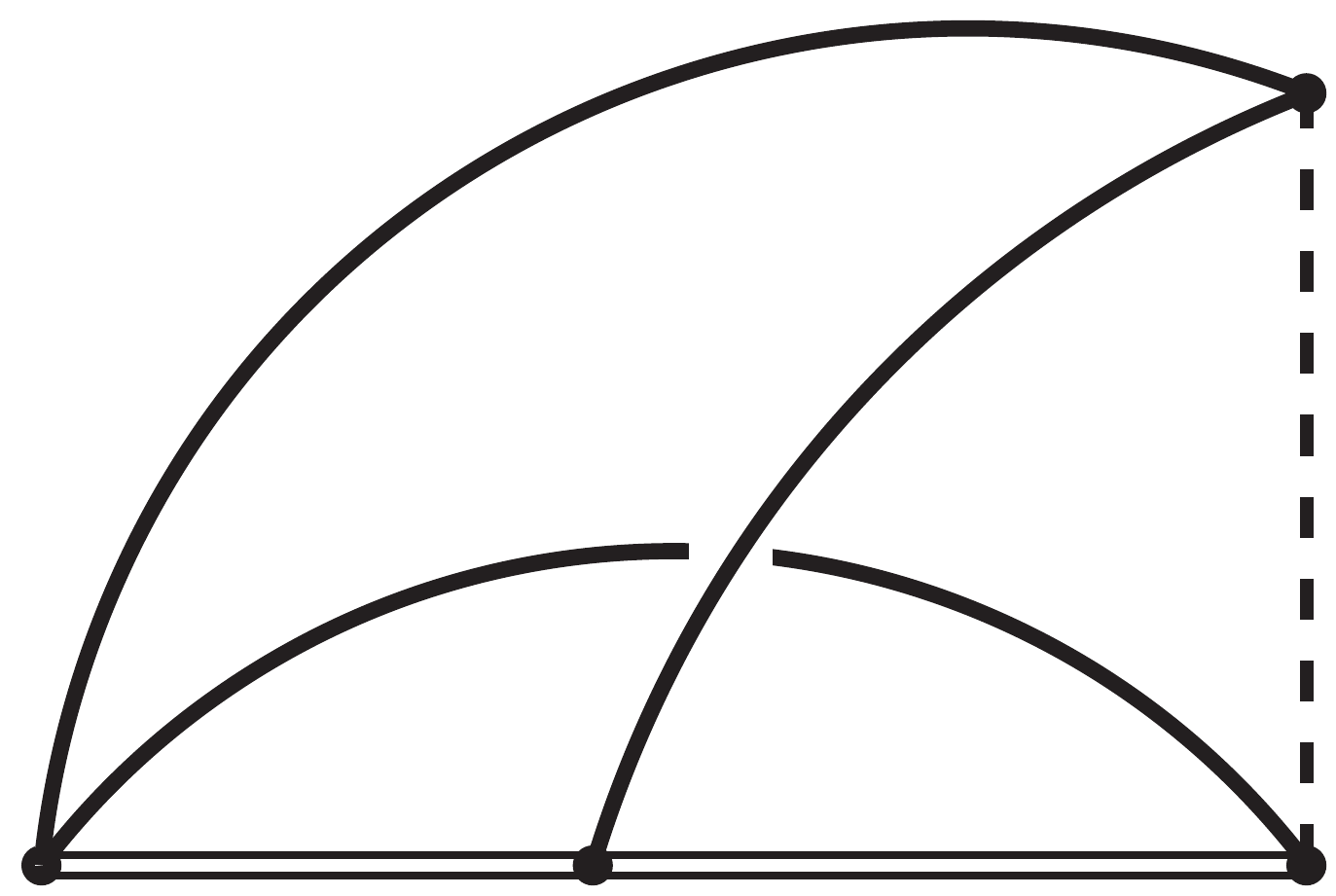}
\label{fig:MI3L21}}
\caption{Diagrammatic representation of the 20 three-loop MIs as described in \fig{MI2L}.}
\label{fig:MI3L}%
\end{figure*}
\subsection{Three-loop integrals}
\label{subsec:three-loop_result}
The vast majority of the three-loop MIs can be mapped to the planar family
\begin{equation}
G(\nu_1, \ldots, \nu_{12})\equiv e^{3\varepsilon \gamma_E}\!\! \int \frac{\dd^d k_1}{i\pi^{d/2}} \frac{\dd^d k_2}{i\pi^{d/2}}\frac{\dd^d k_3}{i\pi^{d/2}} \frac{1}{D_1^{\nu_1} \cdots D_{12}^{\nu_{12}}}\, , \label{eq:def_planar_topo}
\end{equation}
with (propagator) denominators
\begin{align}
\label{eq:Ds3loop}
D_1 = &-\! n \cdot k_1\,, & D_2 = &- \!n \cdot k_2\,, &D_3 = &- \!n \cdot k_3\,, \\
D_4 = &-\!2v\cdot k_1+1 & D_5 = & -\!2v\cdot k_2+1\,, & D_6 = & -\!2v\cdot k_3+1\,, \nn\\
D_7 = &-\! k_1^2\,, & D_8 = & -\! k_2^2\,, &D_9 = &-\! k_3^2\,, \nn \\
D_{10} = &-\!(k_1-k_2)^2 \,, & D_{11} = & -\!(k_2-k_3)^2\,, & D_{12} =& -\!(k_3-k_1)^2\,.\nn
\end{align}
\begin{subequations}
Three MIs can be written as products of one- and two-loop MIs:
\begin{align}
\MI{3}{\MItseven} & = G(0,0,1,1,1,1,1,1,0,0,0,1) = \MI{1}{} \times \MI{2}{c}\,,\\
\MI{3}{\MIseven} & = G(0,0,0,1,1,1,1,1,1,0,0,0) = \bigl(\MI{1}{}\bigr)^3\,,\\
\MI{3}{\MIfive} & = G(0,0,0,0,1,1,1,0,1,1,0,0) =
\MI{1}{} \times \MI{2}{a} \,.
\end{align}
The HQET propagator-type MIs were computed in \rcites{Grozin:2000jv,Grozin:2001fw}
\begin{align}
\MI{3}{\MIone} ={}& G(0,0,0,0,0,1,0,1,0,1,0,1)= e^{3 \varepsilon \gamma_E}\, \Gamma^3\bigl(\tfrac{d}{2}-1\bigr) \, \Gamma(7-3d)\,, \label{eq:MI3L1} \\[0.3cm]
\MI{3}{\MItwo} ={}&
G(0,0,0,0,0,1,1,1,0,0,1,1)
= \dfrac{e^{3 \varepsilon \gamma_E}}{3-d} \frac{\Gamma (11-3 d) \Gamma\bigl(\frac{3 d}{2}-6\big) \Gamma^2\bigl(3-\frac{d}{2}\bigr) }{ \Gamma (5-d) \Gamma^2
(d-2)\Gamma^{-4} \bigl(\frac{d}{2}-1\bigr) }\,, \\[0.1cm]
\MI{3}{\MIthree} ={}& G(0,0,0,0,1,1,0,1,1,1,0,1)
= e^{3 \varepsilon \gamma_E} \frac{\Gamma\bigl(2-\frac{d}{2}\bigr) \Gamma^3\bigl(\frac{d}{2}-1\bigr) \Gamma^2\!\bigl(5-\tfrac{3 d}{2}\bigr) }{(3-d) \Gamma (d-2)}
\\
& \!\times\!
\Biggl\{\Gamma \bigl(\tfrac{3 d}{2}-4\bigr)+ \frac{4 \Gamma (10-3 d) \Gamma (7-2 d) }{(3 d-10) \Gamma
\bigl(12-\frac{7 d}{2}\bigr) \Gamma^2\!\bigl(5-\tfrac{3 d}{2}\bigr) } \nn
\pFq{3}{2}{10-3 d,5-\frac{3 d}{2},5-\frac{3 d}{2}}{12-\frac{7 d}{2},6-\frac{3 d}{2}}{1}\!
\Biggr\},\\[0.3cm]
\MI{3}{\MIfour} ={}& G(0,0,0,0,1,1,1,0,0,1,0,1) = e^{3 \varepsilon \gamma_E} \, \frac{\Gamma (8-3 d) \Gamma^2 (3-d) \Gamma^3 \bigl(\frac{d}{2}-1\bigr)}{\Gamma (6-2 d)}\,.
\end{align}
After integrating out the bubble (two-propagator loop), $\MI{3}{\MInine}$ to $\MI{3}{\MItfour}$ can be written in a form similar to $\MI{2}{c}$ but with one non-integer exponent, and computed using Mellin-Barnes (MB) techniques (see \app{MB}):
\begin{align} \label{eq:MI3L09}
\MI{3}{\MInine} &= G(0,0,1,0,1,1,0,1,0,1,0,1) = e^{3 \varepsilon \gamma_E}\Gamma^2\bigl(\tfrac{d}{2}-1\bigr) \Gamma(9-3d)
\\
& \times \Biggl\{ \frac{2 \Gamma \bigl(\frac{d}{2}-2\bigr) \, }{3 d-10}\pFq{3}{2}{1,8-\frac{5 d}{2},5-\frac{3 d}{2}}{6-\frac{3 d}{2},3-\frac{d}{2}}{1} -\frac{\Gamma \bigl(6-\frac{3 d}{2}\bigr) \Gamma^2 (3-d) \Gamma \bigl(\frac{3 d}{2}-5\bigr)}{\Gamma \bigl(8-\frac{5 d}{2}\bigr)}
\nonumber \\
&\quad -\frac{\Gamma (6-2 d) \Gamma (3-d) \Gamma \bigl(2-\frac{d}{2}\bigr) \Gamma \bigl(\frac{d}{2}-2\bigr) \Gamma (2 d-6)}{\Gamma \bigl(8-\frac{5 d}{2}\bigr) \Gamma
(d-4)} \Biggr\}, \nonumber\\[0.3cm]
\MI{3}{\MIten} &= G(0,0,1,0,1,1,1,0,0,1,0,1) =
e^{3\varepsilon \gamma_E} \frac{\Gamma (9\!-\!3d) \Gamma\bigl(6\!-\!\frac{3 d}{2}\bigr) \Gamma (3\!-\!d) \Gamma\bigl(\frac{3 d}{2}\!-\!5\bigr)}{(7-2 d) \Gamma
\bigl(8-\frac{5 d}{2}\bigr)\Gamma^{-2}\bigl(\frac{d}{2}-1\bigr)} \\
& \times \Biggl\{\frac{(d-3) (2 d-7) }{\Gamma\bigl(\frac{3
d}{2}-3\bigr)\Gamma^{-1}\bigl(\frac{d-4}{2}\bigr)} \pFq{3}{2}{4-d,2-\frac{d}{2},\frac{d}{2}-1}{3-\frac{d}{2},\frac{3 d}{2}-3}{1}
-\frac{\Gamma (9-2 d) \Gamma (d-4)\Gamma^2\! \bigl(2-\frac{d}{2}\bigr) }{4 \Gamma\bigl(6-\frac{3 d}{2}\bigr) \Gamma (4-d) \Gamma\bigl(\frac{3 d}{2}-5\bigr)} \qquad\qquad\qquad
\nonumber \\&\quad
+\frac{\Gamma\bigl(10-\frac{5 d}{2}\bigr) \, }{4 \Gamma\bigl(7-\frac{3 d}{2}\bigr)}\pFq{3}{2}{10-\frac{5 d}{2},8-2 d,5-d}{9-2 d,7-\frac{3 d}{2}}{1}
-\frac{(13 d^2-87 d+146) \Gamma\bigl(8-\frac{5 d}{2}\bigr) }{2 \Gamma \bigl(1-\frac{d}{2}\bigr) \Gamma (d-1) \Gamma^{-1} (2 d-7)}\Biggr\} \,, \nonumber\\[0.5cm]
\MI{3}{\MItone} &= G(0,0,1,0,1,1,1,0,0,1,1,0) = e^{3 \varepsilon \gamma_E} \, \Gamma (9-3 d) \Gamma \bigl(2-\tfrac{d}{2}\bigr) \Gamma^3
\bigl(\tfrac{d}{2}-1\bigr)
\\
& \times \Biggl\{\frac{2}{(d-3) (3 d-10) \Gamma\bigl(2-\frac{d}{2}\bigr)} \pFq{3}{2}{1,7-2 d,5-\frac{3 d}{2}}{6-\frac{3
d}{2},4-d}{1}
+ \frac{\Gamma^2 (3-d)\Gamma (d-2)}{\Gamma (7-2 d) \Gamma
\bigl(\frac{d}{2}-1\bigr)} \nonumber \\
&\quad -\frac{\Gamma\bigl(6-\frac{3 d}{2}\bigr) \Gamma
\bigl(2-\frac{d}{2}\bigr) \Gamma\bigl(\frac{3 d}{2}-5\bigr)}{\Gamma
(7-2 d) \Gamma (d-2)}
\Biggr\}, \nonumber \\[0.3cm]
\MI{3}{\MItfour} &= G(0,0,1,1,1,0,0,1,0,0,1,1) = e^{3 \varepsilon \gamma_E} \, \Gamma \bigl(9-3 d \bigr) \Gamma^3 \bigl(\tfrac{d}{2}-1 \bigr) \\
& \times \Biggl\{ \dfrac{2}{(d-4)^2} \, \pFq{3}{2}{1,4-d,8-\frac{5 d}{2}}{3-\tfrac{d}{2},5-d}{1}-
\dfrac{\Gamma (8-2d)\Gamma^2\bigl(2-\tfrac{d}{2} \bigr) \Gamma(-7+2d)}{\Gamma\bigl(8-\tfrac{5d}{2}\bigr)\Gamma\bigl(-3+\tfrac{3d}{2}\bigr)}
\nonumber \\
&\quad +\frac{\Gamma\bigl( 4-\tfrac{3}{2}d \bigr) \Gamma(4-d) \Gamma\bigl(2-\tfrac{d}{2}\bigr)\Gamma(d-3)}{\Gamma\bigl(8-\tfrac{5d}{2}\bigr)\Gamma\bigl(\tfrac{d}{2}-1\bigr)} \Biggr\}. \nonumber
\end{align}\\[0.2cm]
The $\MI{3}{\MIeight}$ can be computed using the MB approach, but requires a more careful treatment.
This is explained in some detail in \app{FeynParam}:
\begin{flalign}
\MI{3}{\MIeight} &= G(0,0,1,0,1,0,1,0,0,1,1,1)= \frac{3}{2}e^{3 \varepsilon \gamma_E} \Gamma (9-3 d) \Gamma\bigl(\tfrac{d}{2}-2\bigr) (d-4) \label{eq:MI3L08}
\\
& \times \Biggl\{\Gamma^2\! \bigl(\tfrac{d}{2}-2\bigr) \,
\pFq{3}{2}{1,\frac{d}{2}-1,d-3}{3-\tfrac{d}{2},d-2}{1}+2 \dfrac{\Gamma (4-d)}{d-4}\Gamma (d-2) \Gamma \bigl(2-\tfrac{d}{2}\bigr) \Gamma \bigl(\tfrac{3
d}{2}-5\bigr)\Biggr\}. \nonumber &&
\end{flalign}\\[0.2cm]
For the ($\eps$ expansion of the) remaining planar MIs we employed
the method outlined in \app{finite}:
\begin{align}
\MI{3}{\MIsix} ={}& G(0,0,0,1,1,1,0,1,1,1,0,1) = -\frac{\pi ^2}{18 \varepsilon^2} +\frac{1}{\varepsilon} \biggl( \frac{\zeta_3}{3}-\frac{4 \pi ^2}{9} \biggr) \\ &
+ \frac{8 \zeta_3}{3}-\frac{26 \pi ^2}{9}-\frac{49 \pi ^4}{1080} +\varepsilon \biggl(\frac{52 \zeta_3}{3}-\frac{11 \pi ^2 \zeta_3}{36}-\frac{83 \zeta_5}{3}-\frac{160 \pi
^2}{9}-\frac{49 \pi ^4}{135}\biggr)\nonumber \\
&+ \varepsilon^2 \biggl(\frac{320\zeta_3}{3}-\frac{22 \pi ^2\zeta_3}{9}+\frac{130 \zeta_3^2}{3}-\frac{664 \zeta_5}{3}-\frac{968 \pi ^2}{9}-\frac{637 \pi ^4}{270}+\frac{7739 \pi ^6}{25920}\biggr)+\mathcal{O}(\varepsilon^3)\,, \nonumber \\[0.3cm]
\MI{3}{\MItnine} ={}& G(0,1,1,1,1,1,1,0,0,1,0,1) = -\frac{1}{24 \varepsilon^5}-\frac{1}{4 \varepsilon^4} - \frac{1}{\varepsilon^3} \biggl( \frac{3}{2}+\frac{71 \pi ^2}{288} \biggr) \\
& + \frac{1}{\varepsilon^2}\biggl(\frac{25 \zeta_3}{8}-9-\frac{71 \pi ^2}{48} \biggr) + \frac{1}{\varepsilon}\biggl( \frac{75 \zeta_3}{4}-54-\frac{71 \pi ^2}{8}-\frac{1999 \pi ^4}{2304} \biggr)+\frac{225 \zeta_3}{2}\nonumber \\
& +\frac{4285 \pi ^2 \zeta_3}{288}+\frac{4171 \zeta_5}{40}-324-\frac{213 \pi ^2}{4}-\frac{1999
\pi ^4}{384} + \varepsilon \biggl( 675 \zeta_3+\frac{4285 \pi ^2 \zeta_3}{48}
\nonumber \\
&-\frac{5825 \zeta_3^2}{48}+\frac{12513 \zeta_5}{20}-1944-\frac{639
\pi ^2}{2}-\frac{1999 \pi ^4}{64} -\frac{22852649 \pi ^6}{8709120} \biggr) + \mathcal{O}(\varepsilon^2)\,, \nonumber\\[0.3cm]
\end{align}
\newpage
\begin{align}
\MI{3}{\MItthree} ={}& G(0,0,1,0,1,1,1,1,0,1,0,1) = -\frac{1}{16 \varepsilon^4}-\frac{5}{8 \varepsilon^3}-\frac{1}{\varepsilon^2}\biggl(\frac{19}{4}+\frac{125 \pi ^2}{576} \biggr) \\
& +\frac{1}{ \varepsilon} \biggl( \frac{209 \zeta_3}{48}-\frac{65}{2}-\frac{625 \pi ^2}{288}-\frac{11 \pi ^4}{540} \biggr)+ \frac{1045 \zeta_3}{24}-\frac{7 \pi ^2 \zeta_3}{18}+\frac{13 \zeta_5}{2} -211 \nonumber \\
&-\frac{2375 \pi ^2}{144} -\frac{41971 \pi ^4}{69120}+ \varepsilon \biggl( \frac{3971 \zeta_3}{12}+\frac{7055 \pi ^2 \zeta_3}{576}+\frac{25 \zeta_3^2}{2}+\frac{8673 \zeta_5}{80} -1330
\nonumber && \\
&-\frac{8125 \pi ^2}{72} -\frac{192959 \pi ^4}{34560}-\frac{73 \pi ^6}{630} \biggr) + \mathcal{O} (\varepsilon^2)\,, \nonumber \\[0.2cm]
\MI{3}{\MIteight} ={}& G(0,1,1,1,0,1,1,1,0,1,1,1) = \frac{1}{3 \varepsilon^3}+\frac{1}{\varepsilon^2} \biggl( \zeta_3-4+\frac{\pi ^2}{9} \biggr) + \frac{1}{\varepsilon}\biggl(\frac{97}{3}-\frac{23 \zeta_3}{3} \\
&
-\frac{7 \pi ^2}{36}-\frac{73 \pi ^4}{540}\biggr) -\frac{41 \pi ^2 \zeta_3}{12}+31 \zeta_3+\frac{71
\pi ^4}{90}-\frac{26 \pi ^2\!}{9} -220 + \frac{394 \zeta_5}{3} + \varepsilon \biggl( \frac{253 \zeta_3^2}{3} \nonumber \\ &-\frac{215 \zeta_3}{3}-\frac{151 \zeta_3 \pi ^2\!}{12}-\frac{1118 \zeta_5}{3}+\frac{4081}{3} +\frac{1337 \pi ^2\!}{36}-\frac{3913 \pi ^4\!}{1440}-\frac{15697 \pi ^6}{19440} \biggr) + \mathcal{O}(\varepsilon^2)\,, && \nonumber \\[0.2cm]
\MI{3}{\MItsix} ={}& G(0,0,1,1,1,1,0,1,0,1,0,1) = -\frac{1}{96 \varepsilon^4}-\frac{1}{8 \varepsilon^3}-\frac{1}{\varepsilon^2} \biggl( \frac{17}{16}+\frac{71 \pi ^2}{1152} \biggr) \\
& + \frac{1}{\varepsilon} \biggl( \frac{37 \zeta_3}{32}-\frac{125}{16}-\frac{197 \pi ^2}{288} \biggr) + \frac{293 \zeta_3}{24}-\frac{851}{16}-\frac{3205 \pi ^2}{576}-\frac{2159 \pi ^4}{9216} \nonumber \\
&+\varepsilon\biggl( \frac{4621 \zeta_3}{48}+\frac{6985 \pi ^2 \zeta_3}{1152}+\frac{4111 \zeta_5}{160}-\frac{5537}{16}-\frac{22945 \pi
^2}{576}-\frac{85283 \pi ^4}{34560} \biggr) \nonumber
\\ &+ \varepsilon^2 \biggl( \frac{32425 \zeta_3}{48}+\frac{17587 \pi ^2 \zeta_3}{288}-\frac{9761 \zeta_3^2}{192}+\frac{11173 \zeta_5}{40}-\frac{35027}{16}-\frac{153583 \pi ^2}{576} \nonumber \\
& -\frac{1342963 \pi ^4}{69120}-\frac{3853231 \pi ^6}{4976640} \biggr) + \mathcal{O}(\varepsilon^3) \nonumber\, ,\\[0.2cm]
\MI{3}{\MIttwo} ={}& G(0,0,1,0,1,1,1,1,0,0,1,1) = -\frac{1}{48 \varepsilon^4} -\frac{1}{6 \varepsilon^3} - \frac{1}{\varepsilon^2} \biggl( \frac{13}{12}+\frac{29 \pi ^2}{192}\biggr) \\
& + \frac{1}{\varepsilon} \biggl( \frac{103 \zeta_3}{48}-\frac{20}{3}-\frac{29 \pi ^2}{24} \biggr)+ \frac{103 \zeta_3}{6}-\frac{121}{3}-\frac{2639\pi^4}{4608}-\frac{377\pi^2}{18} \nonumber + \varepsilon \biggl( \frac{1339 \zeta_3}{12} \\
& +\frac{825 \pi ^2 \zeta_3}{64} +\frac{15373 \zeta_5}{240}-\frac{728}{3}-\frac{145 \pi
^2}{3}-\frac{2639 \pi ^4}{576} \biggr) + \varepsilon^2 \biggl( \frac{2060 \zeta_3}{3}+\frac{825 \pi ^2 \zeta_3}{8} \nonumber\\
& -\frac{3803 \zeta_3^2}{32} +\frac{15373 \zeta_5}{30}-\frac{4372}{3}-\frac{3509 \pi ^2}{12} -\frac{34307 \pi ^4}{1152}-\frac{305777 \pi ^6}{165888} \biggr) + \mathcal{O}(\varepsilon^3)\,, \nonumber \nonumber \\[0.2cm]
\MI{3}{\MItwenty} ={}& G(0,1,1,1,1,1,1,1,1,1,0,1) = \frac{2}{3 \varepsilon^3}+\frac{1}{\varepsilon^2}\biggl( \frac{5 \pi ^2}{9}-\frac{2 \zeta_3}{3}-\frac{20}{3}\biggr)
\\ &
+\dfrac{1}{\varepsilon} \biggl( \frac{134}{3}-28 \zeta_3-\frac{43 \pi ^2}{18}+\frac{\pi ^4}{5}\biggr) + \frac{478 \zeta_3}{3}-\frac{23 \pi ^2 \zeta_3}{2}-62 \zeta_5-\frac{752}{3}
+7 \pi ^2
\nonumber
+\frac{55 \pi ^4}{36} \\&
+ \varepsilon \biggl(\!\frac{3818}{3} -\frac{2126 \zeta_3}{3}-\frac{334 \pi ^2 \zeta_3}{9}
+\frac{890 \zeta_3^2}{3}-\frac{1318 \zeta_5}{3}
-\frac{355 \pi ^2\!}{18}
-\frac{6217 \pi ^4\!}{720}
+\frac{41317 \pi ^6\!}{34020} \biggr) \nn
\\ & + \mathcal{O}(\varepsilon^2)\,. \nn
\end{align}
The result for $\MI{3}{\MIsix}$ can also be found in \rcite{Grozin:2001fw} to $\mathcal{O}(\varepsilon^0)$.\\
The only non-planar MI can also be computed with the MB method:
\begin{flalign}
\MI{3}{\MIttone}
&= \int \! \frac{\dd^d k_1}{i\pi^{d/2}} \frac{\dd^d k_2}{i\pi^{d/2}}\frac{\dd^d k_3}{i\pi^{d/2}} \dfrac{e^{3\varepsilon \gamma_E}}{(-n\cdot k_3)(-2v\cdot k_1+1)[-2v\cdot (k_1-k_2+k_3)+1](-k_2^2)} && \\ & \times
\dfrac{1}{[-(k_1-k_2)^2][-(k_2-k_3)^2]} =
e^{3\varepsilon \gamma_E}
\frac{2 \Gamma (9-3 d) \Gamma^3 \bigl(\frac{d}{2}-1\bigr) \, }{(d-4)^2}\pFq{3}{2}{1,6-\frac{3 d}{2},4-d}{5-d,3-\frac{d}{2}}{1}. \nonumber \! \! \! \! &&
\end{flalign}
\end{subequations}

\section{Computation of three-loop master integrals}
\label{app:computationMI}
In this appendix we sketch some of the strategies used for the calculation of the two- and three-loop MIs identified in \sec{computation}, the results of which are given in \app{calB_and_MI}.
We perform all computations in the Feynman parameter representation of the integrals (see e.g.\ \rcite{weinzierl2022feynmanintegrals}). Some of the (lower-loop) MIs can be evaluated by straightforward integration over the Feynman parameters (in a suitable order).
Others require additional techniques, which we will briefly discuss using $\MI{3}{\MInine}$ [\,\fig{MI3L09}\,] and $\MI{3}{\MIeight}$ [\,\fig{MI3L08}\,] as examples.

\subsection{Manipulating Feynman parameters}
\label{app:FeynParam}
$\MI{3}{\MIeight}$ is an interesting case because the respective Feynman parameter integrals can be manipulated in a particularly constructive way. Exploiting the projective nature of the Feynman parameter representation we set the parameter associated with the heavy quark propagator to $1$. After performing a straightforward integration, we have
\begin{align}
\MI{3}{\MIeight}=& e^{3\varepsilon \gamma_E} \Gamma (1-\varepsilon ) \Gamma (-1+4 \varepsilon) \int^1_0\!\! \mathd x \int^1_0\!\! \mathd y \int^1_0\!\! \mathd z \int^1_0\!\! \mathd t\, t^{-\varepsilon } x^{-2+3 \varepsilon} y^{-\varepsilon } z^{-\varepsilon }
\\ &
\times
\frac{\bigl[t (2 x yz-2xz-2 x y+x-y z+y)+x (y-zy+z)\bigr]^{1-4 \varepsilon }}{\bigl[t y (3 z-2)-2
t z+t-2 y z+y+z\bigr]^{3-5\varepsilon}}\,, \nn
\end{align}
where we have mapped the integrals from $(0,\infty)$ to $(0,1)$ for convenience. The $z$ integral can be transformed into the form of the $_2F_1$ integral in \eq{hypergeometric}.
After that the $t$ integral can be carried out and we are left with
\begin{equation}
\MI{3}{\MIeight}=e^{3\varepsilon \gamma_E}
\dfrac{\Gamma^3(1-\varepsilon) \Gamma (\varepsilon )}{\Gamma (3-5 \varepsilon )}
\int^1_0\! \mathd x \int^1_0\! \mathd y \int^1_0\! \mathd z\,
\dfrac{(1-y)^{-3+6 \varepsilon} (1-z)^{2-5 \varepsilon } z^{-2+4 \varepsilon} }{x^{2 \varepsilon } y^{2 \varepsilon } (1-y z)^{\varepsilon }(x+y z-xyz)^{1-\varepsilon}}\,.
\end{equation}
We can now do the following substitutions (shown schematically):
\begin{equation}
\int^1_0\!\! \mathd z\! \int^1_0\!\! \mathd y
\xrightarrow{y=y'/z}
\int^1_0\!\! \mathd z \!\int^z_0\!\! \mathd y'
\xrightarrow{}
\int^1_0\!\! \mathd y'\! \int^1_{y'}\!\! \mathd z
\xrightarrow{y''=1-y'\!, \,z=(1-y'')z'}
\!\int^1_0\!\! \mathd y''\! \int^1_{0}\!\! \mathd z'\,.
\end{equation}
Integration over $z'$ yields
\begin{equation}\label{eq:3lg}
\MI{3}{\MIeight}=
e^{3\varepsilon \gamma_E}\dfrac{\Gamma^3 (1-\varepsilon) \Gamma(-2+6\varepsilon)}{\varepsilon }
\int^1_0\! \mathd x \int^1_0\! \mathd y \, x^{-2 \varepsilon } y^{-2 \varepsilon } (x+y-xy)^{-1+\varepsilon}\,.
\end{equation}
The double integral on the RHS is finite. The integrand can thus be safely expanded and integrated order by order in $\eps$.
To obtain the expression in \eq{MI3L08} valid for arbitrary $d$ one can employ the Mellin Barnes transform (with $A=x$, $B=y-xy$ and $\lambda = 1-\varepsilon$) following \app{MB}.

For other integrals, such us as $\MI{3}{\MInine}$, the Feynman parameter integrations can be consecutively and straightfowardly (in a suitable order) integrated
until only two parameter integrals remain without any special manipulation:
\begin{equation}
\MI{3}{\MInine}=
e^{3\varepsilon \gamma_E} \Gamma^3
(1-\varepsilon) \Gamma (-3+6 \varepsilon)
\int^\infty_0\!\! \mathd x \int^\infty_0\!\! \mathd y \,(1+x)^{-2+2 \varepsilon} (1+y)^{2-5 \varepsilon} (1+x+y)^{-1+\varepsilon}. \label{eq:conan_starting_point}
\end{equation}
However, in this case the two-dimensional integral is divergent in the $\eps \to 0$ limit, which prevents a direct order-by-order in $\eps$ evaluation. To obtain an analytic expression other methods like the ones presented in the following have to be applied.

\subsection{Mellin Barnes transform}
\label{app:MB}

For some MIs, like $\MI{2}{c}$, $\MI{3}{\MIttone}$ or (as we have seen) $\MI{3}{\MIeight}$ and $\MI{3}{\MInine}$,
Feynman parameters can be (more or less directly) integrated up to two parameters. The remaining integrals may be carried out with Mellin-Barnes (MB) techniques \cite{mellin1895om,barnes1901vi,barnes1907,barnes1908}
(for a review see \rcite{dubovyk2022mellinbarnesintegralsprimerparticle}).
They are based on the equality
\begin{equation}
(A+B)^{-\lambda}
=
\int_{\cal C} \frac{\mathd s}{2 \pi i} \frac{ \Gamma(s) \Gamma(\lambda-s)}{\Gamma(\lambda)} A^{s-\lambda}B^{-s}\,,
\label{eq:MB}
\end{equation}
where $A$ and $B$ may be polynomials in the Feynman parameters, $\lambda\equiv\lambda(\varepsilon)$ and $\cal C$ is any contour that separate the poles of $ \Gamma(s)$ and $\Gamma(\lambda-s)$. We can choose $A$ and $B$ so that the integral in the two parameters is trivial at the cost of introducing an integral in the complex plane, which can often be computed analytically for any dimension
using Cauchy's residue theorem.

Applying \eq{MB} with $A=x$, $B=(1+y)$ and $\lambda = 1-\varepsilon$ to the factor $(1+x+y)^{-1+\varepsilon}$ in \eq{conan_starting_point} allows integrating over the Feynman parameters:
\begin{equation}
\MI{3}{\MInine} =e^{3\varepsilon \gamma_E} \frac{\Gamma^2 (1-\varepsilon) \Gamma (-3+6 \varepsilon)}{\Gamma (2-2 \varepsilon)}
\int_{\cal C} \frac{\mathd s}{2 \pi i} \frac{\Gamma(s) \Gamma(1-\varepsilon -s) \Gamma (\varepsilon+s) \Gamma (2-3 \varepsilon-s)}{s-3+5 \varepsilon}\,.
\label{eq:to_solve_by_residues}
\end{equation}
The packages \texttt{MB}~\cite{Czakon:2005rk}, \texttt{MBresolve}~\cite{Smirnov:2009up} and \texttt{barnesroutines}~\cite{barnesroutines} allow to rewrite the integral in \eq{to_solve_by_residues} as another integral with the same integrand, but evaluated along a given straight line $s=c+i\,\text{Im}(s)$ [\,with $c \in (0,1)$\,] plus additional terms without any integral involving $\eps$-dependent Gamma functions.
The remaining integral can be computed using Cauchy's residue theorem, and
upon analytic evaluation of the infinite sum of
simple poles (either to the left or to the right of the straight line), we arrive at the result in \eq{MI3L09}.
Note that although it is possible to use the MB representation for more complicated integrals (e.g.\ when more than two Feynman parameters are involved), evaluating the sum over residues can be substantially more difficult since the poles in general depend on the Mellin-Barnes integration variables.

\subsection{Integration by parts}
\label{app:conan}

While dealing with the integration of Feynman parameters, we occasionally end up with expressions of the type
\begin{equation}
J\equiv\!
\int_0^{\infty} \!\!\mathrm{d} z\! \int_0^{\infty}\!\! \mathrm{d} t\,(1+z)^\alpha(1+t)^\beta t^\delta(1+z+t)^\gamma=\!\int_0^1 \!\mathrm{d} x \!\int_0^1 \!\mathrm{d} y\, \frac{(x+y-x y)^\gamma (1-y)^\delta}{y^{2+\beta+\delta+\gamma} x^{2+\alpha+\gamma}}\,,
\label{eq:conan_mapping}
\end{equation}
where $z$ and $t$ are Feynman parameters, and $\alpha$, $\beta$, $\gamma$ and $\delta$ depend linearly on $\varepsilon$. For $\delta=0$, \eq{conan_mapping} is symmetric under the exchange $\alpha\leftrightarrow \beta$. Furthermore, there are overlapping singularities for $x\rightarrow 0$ and $y\rightarrow 0$. It is possible to apply sector decomposition (see \rcite{Heinrich:2008si} for a review) to separate the divergences, but in some cases it is simpler to manipulate the expression as follows. Integrating in $y$, we obtain an hypergeometric function ${}_2F_1$, which we can in turn write as the integral in \eq{hypergeometric}:
\begin{equation}
J =\frac{\Gamma(1+\delta) \Gamma(-1-\beta-\delta-\gamma)}{\Gamma(-\beta) \Gamma(-\gamma)} \int_0^1 \!\mathrm{d} x \! \int_0^1 \!\mathrm{d} y\, \dfrac{(1-y+x y)^{1+\beta+\gamma+\delta} (1-y)^{-1-\gamma}}{y^{1+\beta} x^{3+\alpha+\beta+\delta+\gamma}}\,.
\label{eq:conan_after_hyper_trick}
\end{equation}
After this manipulation, the term $(1-y+xy)^{1+\beta+\gamma+\delta}$ may contain a nested divergence when $x\rightarrow 0$ and $y\rightarrow 1$ depending on the values of the parameters, and the exponents of all remaining factors are modified.
Let us conveniently define
\begin{align}
I(-n+a\varepsilon,-m+b\varepsilon,c,f)\equiv\int_0^1 \!\mathrm{d} x \int_0^1 \!\mathrm{d} y\, \dfrac{(1-y+x y)^{c} (1-y)^{f}}{y^{n-a\varepsilon} x^{m-b\varepsilon}}\,.
\end{align}
Integrating
$x^{-m+b\varepsilon}$ by parts $\ell=\lceil \text{max}(n,m)\rceil$ times we have,
\begin{align}
I(-n+a\varepsilon,-m+b\varepsilon,c,f)=
& \sum_{k=0}^{\ell-1} \frac{(-1)^k \Gamma(1+f)}{(1-m+b \varepsilon)_{k+1}(c+1)_{-k}}\, \frac{\Gamma(1+k-n+a \varepsilon)}{\Gamma(2+k-n+f+a \varepsilon)} \label{eq:conan_after_hyper} \,\\
& +\frac{(-1)^\ell}{(1-m+b \varepsilon)_{\ell}(c+1)_{-\ell}}\, I(\ell-n+a\varepsilon,\ell-m+b\varepsilon,c-\ell,f)\,, \nonumber
\end{align}
where we used Pochhammer symbols \mbox{$(a)_n=\Gamma(a+n)/\Gamma(a)$}. If the integral on the LHS of \eq{conan_after_hyper} has no divergence at $y=1$, which will be the case provided $f+c-m\ge -2$ and $f\ge-1$ in the limit $\varepsilon\to0$,
the integral on the RHS is finite and thus can be safely expanded in $\varepsilon$ before integrating.

We can for instance resume the computation of $\MI{3}{\MInine}$ in \eq{conan_starting_point} with this strategy.
The RHS of \eq{conan_starting_point} is of the form of \eq{conan_mapping} with $\alpha = 2 -5 \varepsilon$, $\beta = -2 + 2\varepsilon $, $\gamma = -1+ \varepsilon$ and $\delta=0$. Hence \eq{conan_after_hyper_trick} translates into
\begin{equation}
\MI{3}{\MInine}=e^{3\varepsilon \gamma_E}\frac{\Gamma (-2+4
\varepsilon )\Gamma (-3+6
\varepsilon) }{ \Gamma (-2+5 \varepsilon) \Gamma^{-2}
(1-\varepsilon )}\,\int^1_0\! \mathd x \int^1_0\! \mathd y\, \frac{ (1-y)^{-\varepsilon } (1-y+xy)^{2-4 \varepsilon }}{y^{3-5 \varepsilon} x^{2-2 \varepsilon}}.
\label{eq:example_after_conan}
\end{equation}%
\newpage

The integral in \eq{example_after_conan} is divergent for $\varepsilon\rightarrow 0$ but can be integrated three times by parts with respect to $x$ using \eq{conan_after_hyper}, yielding

\begin{align}
\MI{3}{\MInine}=&\,e^{3\varepsilon \gamma_E} \frac{\varepsilon (2-19 \varepsilon+107 \varepsilon ^2-294 \varepsilon ^3+312 \varepsilon ^4) }{8 (1-2 \varepsilon )^2
(1-2 \varepsilon-8 \varepsilon ^2)}\Gamma^3(-\varepsilon ) \Gamma (-3+6 \varepsilon)
\\ \nn &
- e^{3\varepsilon \gamma_E}
\frac{2 \Gamma^2(1-\varepsilon ) \Gamma (4 \varepsilon ) \Gamma (-3+6 \varepsilon)}{(1-4 \varepsilon ^2) \Gamma (-2+5 \varepsilon)}
\int^1_0\! \mathd x \int^1_0\! \mathd y\, \frac{(1-y)^{-\varepsilon }+(1-y+xy)^{-1-4 \varepsilon}}{x^{-1-2 \varepsilon}y^{-5 \varepsilon }}\,.
\end{align}
The integral in the last line is already finite and can be expanded and subsequently integrated order by order in $\varepsilon$.
We used the integration by parts method to analytically check the ($\eps$-expanded) results for $\MI{3}{\MInine}$ to $\MI{3}{\MItfour}$ obtained with the MB technique.

\subsection{Quasi-finite integrals}
\label{app:finite}
This method was proposed in \rcite{vonManteuffel:2014qoa} in the context of full QCD Feynman integrals (i.e.\ with only quadratic propagator denominators).
In \rcite{Bruser:2019yjk} it was successfully applied to EFT integrals with the same type of propagator denominators as those in our MIs, see also \rcites{Bruser:2018rad,Liu:2020wmp}.
Here we only briefly review the approach and refer to those references for more details.

We start by identifying quasi-finite integrals in the same sector as the MI we want to compute, i.e.\ where the exact same set of $D_i$ occurs with positive powers in the denominator.
To this end we employ a dedicated routine implemented in \texttt{Reduze2}~\cite{vonManteuffel:2012np}.
By definition the quasi-finite integrals are free of $1/\eps^n$ divergences associated with endpoint singularities of the corresponding Feynman parameter integrals.%
\footnote{In principle, divergences can be present in the pre-factor of the Feynman parameter representation or arise away from the endpoints, but this does not happen in the cases we consider here.}
This may require that the quasi-finite integral is evaluated in $d\pm \Delta$ dimensions, where $\Delta$ is a (small) even integer.
We can however always relate the $d$-dimensional MI to a suitable $(d\pm \Delta)$-dimensional quasi-finite integral
via IBP and dimensional recurrence relations, as described in \sec{computation}.
Besides the MI and its quasi-finite counterpart this linear relation
often includes only lower-sector MIs (i.e.\ with less propagator denominators) which are simpler to compute
(e.g.\ iteratively with the same method).
In case there are yet unknown same-sector MIs involved, one has to repeat the procedure for those choosing different quasi-finite integrals and finally solve the corresponding system of equations.
What remains to be done is to evaluate the quasi-finite integrals.
In order to do so, we can safely expand their integrands to sufficiently high order in $\eps$ and perform the integrations order-by-order analytically using \texttt{HyperInt}~\cite{Panzer:2014caa}.

The described method allows to obtain analytic results for all our MIs to (more than) sufficiently high order in $\eps$.
Returning to the (simple) case of $\MI{3}{\MInine}$, we found an integral in the same sector that is finite in four dimensions:
\begin{equation}
F
\equiv
G(0, 0 , 2 , 0 , 2 , 3 , 0 , 1 , 0 , 1 , 0 , 1)
=
\frac{\pi ^2}{12}-\frac{1}{2}+\varepsilon \biggl(\frac{3}{2}-\frac{3 \zeta_3}{2}-\frac{\pi
^2}{12}\biggr)+\mathcal{O}(\varepsilon^2)\,.
\label{eq:ex_solve_quasi}
\end{equation}
\newpage
IBP reduction yields the relation
\begin{align}
\MI{3}{\MInine}
=&
f_{F}(d) \,F + f_{\MIone}(d) \,\MI{3}{\MIone}\,, \label{eq:Q_IBP}
\end{align}
where $f_{F}$ and $f_{\MIone}$ are rational functions depending on $d=4-2\eps$ of $\ord(\eps^0)$
and $\ord(\eps^{-2})$, respectively.
With \eq{ex_solve_quasi} and \eq{MI3L1} (expanded in $\eps$) we obtain the result for $\MI{3}{\MInine}$ to $\ord(\eps)$ in agreement with \eq{MI3L09}.

\bibliography{BHQET}
\bibliographystyle{JHEP}

\end{document}